# A New Sample of Mid-Infrared Bright, Long-Period Mira Variables from the MACHO Galactic Bulge Fields

Klaus Bernhard[1,3], Stefan Uttenthaler[2], Stefan Hümmerich[1,3]

1) Bundesdeutsche Arbeitsgemeinschaft für Veränderliche Sterne e.V. (BAV)
2) Kuffner Observatory, Johann Staud-Straße 10, 1160 Wien, Austria
3) American Association of Variable Star Observers (AAVSO)

September 2016

**Abstract:** *Mid-infrared bright objects in the direction of the Galactic Bulge were investigated using time series photometry from the MACHO data archive, which led to the discovery of a large number of long-period variables. Among these, a total of 192 bona-fide Mira variables was identified, which – to the best of our knowledge – are reported here for the first time. Together with the results from our previous investigations, we thereby bring the number of Mira variables found in the MACHO Galactic Bulge fields to a new total of 1286 stars. Light curves, folded light curves and summary data for all new Mira variables are presented and their properties in colour-colour, period-colour and period-magnitude space are investigated. In agreement with our expectations, the present sample of mid-infrared bright objects is composed mostly of luminous, long-period variables.*

## 1. Introduction

Mira stars, named after their bright prototype Mira ('the wonderful', o Ceti), are radially pulsating variables exhibiting large photometric amplitudes and periods on the order of $80 \lesssim P(d) \lesssim 1000$, although stars with longer periods of up to 2000 days exist [1]. By definition, Mira stars show amplitudes of $\Delta V \geq 2.5$ mag [2], although this cut-off is to some extent arbitrary and results in physically similar stars being classified as semi-regular variables (GCVS-type SRA). Observed light amplitudes in the infrared are much smaller (e.g. $\Delta I \geq 0.8$ or 0.9 mag; $\Delta K \geq 0.4$ mag [3]).

Miras are low- and intermediate-mass stars in an advanced, short-lived (~2 x $10^5$ years) state and populate the tip of the asymptotic giant branch (AGB) [4]. They are late-type stars with spectra indicative of strong molecular absorption features (e.g. TiO, ZrO, CN) and emission lines that result from pulsation-related shock waves. Thus, they are found at spectral types Me (oxygen-rich atmosphere), Ce (carbon-rich atmosphere), or, more rarely, Se (intermediate) [3]. The Mira phase is characterised by heavy mass loss, which is why Miras often show considerable circumstellar extinction because of thick dust shells (e.g. [5]). This holds true especially for the longer-period objects.

Mira variables have been shown to follow a distinct period luminosity relation [6]. Because of their luminosity, they are important distance indicators and tools for investigating Galactic structure (e.g. [7]). It is therefore important to increase the sample size of known Mira variables. Our own efforts in this respect [8,9,10] have led to the identification of 1094 Mira variables in the Massive Compact Halo Object (MACHO) project data archive.

In the present investigation, we report on the discovery of an additional 192 Mira variables in the MACHO data archive that have been found by a different methodological approach. Observations and target selection are described in Section 2, data are analysed in Section 3. Results are presented and discussed in Section 4, and we conclude in Section 5.

## 2. Observations and Target Selection

### 2.1 The Massive Compact Halo Object (MACHO) Project

Aim of the MACHO Project was the search for dark matter in the form of massive compact halo objects, so called 'MACHOs'. To this end, millions of stars were monitored in the Magellanic Clouds and the Galactic Bulge in order to search for gravitational microlensing events caused by the – otherwise invisible – MACHOs [11]. As a by-product, thousands of variable stars were discovered in the resulting photometric data.

Observations were carried out between 1992 and 2000 with the 1.27m Great Melbourne Telescope situated at Mount Stromlo in Australia. Using a dichroic beam-splitter, all observations were taken simultaneously through the non-standard MACHO blue filter (~4500-6300 Å; hereafter MACHO *B*-band) and MACHO red filter (~6300-7600 Å; hereafter MACHO *R*-band) with a combination of eight 2048*2048 CCD cameras [12]. For more information on the MACHO project, the reader is referred to [11,12]. MACHO observations are available online through the MACHO Project data archive[1].

### 2.2 Target Selection

In our previous searches for Mira variables in the MACHO data archive, objects whose MACHO *R*-band light curves show a larger deviation than that of other stars of similar magnitude were selected and visually inspected in order to find suitable candidates. For the present investigation, a different methodological approach was taken. Mid-infrared bright objects in the direction of the Galactic Bulge[2] were chosen using observations from the Wide-field Infrared Survey Explorer (WISE), which surveyed the whole sky in the four infrared bands *W1*, *W2*, *W3*, and *W4*, which are centered at 3.4, 4.6, 12, and 22 μm, respectively [13]. Only objects with *W4* ≤ 4 mag were selected for the construction of an initial sample of candidate variable stars. This cut-off was imposed because these *W4*-bright objects appear as a general 'nuisance' in studies of star-formation regions towards the Galactic bulge (J. Alves, private communication). We also expect that with this cut-off we can identify objects with a clear shell signal that are likely to correspond to dust-veiled AGB stars.

## 3. Data Analysis

Light curves of our candidate stars were downloaded from the MACHO data archive and MACHO instrumental magnitudes were transformed on to the Kron-Cousins system by using equation (2) of [12]. The light curves were visually inspected, which led to the discovery of 1169 clearly variable objects (mostly large amplitude semi-regular and Mira variables). Doubtful cases and stars whose variability is obviously caused by instrumental artifacts (mostly blending issues) were rejected.

In order to separate Miras from semi-regular variables, stars with an amplitude > 2 mag ($R_C$) were selected (cf. [9,10]). Objects exhibiting significant changes in amplitude, mean magnitude and / or period suggesting semi-regularity were subsequently rejected. The sample of Mira variables was cross-matched with the 2MASS Catalog [15], from which we derived astrometric positions and near-infrared color indices. Each object was checked against the VizieR service [16] and the AAVSO International Variable Star Index (VSX; [17]) for any information in variability catalogues about the existence of a Mira star at the given position. Known Mira variables were dropped from the sample. In total, we identified 192 new bona-fide Mira variables.

---

[1] http://macho.anu.edu.au/
[2] Centre coordinates of the MACHO Galactic Bulge fields are found at http://macho.nci.org.au/Macho_fields.html.

## 4. Results

### 4.1 The New Sample

Following the methodology outlined above, a total of 192 Mira variables in the direction of the Galactic Bulge were found in the MACHO data archive. To the best of our knowledge, these Miras are reported here for the first time. Light curves, folded light curves and summary data for all new Mira variables are presented in the Appendix (Table 1 and Figure 6).

### 4.2 Statistical Analyses

In the following subsections, statistical properties of the present sample of mid-infrared bright Mira variables ($N$ = 192) are investigated and compared to the properties of the sample of Mira variables from the MACHO Galactic Bulge fields presented in [8,9,10] ($N$ = 1094). For ease of use, the present sample is referred to in the following as the MIBR (**m**id-**i**nfrared **br**ight) sample.

#### 4.2.1 Period Distribution

A comparison of the period distribution of the MIBR sample ($N$ = 192) with the sample of Mira variables presented in [8,9,10] ($N$ = 1094) is shown in Figure 1 (cf. also Figure 6 in [10]). It becomes obvious that the MIBR sample contains considerably more Miras of longer period, particularly in the range 350 ≤ P(d) ≤ 500. Several Miras of the sample show periods close to one year, which results in poor phase coverage that may pose problematic for standard period search algorithms. However, in our previous searches for Mira variables in the MACHO database [8,9,10], suitable candidates were chosen by visual inspection only. We therefore expect no bias in the discovery of variables with periods close to one year in our previous samples and are confident that the excess of long-period Miras in the MIBR sample is significant and an intrinsic characteristic of the sample.

An excess of long-period Miras is to be expected as the present sample is exclusively made up of mid-infrared bright stars with *W4* ≤ 4 mag (cf. Section 2.2), which correspond to highly reddened objects. It is a well-known fact that long-period (log(P) ≳ 2.5; cf. e.g. [19]) Miras are prone to exhibiting considerable colour excess due to circumstellar dust shells [19]; thus, very red Miras are usually also long-period Miras. This phenomenon is also obvious in the period-colour diagram presented below (cf. Figure 3).

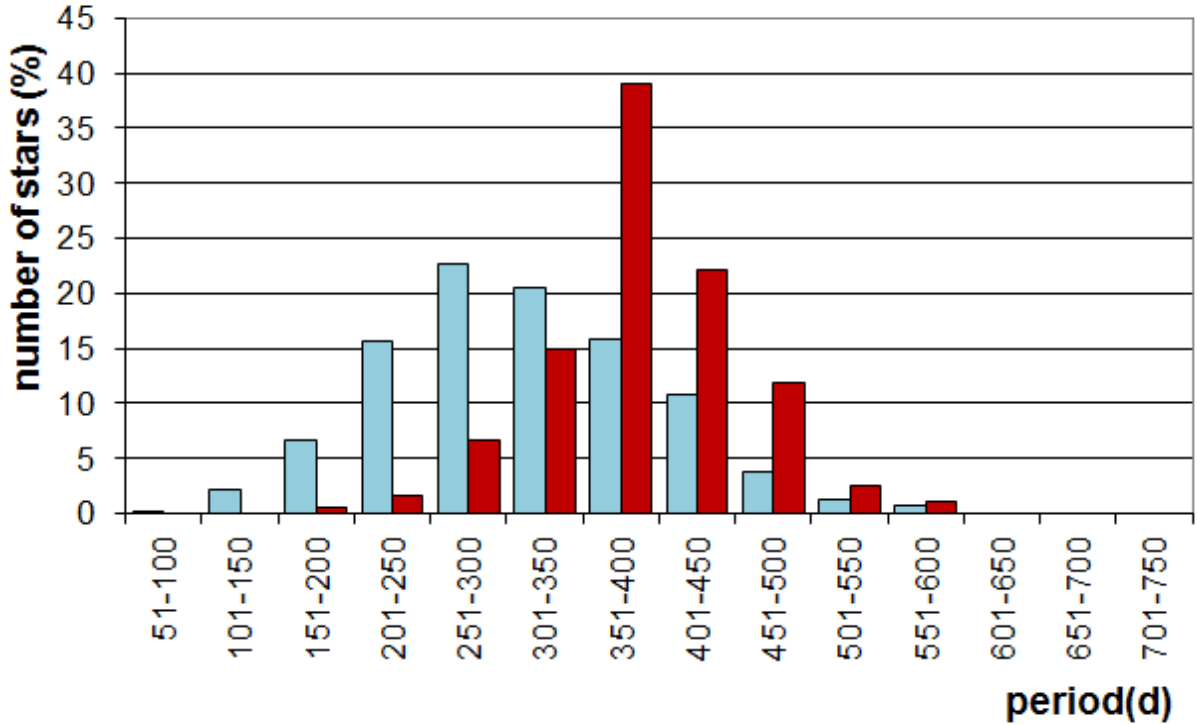

Figure 1 – Period distribution of Mira variables found in the MACHO Galactic Bulge fields, based on the MIBR sample (red; $N = 192$) and the samples presented in [8,9,10] (blue, $N = 1094$).

*4.2.2 Colour-Magnitude Diagram*

A colour-magnitude diagram, based on 2MASS photometry, is presented in Figure 2. 2MASS observations generally consist of six consecutive exposures for a total integration time of 7.8 seconds [15]. Most objects were visited only once during the course of the survey, which also applies to the Miras of the MIBR sample. Some scatter, therefore, would be expected due to the unknown pulsational phase at which the 2MASS observations were taken.

However, the observed amplitudes of Mira variables in the $K_s$-band are relatively small (cf. Section 1), and the scatter introduced by the single-epoch measurements is very small in comparison to the distribution of brightness in Figure 2. This also holds true for the $(H-K_s)$ colour index derived from 2MASS and employed in Figure 3. Thus, the scatter introduced by the single-epoch measurements is negligible in this context and does not preclude us from drawing conclusions from Figures 2 and 3.

Note the 'red tail' of Miras with $(H-K_s) \geq 1$ that extends to $(H-K_s) \sim 1.5$, which is present in both samples. The presence of this feature confirms the findings of [19] (cf. in particular their Figure 7) and is most obvious in the colour-magnitude diagram for the sample of Mira stars from the OGLE-III Catalog of Long-Period Variables (LPVs) in the Galactic Bulge [20], which is presented in Figure 7 of [10]. Apparently, Miras with $(H-K_s) \geq 1$ become fainter in the $K_s$ band with increasing $(H-K_s)$, which is likely caused by circumstellar extinction due to dust (e.g. [21]). As expected, the Miras of the MIBR sample are mostly situated at the red $((H-K_s) \gtrsim 0.60$ mag) and bright $(K_s \lesssim 7.5$ mag) end of the 'main clump' of Mira variables in the diagram.

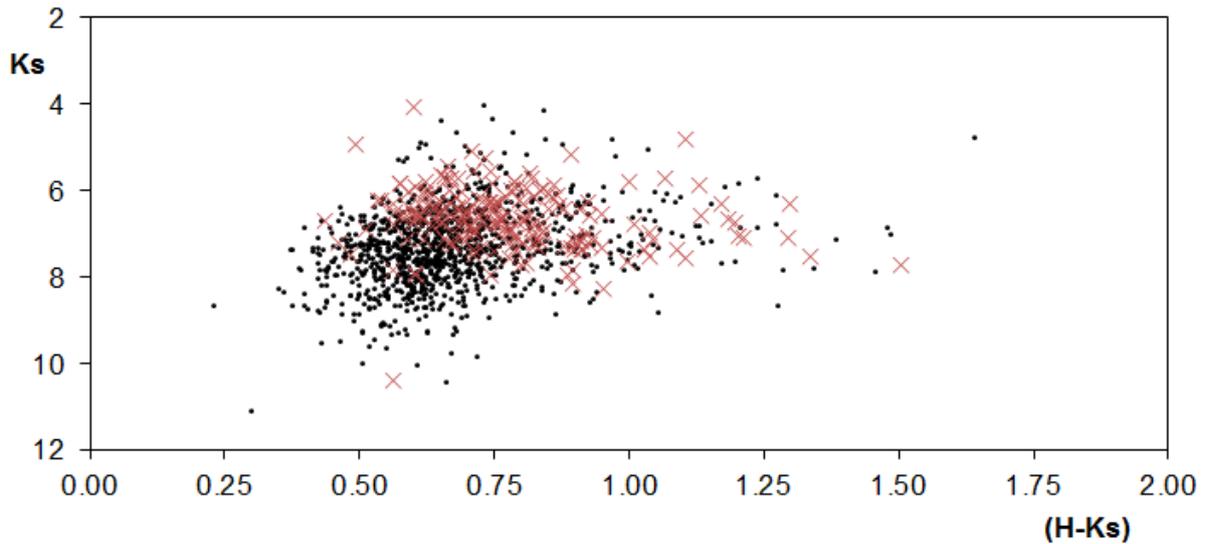

Figure 2 - 2MASS *(H-K$_s$)* vs. *K$_s$* diagram, illustrating the positions of the Mira variables of the MIBR sample (red crosses; *N* = 192) and the samples presented in [8,9,10] (black dots, *N* = 1094).

*4.2.3 Period-Colour Diagram*

A period-colour diagram is given in Figure 3, which confirms that Miras of longer period have larger *(H-K$_s$)* values and hence redder colours. This agrees with earlier findings from OGLE data [19, in particular their Fig. 10], where a significant increase or even a step in *(H-K$_s$)* colour is reported at log(*P*) ~ 2.6. This is likely caused by colour excess due to circumstellar dust that is observed for Miras with periods longer than this value (cf. Section 4.2.1 and [19]). The Miras of the MIBR sample are nearly exclusively found among the long-period objects with log(P) ≳ 2.5.

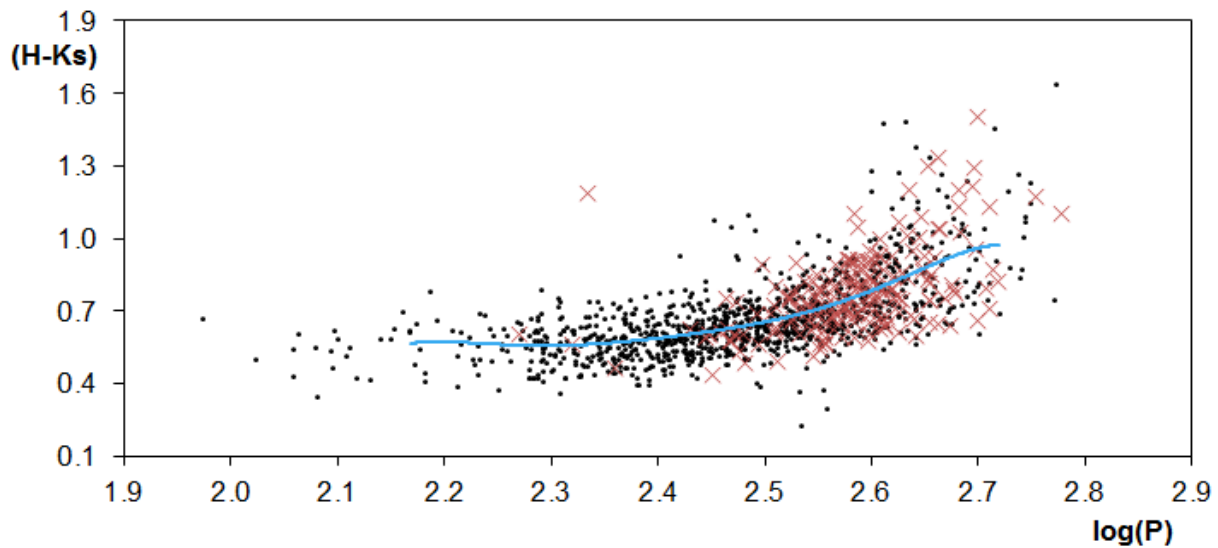

Figure 3 – Log(*P*) vs. 2MASS *(H-K$_s$)* diagram, illustrating the positions of the Mira variables of the MIBR sample (red crosses; *N* = 192) and the samples presented in [8,9,10] (black dots, *N* = 1094). The blue line indicates the moving average. Note the increase in *(H-K$_s$)* colour at log(*P*) ~ 2.6.

*4.2.4 Period-Magnitude Diagram*

Variable red giant stars occupy several well-known sequences in period-luminosity space [22]. Miras and Mira-like semi-regular variables occupy what is commonly referred to as sequence C and are well separated from semi-regular variables of smaller amplitude (like e.g. the so-called OSARG (**O**GLE **S**mall **A**mplitude **R**ed **G**iant) variables). We have investigated the distribution of the MACHO Mira samples in the near-infrared period-magnitude diagram, using 2MASS $K_s$ photometry, which – for red giant stars in the Galactic Bulge – serves as a reasonable proxy for absolute magnitude (Figure 4). Again, as expected, the Miras of the MIBR sample are situated almost exclusively at the bright, long-period end.

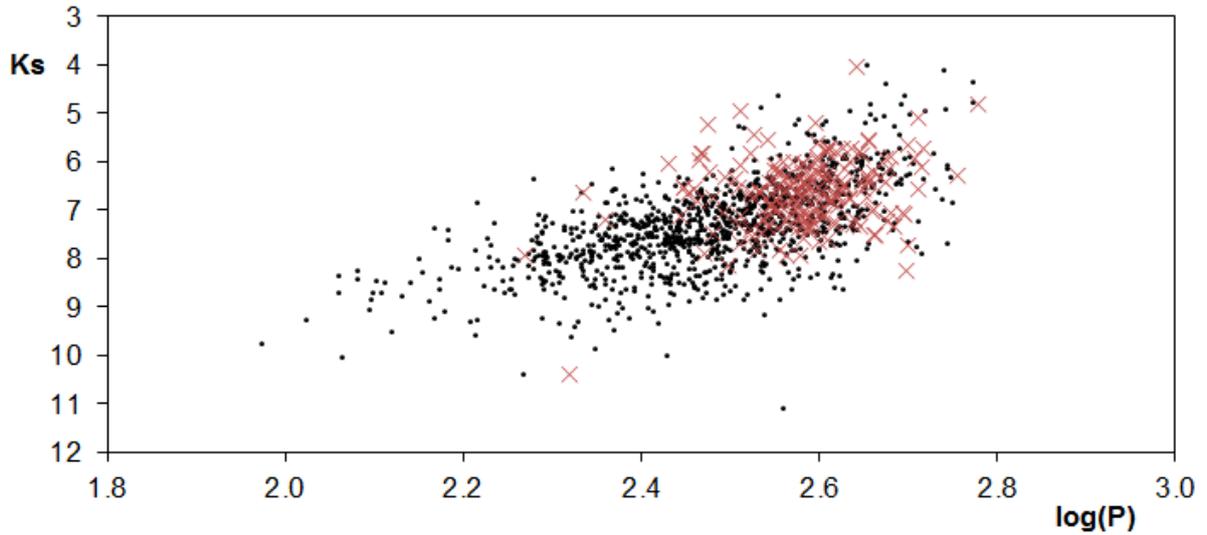

Figure 4 – Log(*P*) vs. 2MASS $K_s$ diagram, illustrating the positions of the Mira variables of the MIBR sample (red crosses; $N = 192$) and the samples presented in [8,9,10] (black dots, $N = 1094$).

Although there is considerable scatter due to reddening (line-of-sight extinction, circumstellar extinction), the general trend is clearly visible and our result is in excellent agreement with the findings in the literature (cf. e.g. [20], especially their Figure 5, and [22]).

At very strong circumstellar extinction, even the near-IR $K_s$ band loses its power as luminosity indicator of red giant stars. In order to reduce the effects of extinction, we have constructed a period-magnitude diagram for the MIBR sample that is based on the reddening-independent Wesenheit index $W_{JK}$ (e.g. [20]), the result of which is shown in Figure 5. The Wesenheit $W_{JK}$ index is defined as

$$W_{JK} = K - 0.686(J - K). \tag{1}$$

This helps to reduce the scatter and narrow down the sequence, although some clear outliers remain that have a severe effect on the linear regression fit. Nevertheless, sequence C is clearly visible in the plot.

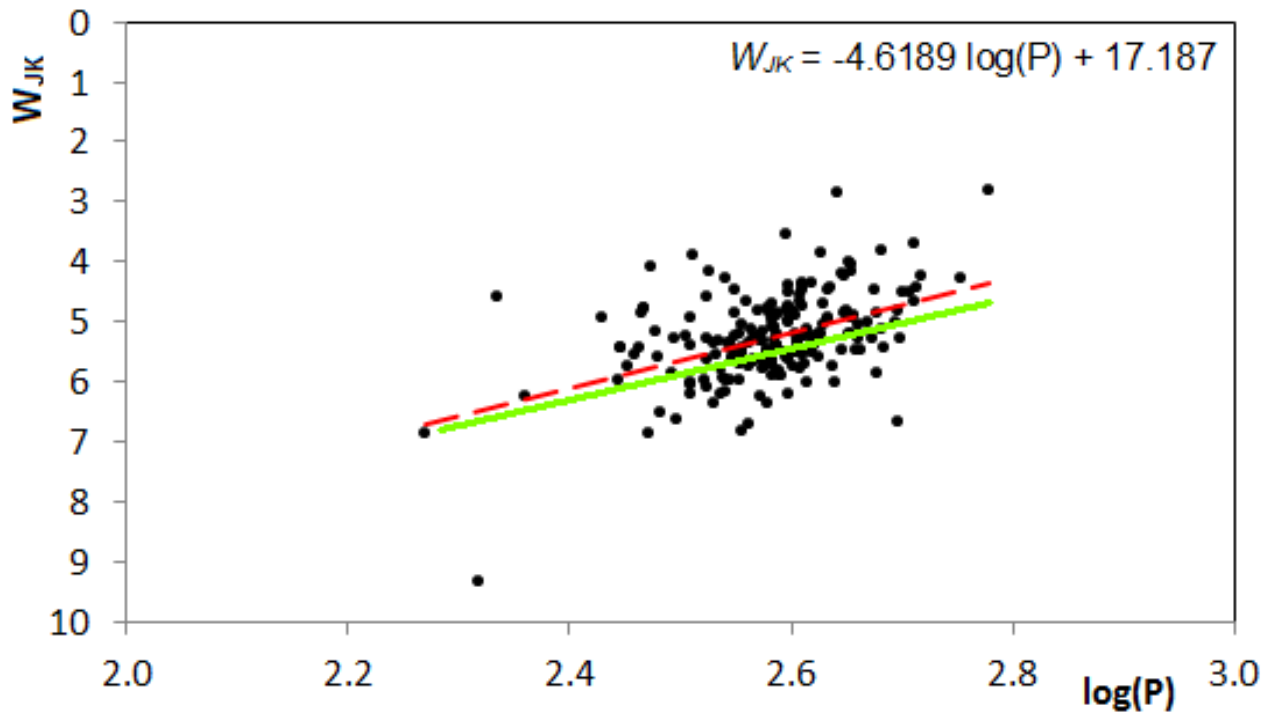

Figure 5 – Log(*P*) vs. $W_{JK}$ diagram, illustrating the positions of the Mira variables of the MIBR sample (red crosses; *N* = 192). The dashed red line is a linear fit to the data; the solution is reproduced in the upper right of the diagram. The green line roughly indicates the position of sequence C based on an approximate fit to the data in Fig. 5 of [20].

## 5. Conclusion

We have investigated mid-infrared bright objects (WISE *W4* ≤ 4 mag) in the direction of the Galactic Bulge using time series photometry from the MACHO data archive, which led to the discovery of a large number of long-period variables. A total of 192 bona-fide Mira variables was found, which have not been included in existing variability catalogues, and which are – to the best of our knowledge – announced here for the first time. Together with the results from our previous investigations [8,9,10], we bring the number of Mira variables found in the MACHO Galactic Bulge fields to a new total of 1286 stars.

We present light curves, folded light curves and summary data for all new Mira variables and investigate their properties in colour-colour, period-colour and period-magnitude space. As expected, the present sample of mid-infrared bright objects is composed mostly of luminous, long-period Miras.

**Acknowledgements**

We thank Prof. João Alves (University of Vienna) for discussions on mid-infrared bright objects in the direction of the Pipe Nebula which provided the stimulus that instigated this research. This paper utilizes public domain data obtained by the MACHO Project, jointly funded by the US Department of Energy through the University of California, Lawrence Livermore National Laboratory under contract No. W-7405-Eng-48, by the National Science Foundation through the Center for Particle Astrophysics of the University of California under cooperative agreement AST-8809616, and by the Mount Stromlo and Siding Spring Observatory, part of the Australian National University. This research has also made use of the SIMBAD and VizieR databases operated at the Centre de Données Astronomiques (Strasbourg) in France. Furthermore, data products from the Two Micron All Sky Survey were employed, which is a joint project of the University of Massachusetts and the Infrared Processing and Analysis Center/California Institute of Technology, funded by the National Aeronautics and Space Administration and the National Science Foundation.

**References**

[1] Whitelock, P., Menzies, J., Feast, M., et al. 1994, MNRAS, 267, 711

[2] Samus, N. N., Durlevich, O. V., Kazarovets, E. V., et al. 2007-2016, General Catalogue of Variable Stars, VizieR On-line Catalog (http://cdsarc.u-strasbg.fr/viz-bin/Cat?B/gcvs)

[3] Whitelock, P. 2012, Ap&SS, 341, 123

[4] Whitelock, P., Marang, F., Feast, M. 2000, MNRAS, 319, 728

[5] Ita, Y., Matsunaga, N. 2011, MNRAS, 412, 2345

[6] Whitelock, P., Feast, M., van Leeuwen, F. 2008, MNRAS, 386, 313

[7] Catchpole, R., Whitelock, P., Feast, M. 2016, MNRAS, 455, 2216

[8] Bernhard, K. 2011, PZP, 11, 12

[9] Huemmerich, S., Bernhard, K. 2012, OEJV, 149, 1

[10] Bernhard, K., Huemmerich, S., 2013, OEJV, 159, 1

[11] Alcock, C., Allsman, R., Alves, D., et al. 1997, ApJ, 486, 697

[12] Alcock, C., Allsman, R., Alves, D., et al. 1999, PASP, 111, 1539

[13] Wright, E.L., Eisenhardt, P.R.M., Mainzer, A.K., et al. 2010, AJ, 140, 1868

[14] Lenz, P., Breger, M. 2005, CoAst, 146, 53

[15] Skrutskie, M. F., Cutri, R. M., Stiening, R., et al. 2006, AJ, 131, 1163

[16] Ochsenbein, F., Bauer, P., Marcout, J. 2000, A&AS, 143, 23 (VizieR)

[17] Watson, C. L. 2006, SASS, 25, 47 (AAVSO International Variable Star Index)

[18] Zijlstra, A., Bedding, T. 2002, JAVSO, 31, 2

[19] Matsunaga, N., Fukushi, H., Nakada, Y. 2005, MNRAS, 364, 117

[20] Soszyński, I., Udalski, A., Szymański, M. K., et al. 2013, AcA, 63, 21

[21] Fraser, O. J. 2008, PhDT, Proquest / UMI, AAT 3328397

[22] Wood, P. R., Alcock, C., Allsman, R. A., et al. 1999, in IAU Symp. 191, Asymptotic Giant Branch Stars, ed. T. Le Bertre, A. Lebre, & C. Waelkens (Cambridge: Cambridge Univ. Press), 151

## Appendix

Table 1: Essential data for the new sample of 192 Mira variables identified in the MACHO data archive. Positional information and *(J-K_s)* indices were derived from the 2MASS catalogue.

| MACHO ID | RA (J2000) | DEC (J2000) | 2MASS ID | J-K_s | P(d) | Epoch (max) | Max (Rc) | Min (Rc) |
|---|---|---|---|---|---|---|---|---|
| 403.47547.34 | 17 54 56.370 | -29 48 33.53 | 17545637-2948335 | 1.639 | 292 | 2451041 | 11.2 | 15.2 |
| 402.47682.580 | 17 55 25.487 | -28 48 10.67 | 17552548-2848106 | 2.046 | 382 | 2451220 | <14.2 | 17.6 |
| 402.47676.192 | 17 55 33.638 | -29 11 30.60 | 17553363-2911305 | 1.837 | 380 | 2451241 | 13.2 | 17.0 |
| 402.47741.111 | 17 55 46.512 | -28 51 10.54 | 17554651-2851105 | 2.154 | 314 | 2451432 | 13.1 | 18.0 |
| 402.47745.888 | 17 55 49.681 | -28 35 14.71 | 17554968-2835147 | 2.650 | 406 | 2450174 | 16.4 | 19.7 |
| 402.47864.438 | 17 56 21.990 | -28 39 59.89 | 17562192-2839598 | 1.825 | 412 | 2450547 | 14.8 | 17.6 |
| 403.47853.1588 | 17 56 26.667 | -29 25 41.71 | 17562666-2925417 | 2.520 | 387 | 2449976 | 15.0 | 17.8 |
| 401.47877.236 | 17 56 26.781 | -27 50 12.92 | 17562678-2750129 | 2.449 | 369 | 2451220 | 16.0 | 19.6 |
| 403.47906.1195 | 17 56 37.073 | -29 52 33.25 | 17563707-2952332 | 2.191 | 417 | 2450907 | 15.0 | 18.9 |
| 403.47909.1283 | 17 56 40.307 | -29 40 37.18 | 17564030-2940371 | 2.264 | 450 | 2449942 | 14.4 | 16.7 |
| 401.48053.1848 | 17 57 09.327 | -28 05 43.88 | 17570932-2805438 | 2.306 | 379 | 2450040 | 15.1 | 18.6 |
| 401.48050.561 | 17 57 09.307 | -28 15 22.60 | 17570930-2815226 | 2.112 | 339 | 2451444 | 14.4 | 17.3 |
| 402.48099.17 | 17 57 31.212 | -29 00 04.53 | 17573121-2900045 | 2.017 | 452 | 2450182 | 13.1 | 15.4 |
| 402.48096.271 | 17 57 32.987 | -29 11 00.12 | 17573298-2911001 | 2.270 | 397 | 2451463 | 14.8 | 17.7 |
| 118.18270.862 | 17 58 56.439 | -30 05 10.29 | 17585643-3005102 | 2.024 | 475 | 2451326 | 14.2 | 17.8 |
| 118.18796.636 | 18 00 05.315 | -29 43 28.64 | 18000531-2943286 | 1.741 | 352 | 2450646 | 12.7 | 16.7 |
| 176.18832.33 | 18 00 17.705 | -27 17 36.55 | 18001770-2717365 | 1.954 | 363 | 2451344 | 14.0 | 16.4 |
| 118.18921.2024 | 18 00 32.809 | -29 59 44.55 | 18003280-2959445 | 2.963 | 461 | 2450345 | 15.5 | 19.5 |
| 176.18955.1440 | 18 00 33.529 | -27 47 05.09 | 18003352-2747050 | 2.506 | 382 | 2450534 | 15.6 | 19.5 |
| 176.19089.44 | 18 00 52.958 | -27 30 19.14 | 18005295-2730191 | 2.260 | 452 | 2449816 | 13.0 | 18.1 |
| 176.19352.3590 | 18 01 16.258 | -27 19 12.75 | 18011625-2719127 | 2.264 | 382 | 2450517 | 15.2 | 18.6 |
| 176.19481.2463 | 18 01 41.482 | -27 22 22.05 | 18014148-2722220 | 2.381 | 401 | 2451310 | 15.8 | 19.5 |
| 104.20384.163 | 18 03 50.474 | -27 50 42.30 | 18035047-2750422 | 1.684 | 359 | 2449054 | <12.9 | 18.0 |
| 120.21002.18 | 18 05 19.463 | -29 56 51.69 | 18051946-2956516 | 1.586 | 360 | 2449061 | 11.8 | 15.8 |
| 120.21395.12 | 18 06 18.306 | -29 44 11.89 | 18061830-2944118 | 1.055 | 374 | 2450259 | 13.1 | >15.7 |
| 121.21382.110 | 18 06 23.626 | -30 36 21.13 | 18062362-3036211 | 1.630 | 405 | 2451305 | 12.0 | 17.0 |
| 179.21577.71 | 18 06 37.262 | -26 19 25.07 | 18063726-2619250 | 1.954 | 342 | 2450505 | 13.5 | 16.3 |
| 121.21517.25 | 18 06 43.807 | -30 18 38.00 | 18064380-3018380 | 1.723 | 410 | 2449156 | 13.0 | 15.3 |
| 124.21640.914 | 18 06 47.617 | -30 47 15.72 | 18064761-3047157 | 1.832 | 408 | 2449109 | 12.1 | 16.8 |
| 179.21712.731 | 18 06 51.813 | -25 59 21.91 | 18065181-2559219 | 2.205 | 430 | 2451305 | 14.1 | 17.7 |
| 124.21632.42 | 18 06 51.930 | -31 16 45.67 | 18065193-3116456 | 1.591 | 410 | 2450247 | 12.5 | 15.9 |
| 120.21784.522 | 18 07 07.436 | -29 49 13.39 | 18070743-2949133 | 2.452 | 380 | 2450018 | 13.7 | 16.7 |
| 121.21777.223 | 18 07 09.804 | -30 17 39.49 | 18070980-3017394 | 1.799 | 366 | 2450664 | 13.2 | 17.0 |
| 124.21894.182 | 18 07 24.097 | -31 10 43.65 | 18072409-3110436 | 1.933 | 391 | 2449843 | 14.4 | >17.8 |
| 120.21911.3312 | 18 07 33.449 | -30 01 00.98 | 18073344-3001009 | 1.593 | 320 | 2449900 | 10.6 | 14.8 |
| 180.22499.1546 | 18 08 45.662 | -25 27 45.86 | 18084566-2527458 | 2.270 | 412 | 2451251 | 15.8 | 18.8 |
| 121.22423.17 | 18 08 48.124 | -30 31 45.03 | 18084812-3031450 | 1.617 | 324 | 2449105 | 11.7 | 15.7 |
| 110.22582.55 | 18 09 01.782 | -28 35 48.48 | 18090178-2835484 | 1.843 | 336 | 2449843 | 11.8 | 17.5 |
| 121.22555.10 | 18 09 02.714 | -30 26 14.82 | 18090271-3026148 | 1.763 | 438 | 2451384 | 10.5 | 14.2 |
| 178.22751.265 | 18 09 15.219 | -26 01 03.54 | 18091521-2601035 | 2.560 | 446 | 2450273 | 15.1 | 18.3 |
| 110.22708.392 | 18 09 16.297 | -28 53 02.57 | 18091629-2853025 | 2.130 | 395 | 2449988 | 14.0 | 18.3 |
| 127.22793.24 | 18 09 44.504 | -31 53 52.23 | 18094450-3153522 | 2.007 | 389 | 2449153 | 13.1 | 16.8 |
| 180.23019.926 | 18 10 02.169 | -25 31 12.57 | 18100216-2531125 | 2.699 | 443 | 2450601 | 15.8 | 20.0 |
| 180.23151.26 | 18 10 12.393 | -25 20 05.55 | 18101239-2520055 | 1.730 | 366 | 2450942 | 12.6 | >17.0 |
| 125.23200.58 | 18 10 33.182 | -30 44 33.30 | 18103318-3044333 | 1.365 | 351 | 2450715 | 10.9 | 15.6 |
| 102.23380.29 | 18 10 56.783 | -27 26 58.29 | 18105678-2726582 | 1.614 | 269 | 2449109 | 12.5 | 18.0 |
| 122.23464.38 | 18 11 04.391 | -30 31 12.28 | 18110439-3031122 | 1.693 | 459 | 2450657 | 13.2 | 15.6 |
| 167.23649.338 | 18 11 30.270 | -26 50 25.20 | 18113026-2650252 | 1.807 | 323 | 2449179 | 12.7 | 16.3 |
| 178.23785.139 | 18 11 41.056 | -26 25 36.51 | 18114105-2625365 | 1.736 | 375 | 2449950 | 12.9 | 16.7 |
| 116.23736.60 | 18 11 43.594 | -29 43 30.48 | 18114359-2943304 | 2.140 | 522 | 2450580 | 12.4 | 15.5 |
| 167.23774.170 | 18 11 44.238 | -27 09 24.44 | 18114423-2709244 | 1.857 | 387 | 2451281 | <12.8 | 15.9 |
| 122.23730.28 | 18 11 44.212 | -30 06 30.15 | 18114421-3006301 | 2.009 | 402 | 2449154 | 13.5 | 15.8 |
| 167.23776.107 | 18 11 48.762 | -27 01 32.15 | 18114876-2701321 | 1.920 | 355 | 2451361 | 13.6 | >15.8 |
| 167.23776.677 | 18 11 53.409 | -27 01 19.42 | 18115340-2701194 | 2.249 | 387 | 2451280 | 13.9 | 17.2 |
| 167.23905.62 | 18 12 01.280 | -27 06 42.62 | 18120128-2706426 | 1.481 | 288 | 2451313 | 10.8 | 14.6 |
| 111.23878.37 | 18 12 05.619 | -28 52 46.27 | 18120561-2852462 | 1.651 | 324 | 2451313 | 11.0 | 14.9 |
| 111.23873.16 | 18 12 08.983 | -29 12 38.54 | 18120898-2912385 | 1.551 | 325 | 2451351 | 11.6 | 13.6 |
| 161.24051.211 | 18 12 13.207 | -26 02 25.44 | 18121320-2602254 | 2.274 | 389 | 2449201 | 13.2 | 17.5 |
| 116.24126.2803 | 18 12 47.280 | -29 39 59.43 | 18124728-2939594 | 2.047 | 333 | 2450932 | 13.7 | 17.2 |
| 167.24294.33 | 18 12 53.100 | -27 08 05.99 | 18125309-2708059 | 1.655 | 354 | 2449005 | <12.2 | 15.8 |
| 116.24255.712 | 18 12 55.466 | -29 45 17.80 | 18125546-2945177 | 2.294 | 444 | 2449926 | 15.2 | 18.5 |
| 167.24426.144 | 18 13 11.870 | -27 01 05.39 | 18131186-2701053 | 2.065 | 426 | 2450301 | 13.6 | 15.8 |
| 116.24392.970 | 18 13 17.421 | -29 16 39.00 | 18131742-2916390 | 2.276 | 435 | 2449234 | 14.3 | 17.4 |
| 116.24392.1337 | 18 13 25.544 | -29 16 07.94 | 18132554-2916079 | 1.943 | 384 | 2451250 | <14.1 | 17.2 |
| 167.24559.794 | 18 13 43.036 | -26 49 41.18 | 18134303-2649411 | 2.109 | 510 | 2449224 | 14.4 | 17.9 |
| 161.24825.451 | 18 14 02.530 | -26 26 51.95 | 18140252-2626519 | 1.909 | 375 | 2450875 | 14.4 | 16.8 |
| 307.35039.28 | 18 14 27.622 | -23 47 45.08 | 18142762-2347450 | 2.122 | 374 | 2450020 | 13.6 | 16.7 |
| 307.35035.93 | 18 14 27.787 | -24 03 24.48 | 18142778-2403244 | 2.501 | 403 | 2449964 | 14.4 | 17.3 |
| 177.24974.49 | 18 14 29.950 | -25 08 32.25 | 18142994-2508322 | 2.075 | 382 | 2450186 | 12.8 | 16.4 |
| 306.35055.307 | 18 14 32.511 | -22 46 38.09 | 18143251-2246380 | 2.363 | 395 | 2451247 | <14.1 | 18.5 |
| 306.35044.66 | 18 14 32.753 | -23 27 40.24 | 18143275-2327402 | 1.986 | 417 | 2450943 | 14.4 | 18.2 |
| 307.35044.281 | 18 14 32.753 | -23 27 40.24 | 18143275-2327402 | 1.986 | 417 | 2450527 | 14.4 | 18.2 |
| 177.24975.162 | 18 14 33.675 | -25 04 21.14 | 18143367-2504211 | 2.310 | 408 | 2451220 | 13.9 | 17.0 |
| 161.25086.27 | 18 14 46.626 | -26 21 23.13 | 18144662-2621231 | 1.520 | 293 | 2450643 | 11.1 | 14.8 |
| 162.25092.4373 | 18 14 47.374 | -25 59 40.20 | 18144737-2559401 | 1.936 | 386 | 2450016 | 12.1 | 16.1 |
| 306.35220.158 | 18 14 49.579 | -22 57 05.53 | 18144957-2257055 | 1.951 | 451 | 2450885 | 14.4 | 17.0 |
| 305.35409.346 | 18 14 54.724 | -21 34 10.83 | 18145472-2134108 | 2.331 | 354 | 2450024 | <14.9 | 18.2 |
| 307.35373.38 | 18 14 55.322 | -23 56 13.37 | 18145532-2356133 | 2.141 | 380 | 2449892 | 14.3 | 17.0 |
| 306.35383.89 | 18 14 59.089 | -23 17 32.84 | 18145908-2317328 | 1.960 | 334 | 2451089 | <13.6 | 16.4 |
| 307.35373.337 | 18 15 08.104 | -23 55 37.15 | 18150810-2355371 | 1.825 | 361 | 2450038 | <14.0 | 16.8 |
| 161.25218.1431 | 18 15 08.941 | -26 13 45.78 | 18150894-2613457 | 2.078 | 416 | 2451355 | 15.0 | 18.9 |
| 305.35571.2200 | 18 15 18.605 | -21 56 46.40 | 18151860-2156463 | 2.923 | 600 | 2451231 | 16.0 | 20.7 |
| 305.35571.2 | 18 15 19.640 | -21 55 31.18 | 18151964-2155311 | 1.811 | 380 | 2450665 | 12.9 | 15.8 |
| 168.25334.4601 | 18 15 23.370 | -27 09 58.94 | 18152336-2709589 | 2.182 | 453 | 2450954 | 14.2 | 16.9 |
| 162.25341.218 | 18 15 26.484 | -26 40 17.99 | 18152648-2640179 | 2.295 | 379 | 2450835 | <13.8 | 16.3 |
| 162.25346.3115 | 18 15 26.897 | -26 21 04.28 | 18152689-2621042 | 2.726 | 383 | 2449214 | 13.9 | 17.6 |
| 159.25486.509 | 18 15 40.334 | -25 42 41.96 | 18154033-2542419 | 2.796 | 483 | 2450596 | 16.2 | 19.1 |
| 168.25462.4818 | 18 15 40.696 | -27 18 47.64 | 18154069-2718476 | 2.787 | 514 | 2450305 | 15.8 | 20.5 |
| 306.35887.19 | 18 15 43.694 | -23 17 21.02 | 18154369-2317210 | 1.681 | 278 | 2451276 | 12.8 | 15.7 |
| 306.35886.142 | 18 15 47.607 | -23 19 58.36 | 18154760-2319583 | 2.021 | 514 | 2450705 | 12.8 | 16.4 |
| 148.25549.18 | 18 16 03.144 | -30 11 05.51 | 18160314-3011055 | 1.590 | 395 | 2450200 | 12.2 | >15.9 |
| 168.25595.4518 | 18 16 07.891 | -27 05 31.82 | 18160789-2705318 | 1.714 | 298 | 2451025 | 11.6 | 14.1 |
| 304.36069.723 | 18 16 09.222 | -22 20 06.34 | 18160922-2220063 | 3.296 | 497 | 2451387 | 17.9 | 20.7 |
| 162.25736.298 | 18 16 12.046 | -26 21 12.02 | 18161204-2621120 | 2.468 | 459 | 2449104 | 16.0 | 21.5 |
| 304.36240.3389 | 18 16 19.901 | -22 07 23.65 | 18161990-2207236 | 3.338 | 459 | 2450571 | 17.2 | 20.5 |
| 306.36229.168 | 18 16 27.221 | -22 52 58.07 | 18162722-2252580 | 1.879 | 433 | 2450884 | 14.5 | 17.9 |
| 155.25788.48 | 18 16 31.518 | -31 33 11.67 | 18163151-3133116 | 1.436 | 422 | 2450641 | 11.6 | 15.5 |
| 159.25873.271 | 18 16 31.678 | -25 55 16.35 | 18163167-2555163 | 1.998 | 394 | 2450882 | 13.4 | 15.8 |
| 159.25873.42 | 18 16 33.287 | -25 54 47.36 | 18163328-2554473 | 1.578 | 350 | 2450105 | 12.6 | 15.7 |
| 306.36388.17 | 18 16 35.649 | -23 27 19.34 | 18163564-2327193 | 1.563 | 280 | 2451341 | 11.2 | 15.5 |
| 306.36398.31 | 18 16 37.263 | -22 50 33.85 | 18163726-2250338 | 1.668 | 323 | 2449923 | 12.8 | 16.2 |
| 159.25880.24 | 18 16 41.343 | -25 25 22.86 | 18164134-2525228 | 1.789 | 365 | 2449115 | 13.0 | >16.3 |
| 152.25795.34 | 18 16 46.300 | -31 04 52.74 | 18164630-3104527 | 1.685 | 396 | 2450623 | 12.5 | >16.7 |
| 162.25992.77 | 18 16 50.900 | -26 35 55.69 | 18165090-2635556 | 1.695 | 396 | 2450240 | 13.2 | 15.6 |
| 162.26122.17 | 18 17 08.790 | -26 38 51.09 | 18170878-2638510 | 1.833 | 348 | 2449891 | 10.7 | >13.3 |
| 159.26142.374 | 18 17 13.971 | -25 16 58.36 | 18171397-2516583 | 2.029 | 353 | 2450050 | <14.2 | 17.6 |
| 159.26133.30 | 18 17 14.354 | -25 52 51.90 | 18171435-2552518 | 1.715 | 376 | 2450577 | 12.4 | >16.0 |
| 148.26069.278 | 18 17 21.573 | -30 11 07.97 | 18172157-3011079 | 2.155 | 324 | 2449888 | 13.8 | 17.6 |
| 159.26270.62 | 18 17 29.061 | -25 26 22.79 | 18172906-2526227 | 1.906 | 407 | 2451291 | 12.5 | 15.9 |
| 168.26243.224 | 18 17 33.106 | -27 14 48.52 | 18173310-2714485 | 2.495 | 406 | 2450909 | 14.6 | 17.2 |
| 309.36913.38 | 18 17 34.325 | -22 04 44.67 | 18173432-2204446 | 2.400 | 518 | 2449919 | 14.3 | 16.5 |
| 168.26244.769 | 18 17 34.989 | -27 11 09.98 | 18173498-2711099 | 2.841 | 480 | 2450666 | 15.9 | 19.5 |
| 168.26243.222 | 18 17 36.617 | -27 14 54.66 | 18173661-2714546 | 1.769 | 396 | 2451331 | 12.9 | 16.7 |
| 311.36886.124 | 18 17 38.272 | -23 52 19.41 | 18173827-2352194 | 1.761 | 452 | 2450617 | 13.6 | 15.6 |
| 159.26398.18 | 18 17 42.579 | -25 34 27.67 | 18174257-2534276 | 2.053 | 345 | 2449085 | <13.4 | 17.0 |
| 310.37068.189 | 18 17 46.105 | -22 57 07.48 | 18174610-2257074 | 2.164 | 365 | 2449971 | 12.1 | 16.4 |
| 311.37052.34 | 18 17 47.594 | -24 02 31.07 | 18174759-2402310 | 1.653 | 302 | 2451026 | 12.6 | 15.8 |
| 310.37062.2051 | 18 17 51.027 | -23 22 37.44 | 18175102-2322374 | 2.978 | 494 | 2450640 | 15.5 | 19.4 |
| 159.26395.19 | 18 17 51.117 | -25 43 47.84 | 18175111-2543478 | 1.613 | 335 | 2449827 | 11.6 | 15.6 |
| 309.37079.321 | 18 17 52.604 | -22 13 10.51 | 18175260-2213105 | 2.356 | 442 | 2450973 | 12.9 | 17.7 |
| 308.37091.896 | 18 17 53.054 | -21 25 00.44 | 18175305-2125004 | 3.001 | 480 | 2450222 | 17.1 | 20.6 |
| 309.37241.101 | 18 18 07.106 | -22 38 04.59 | 18180710-2238045 | 1.828 | 344 | 2449775 | <13.1 | 16.9 |
| 311.37219.401 | 18 18 07.710 | -24 03 58.06 | 18180771-2403580 | 1.945 | 399 | 2451289 | 14.0 | 16.6 |
| 311.37225.69 | 18 18 13.858 | -23 42 04.49 | 18181385-2342044 | 1.792 | 346 | 2449787 | 12.6 | 16.7 |
| 309.37413.667 | 18 18 20.084 | -22 20 15.91 | 18182008-2220159 | 2.652 | 430 | 2450159 | 15.7 | 18.8 |
| 310.37400.99 | 18 18 22.842 | -23 11 54.96 | 18182284-2311549 | 1.712 | 348 | 2449770 | <12.9 | 16.6 |
| 148.26588.65 | 18 18 26.748 | -30 12 43.81 | 18182674-3012438 | 1.881 | 423 | 2450577 | 14.0 | 16.7 |
| 308.37421.1264 | 18 18 29.231 | -21 49 33.62 | 18182923-2149336 | 2.379 | 378 | 2450030 | 15.4 | 18.8 |
| 309.37411.685 | 18 18 30.771 | -22 28 41.83 | 18183077-2228418 | 2.952 | 568 | 2450295 | 16.4 | 18.4 |
| 163.26649.58 | 18 18 31.640 | -26 10 03.00 | 18183164-2610030 | 1.519 | 300 | 2449877 | 11.5 | 15.3 |
| 152.26706.50 | 18 18 40.928 | -31 02 39.52 | 18184092-3102395 | 1.651 | 374 | 2450246 | 13.0 | >15.7 |
| 308.37594.263 | 18 18 41.073 | -21 30 47.42 | 18184107-2130474 | 2.387 | 444 | 2449931 | 15.9 | 18.7 |
| 152.26703.17 | 18 18 45.428 | -31 13 01.79 | 18184542-3113017 | 1.677 | 501 | 2450974 | 11.7 | 16.0 |
| 148.26717.152 | 18 18 50.816 | -30 19 33.11 | 18185081-3019331 | 1.895 | 467 | 2450671 | 12.2 | 15.4 |
| 163.26778.30 | 18 18 51.107 | -26 14 22.19 | 18185110-2614221 | 2.111 | 387 | 2449225 | 13.4 | 16.4 |
| 311.37723.52 | 18 18 56.779 | -24 04 19.62 | 18185677-2404196 | 1.712 | 384 | 2449933 | 13.6 | 15.6 |
| 308.37754.2956 | 18 19 01.968 | -21 59 38.43 | 18190196-2159384 | 1.838 | 334 | 2451402 | <12.6 | 15.8 |
| 160.26916.12 | 18 19 05.656 | -25 41 03.57 | 18190565-2541035 | 1.955 | 400 | 2450682 | 11.2 | 15.1 |
| 309.37913.118 | 18 19 08.668 | -22 35 32.22 | 18190866-2235322 | 2.275 | 396 | 2450016 | 13.6 | 16.6 |
| 160.27052.8 | 18 19 18.230 | -25 16 25.10 | 18191822-2516251 | 1.786 | 364 | 2450203 | 12.3 | >14.8 |
| 308.37927.53 | 18 19 19.688 | -21 42 43.17 | 18191968-2142431 | 2.320 | 498 | 2449933 | 15.1 | 19.4 |
| 163.27039.1305 | 18 19 20.067 | -26 09 08.93 | 18192006-2609089 | 1.876 | 358 | 2451090 | <13.6 | 16.3 |
| 309.38080.54 | 18 19 26.324 | -22 39 02.53 | 18192632-2239025 | 2.151 | 397 | 2449952 | 13.9 | 15.8 |
| 309.38082.94 | 18 19 29.969 | -22 32 30.87 | 18192996-2232308 | 2.026 | 402 | 2449971 | 12.9 | 16.7 |
| 163.27032.2748 | 18 19 30.688 | -26 37 23.72 | 18193068-2637237 | 1.622 | 356 | 2451384 | 12.3 | 16.6 |
| 309.38085.49 | 18 19 32.294 | -22 19 04.32 | 18193229-2219043 | 1.781 | 340 | 2450110 | 12.3 | 16.4 |
| 310.38237.86 | 18 19 41.519 | -23 25 30.15 | 18194151-2325301 | 1.678 | 360 | 2449775 | 12.2 | 16.1 |
| 308.38599.62 | 18 20 17.077 | -21 42 22.94 | 18201707-2142229 | 2.126 | 397 | 2449970 | 14.9 | 18.2 |
| 160.27441.217 | 18 20 17.758 | -25 23 28.12 | 18201775-2523281 | 2.118 | 426 | 2450661 | 13.3 | 16.6 |
| 310.38742.186 | 18 20 34.835 | -23 20 58.96 | 18203483-2320589 | 2.093 | 369 | 2451180 | <14.6 | 17.2 |
| 163.27555.232 | 18 20 36.281 | -26 26 48.06 | 18203628-2626480 | 2.021 | 357 | 2449770 | 13.7 | 16.3 |
| 311.38732.104 | 18 20 36.475 | -24 00 20.43 | 18203647-2400204 | 2.108 | 375 | 2451210 | <14.0 | 16.0 |
| 160.27572.1769 | 18 20 43.324 | -25 17 33.96 | 18204332-2517339 | 1.976 | 291 | 2451290 | 14.9 | 18.7 |
| 160.27699.302 | 18 20 46.220 | -25 28 59.84 | 18204622-2528598 | 2.069 | 395 | 2449959 | 12.0 | 15.7 |
| 142.27644.69 | 18 20 49.104 | -29 09 37.98 | 18204910-2909379 | 1.698 | 311 | 2449901 | 11.7 | 15.1 |
| 142.27779.83 | 18 21 02.770 | -28 49 01.10 | 18210276-2849011 | 1.734 | 422 | 2450599 | 13.3 | 16.9 |
| 153.27751.109 | 18 21 09.507 | -30 40 26.89 | 18210950-3040268 | 1.547 | 360 | 2449567 | 12.8 | 16.0 |
| 142.27900.1700 | 18 21 24.929 | -29 24 12.56 | 18212492-2924125 | 1.550 | 209 | 2450606 | 15.8 | 19.0 |
| 142.27906.35 | 18 21 28.061 | -29 01 48.46 | 18212806-2901484 | 1.576 | 341 | 2449540 | 11.8 | <15.1 |
| 142.27907.80 | 18 21 29.408 | -28 58 17.41 | 18212940-2858174 | 2.177 | 416 | 2449592 | 13.3 | 16.5 |
| 146.27895.17 | 18 21 31.692 | -29 46 08.71 | 18213169-2946087 | 1.649 | 371 | 2450596 | 13.3 | <15.9 |
| 146.27899.25 | 18 21 36.608 | -29 27 50.58 | 18213660-2927505 | 1.730 | 436 | 2450224 | 13.2 | 17.4 |
| 153.28008.17 | 18 21 40.503 | -30 52 53.03 | 18214050-3052530 | 2.706 | 423 | 2450295 | 13.7 | 16.9 |
| 156.28515.180 | 18 23 07.692 | -31 46 44.23 | 18230769-3146442 | 3.342 | 450 | 2449505 | 16.1 | 19.5 |
| 136.28693.40 | 18 23 11.439 | -28 33 56.24 | 18231143-2833562 | 1.488 | 296 | 2449919 | 11.2 | 15.5 |
| 146.28675.49 | 18 23 17.236 | -29 43 46.52 | 18231723-2943465 | 1.373 | 283 | 2449592 | 12.0 | 15.6 |
| 137.29336.35 | 18 24 56.017 | -29 00 26.49 | 18245601-2900264 | 1.628 | 347 | 2451430 | 12.5 | 15.6 |
| 143.29457.8 | 18 25 11.231 | -29 36 46.90 | 18251123-2936469 | 1.703 | 372 | 2450607 | 12.5 | <15.3 |
| 143.29845.51 | 18 25 55.396 | -29 47 26.72 | 18255539-2947267 | 1.827 | 408 | 2451334 | 12.4 | 15.7 |
| 143.30111.38 | 18 26 34.461 | -29 21 02.14 | 18263446-2921021 | 1.401 | 229 | 2450240 | 11.4 | 15.2 |
| 143.30110.14 | 18 26 36.115 | -29 23 54.74 | 18263611-2923547 | 1.677 | 472 | 2449462 | 13.0 | 16.0 |
| 143.30111.25 | 18 26 49.095 | -29 22 33.65 | 18264909-2922336 | 1.358 | 303 | 2450609 | 12.5 | 16.0 |
| 132.30519.13 | 18 27 39.540 | -28 10 52.19 | 18273954-2810521 | 1.832 | 477 | 2451303 | 11.7 | 15.2 |
| 132.30776.12 | 18 28 10.265 | -28 20 00.39 | 18281026-2820003 | 1.505 | 312 | 2450635 | 11.4 | 15.2 |
| 138.30903.43 | 18 28 32.977 | -28 32 07.03 | 18283297-2832070 | 2.022 | 381 | 2451260 | <13.3 | 15.6 |
| 132.31039.35 | 18 28 53.158 | -28 08 12.36 | 18285315-2808123 | 1.560 | 186 | 2450973 | 13.7 | 18.1 |
| 132.31298.385 | 18 29 21.718 | -28 11 45.55 | 18292171-2811455 | 2.508 | 416 | 2450647 | 15.6 | 19.7 |
| 303.44071.416 | 18 29 24.623 | -15 16 56.18 | 18292462-1516561 | 3.021 | 216 | 2450670 | 17.4 | 21.6 |
| 144.31804.36 | 18 30 33.284 | -29 08 30.76 | 18303328-2908307 | 1.875 | 352 | 2449395 | <13.2 | 15.8 |
| 303.45254.157 | 18 31 16.914 | -14 50 22.75 | 18311691-1450227 | 2.040 | 403 | 2449789 | 14.3 | >16.8 |
| 139.32207.1904 | 18 31 32.539 | -28 18 10.58 | 18313253-2818105 | 3.537 | 500 | 2450230 | 16.7 | 20.4 |
| 139.32203.1597 | 18 31 37.023 | -28 32 17.35 | 18313702-2832173 | 2.633 | 431 | 2450970 | 15.7 | 19.6 |
| 301.45610.67 | 18 31 50.324 | -13 28 45.18 | 18315032-1328451 | 2.134 | 473 | 2451301 | 14.5 | 17.6 |
| 303.45583.155 | 18 32 02.377 | -15 15 06.15 | 18320237-1515061 | 2.224 | 456 | 2451310 | 16.1 | 20.0 |
| 301.45778.398 | 18 32 02.568 | -13 27 40.75 | 18320256-1327407 | 2.201 | 351 | 2451103 | 14.1 | 17.1 |
| 303.45752.231 | 18 32 14.204 | -15 11 24.71 | 18321420-1511247 | 2.122 | 335 | 2451401 | 14.6 | 18.3 |
| 139.32728.13 | 18 32 40.682 | -28 14 59.07 | 18324068-2814590 | 1.571 | 383 | 2451302 | <11.8 | 15.7 |
| 139.32850.23 | 18 33 00.154 | -28 44 01.19 | 18330015-2844011 | 1.641 | 396 | 2451302 | 12.0 | 15.2 |
| 139.32982.4362 | 18 33 22.698 | -28 37 39.05 | 18332269-2837390 | 1.832 | 386 | 2451302 | 13.2 | 16.2 |
| 134.33253.632 | 18 33 57.541 | -27 52 55.45 | 18335754-2752554 | 1.484 | 360 | 2449611 | 12.3 | 15.6 |
| 134.33381.3349 | 18 34 17.318 | -27 59 44.34 | 18341731-2759443 | 1.491 | 357 | 2449541 | 12.1 | 15.6 |
| 134.33385.120 | 18 34 21.530 | -27 46 10.28 | 18342152-2746102 | 1.798 | 369 | 2449400 | <13.2 | 15.9 |

Figure 6 – Light curves (left panels) and folded light curves (right panels) of the 192 Mira variables identified in this work, based on MACHO $R_C$ data. Data have been folded with the periods listed in Table 1.

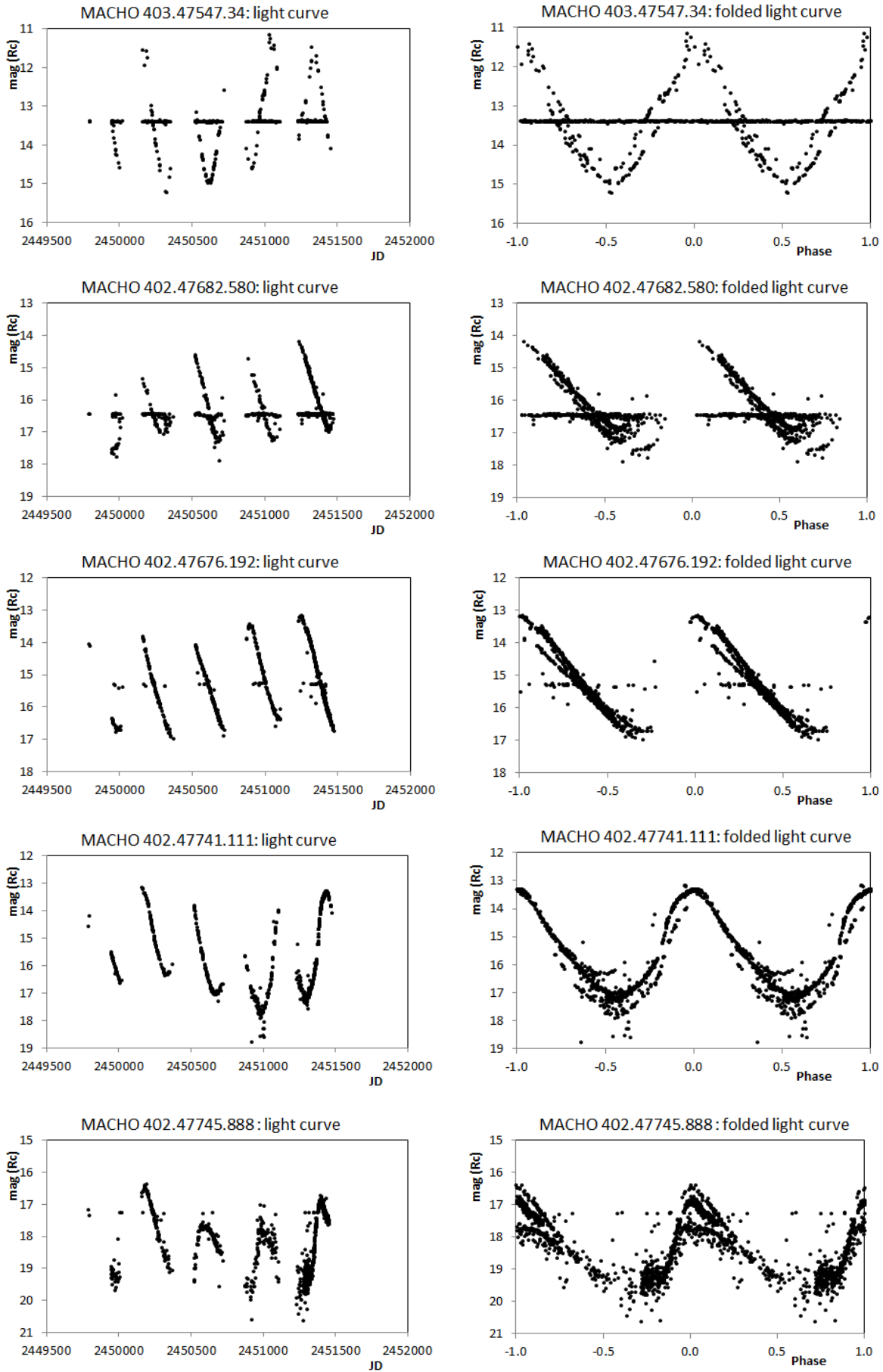

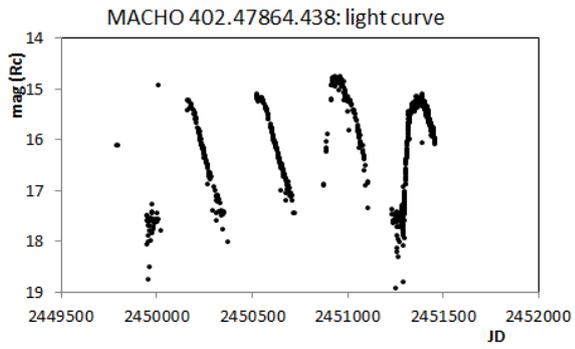
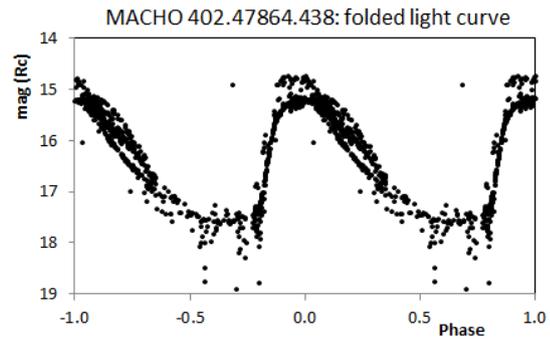
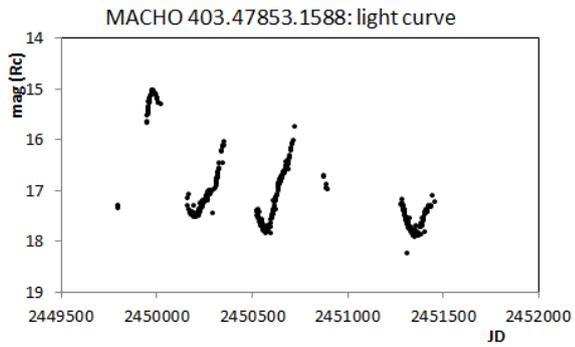
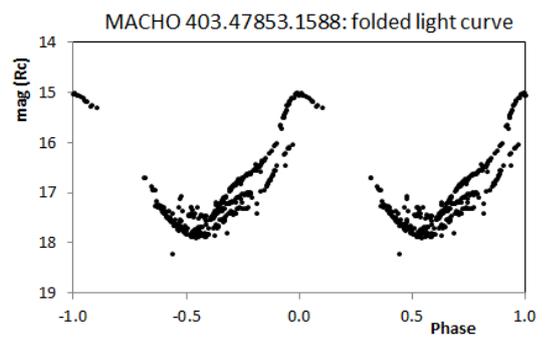
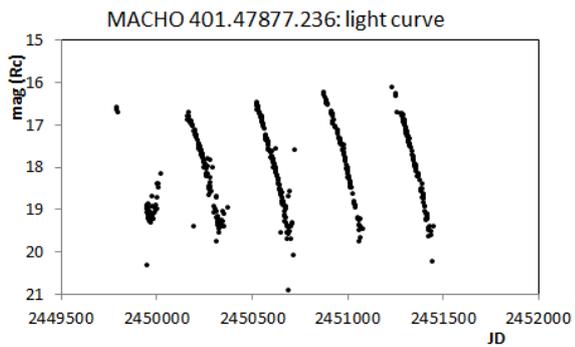
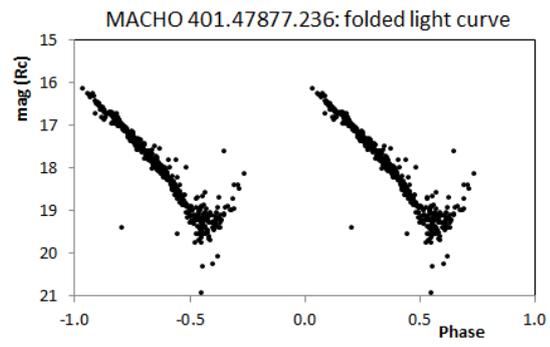
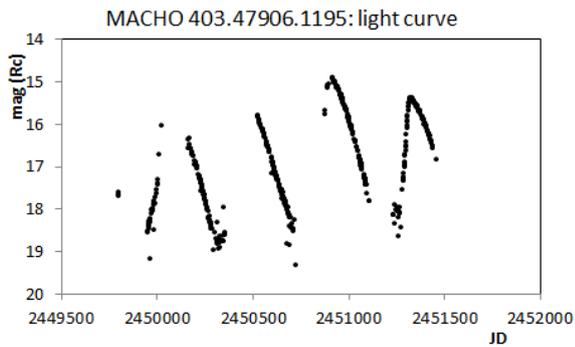
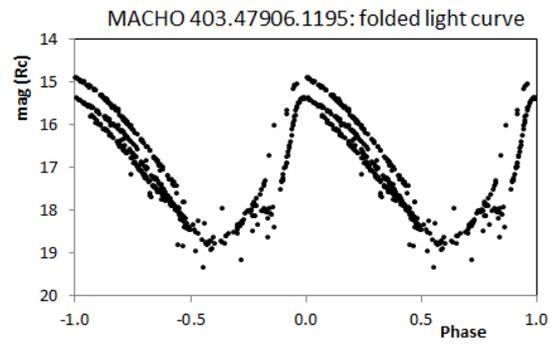
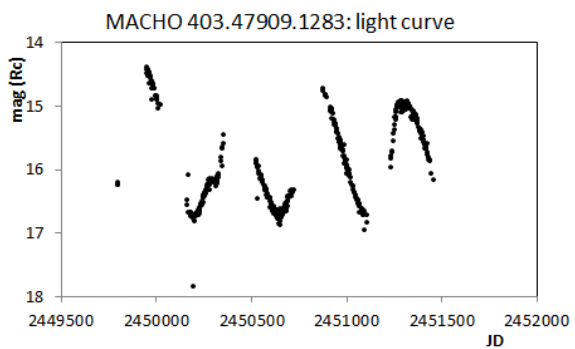
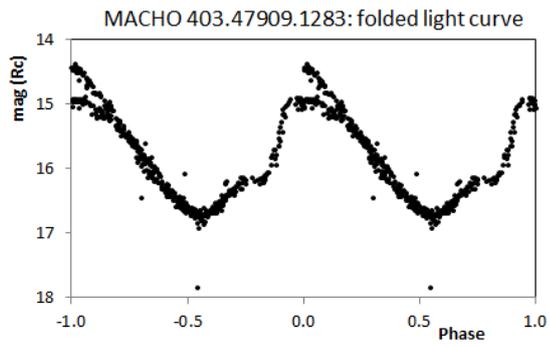

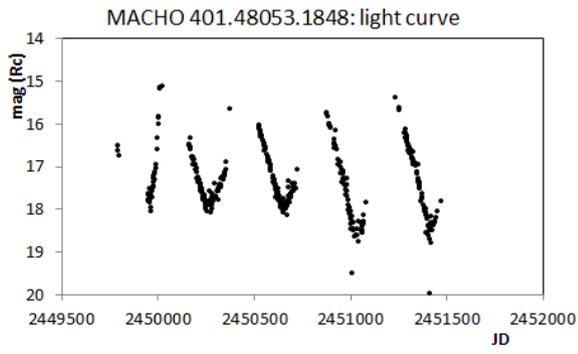
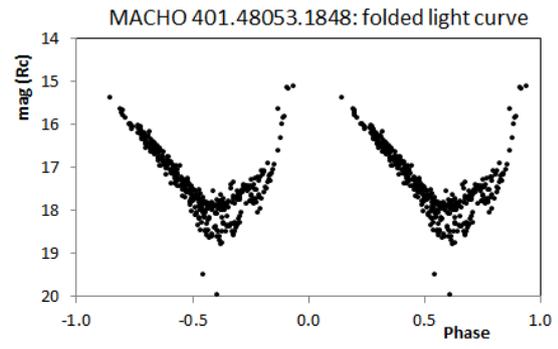
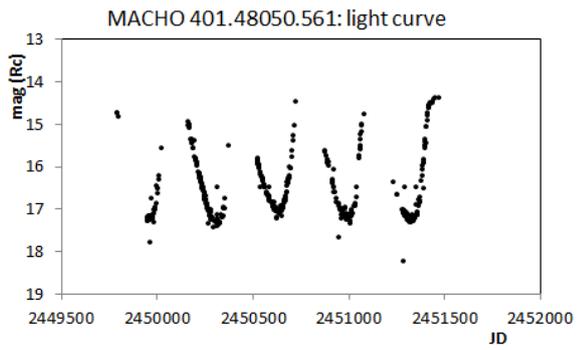
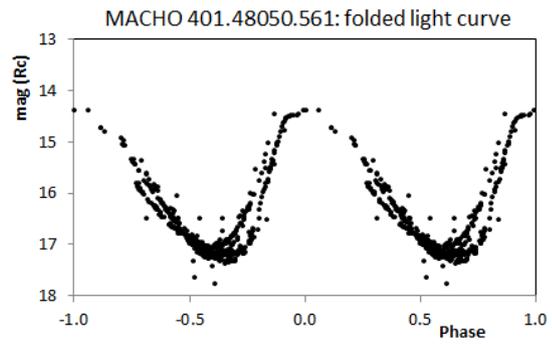
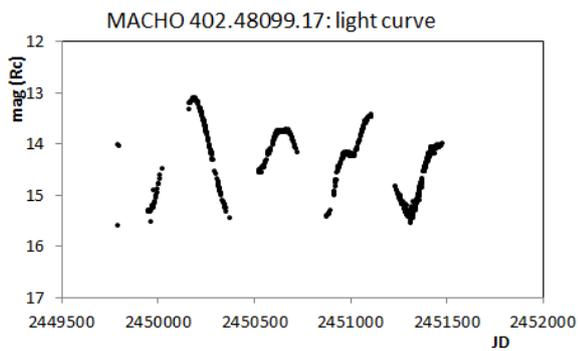
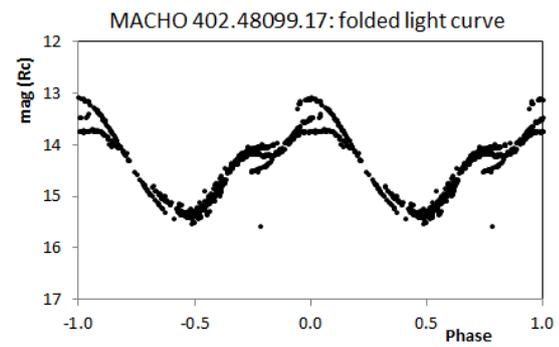
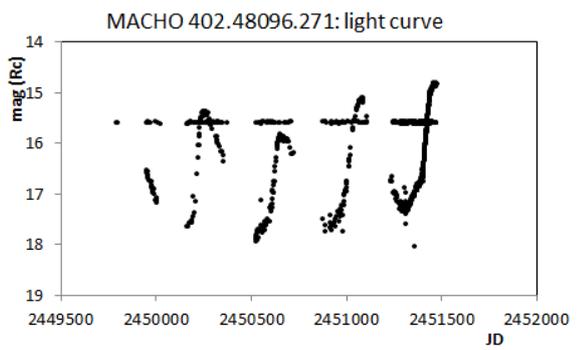
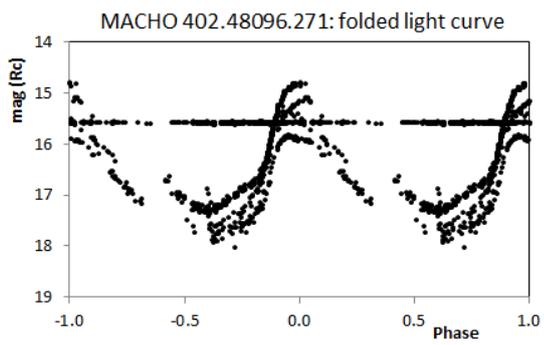
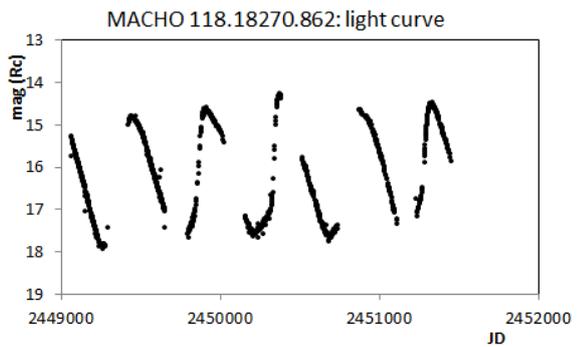
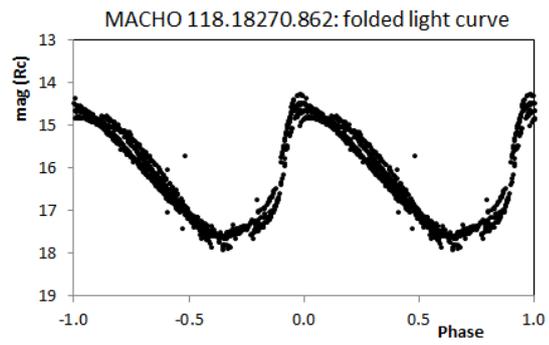

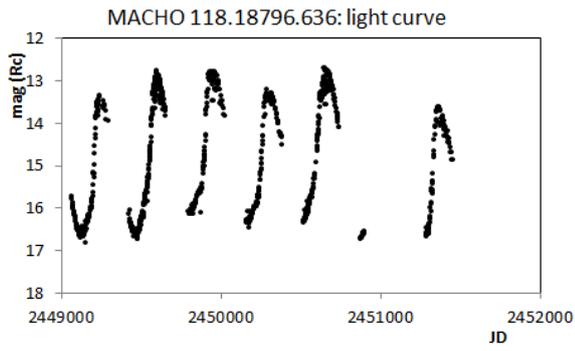
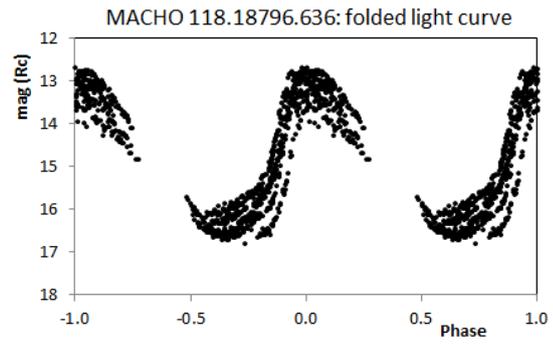
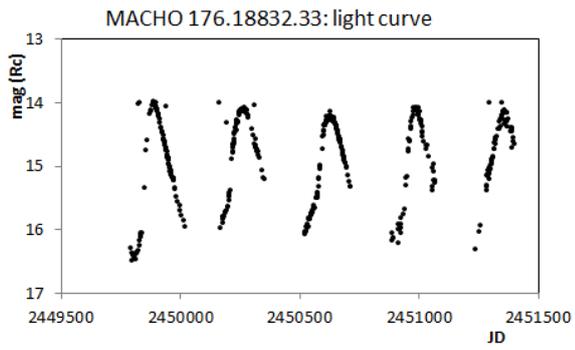
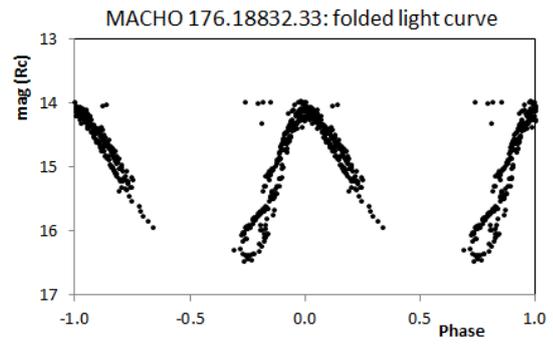
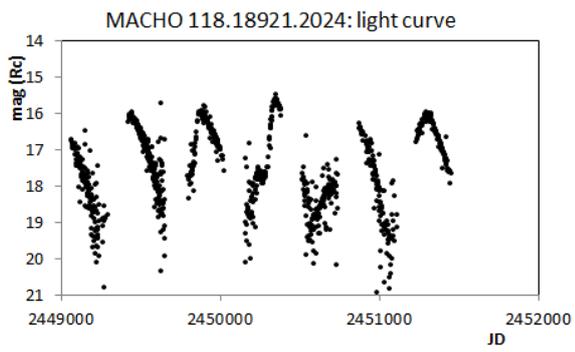
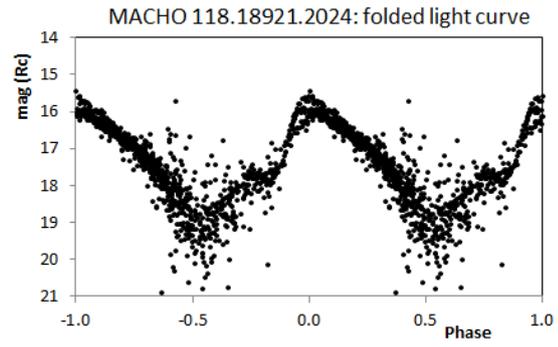
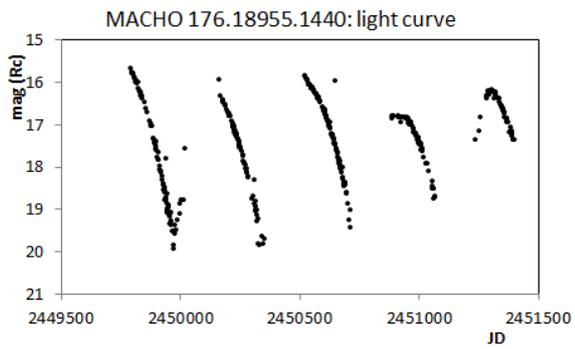
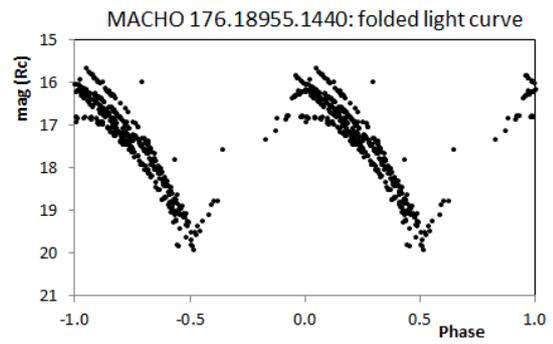
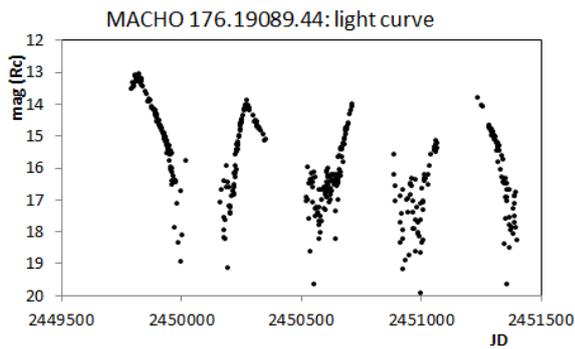
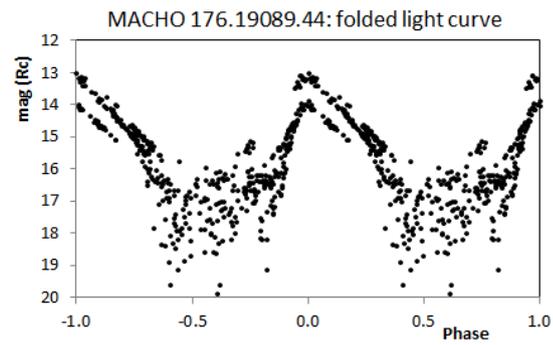

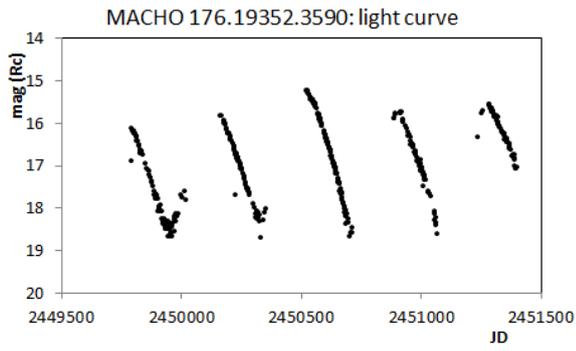
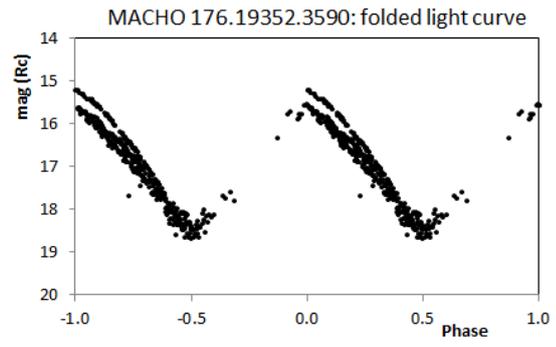
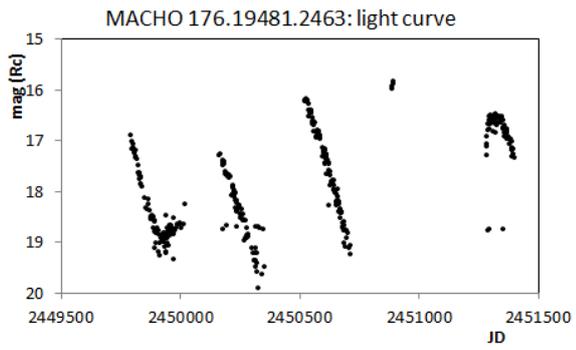
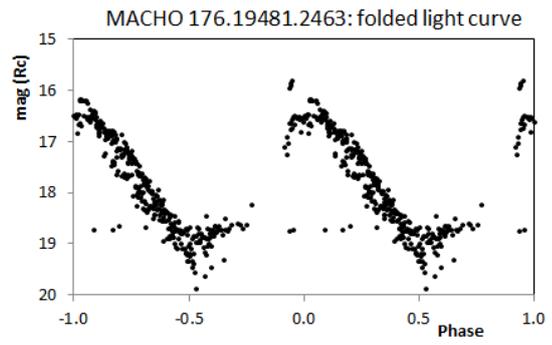
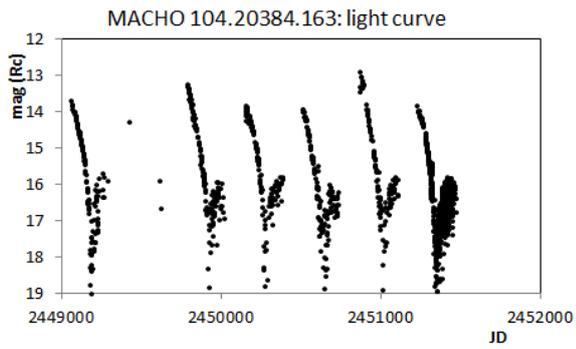
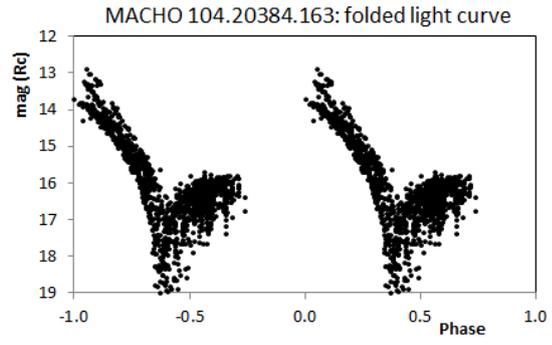
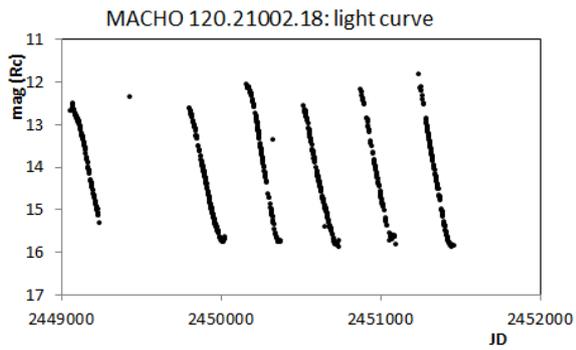
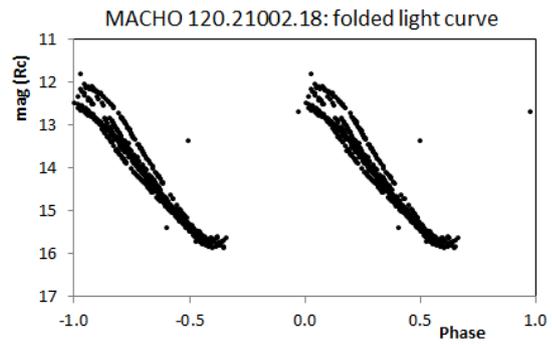
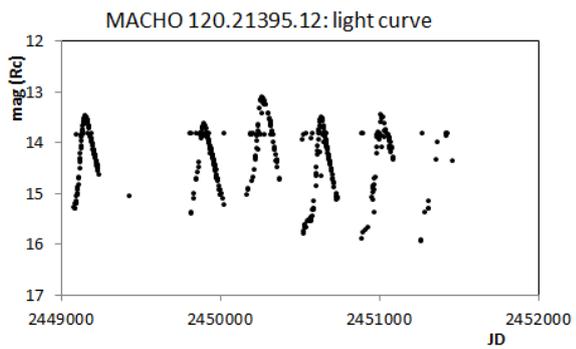
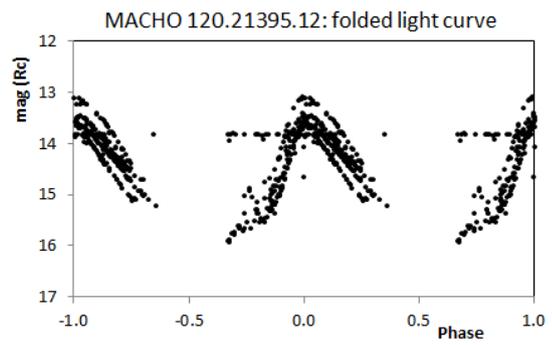

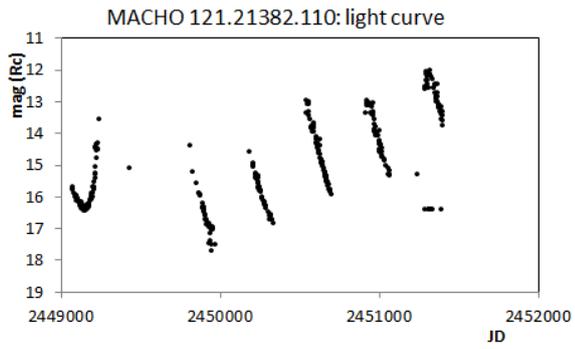
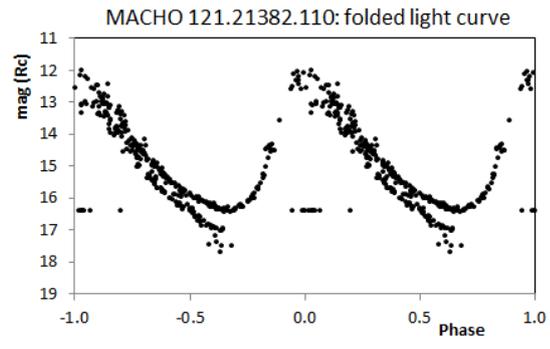
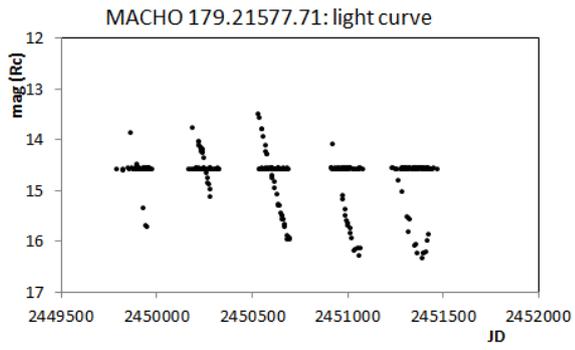
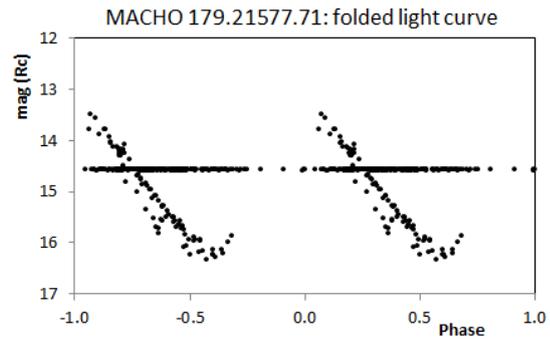
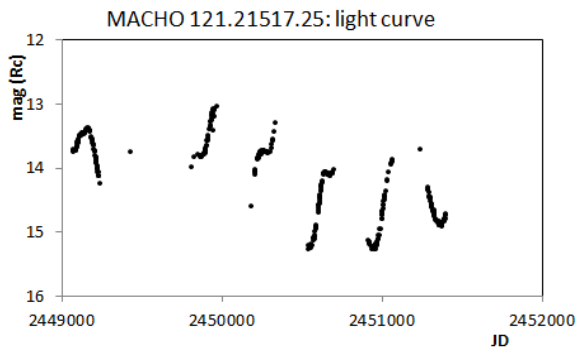
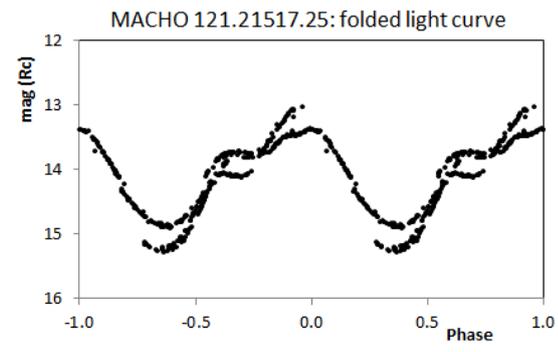
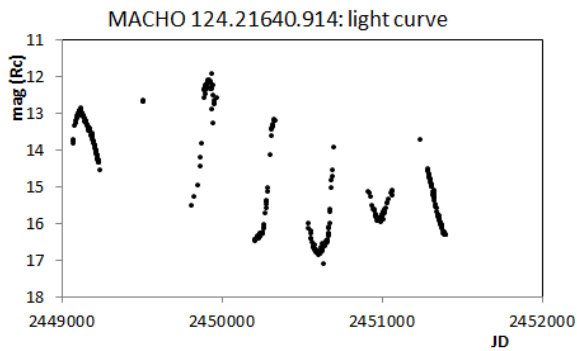
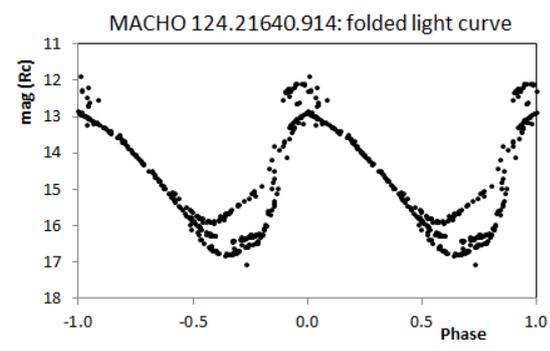
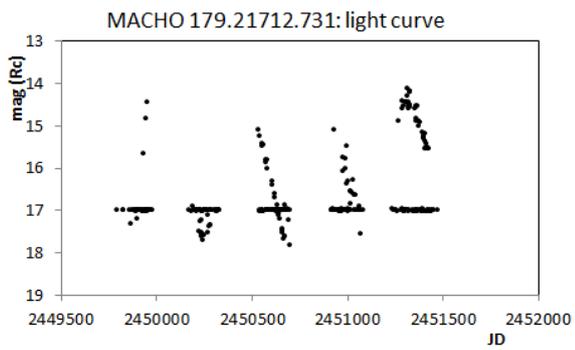
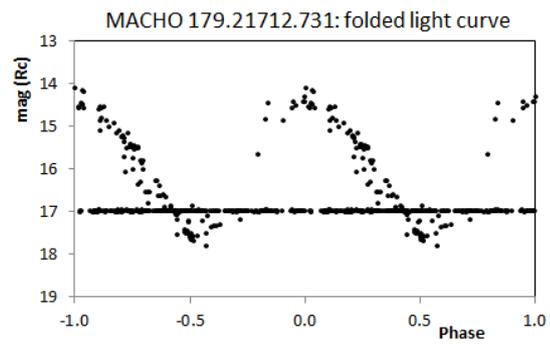

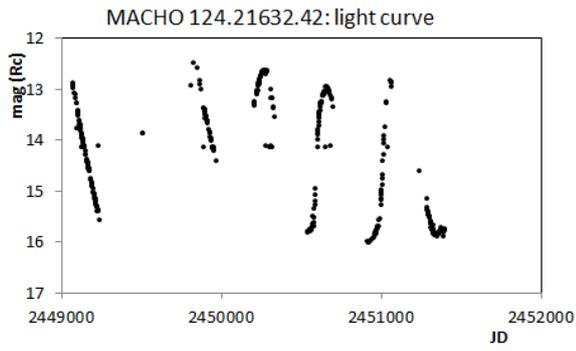
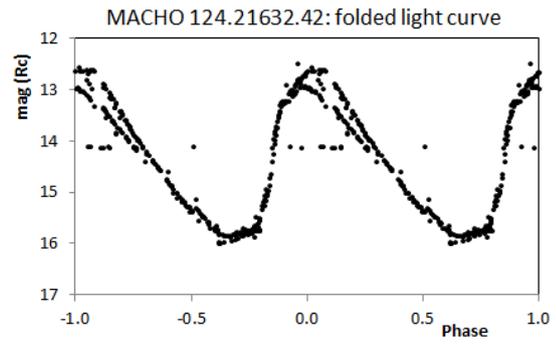
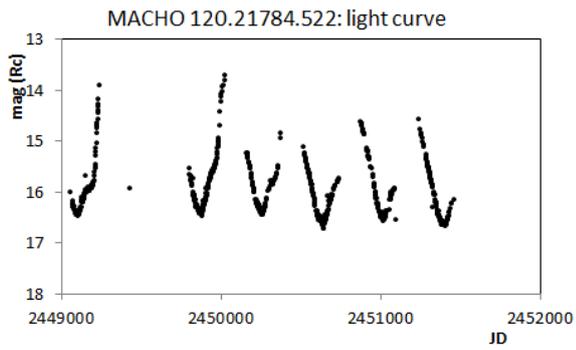
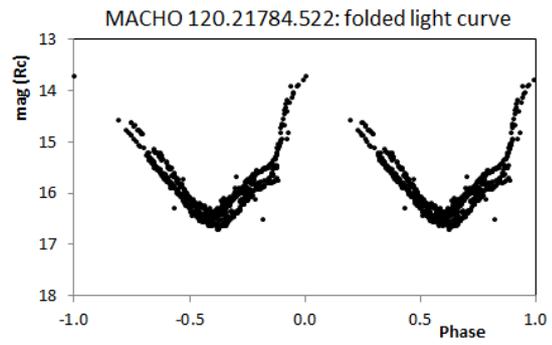
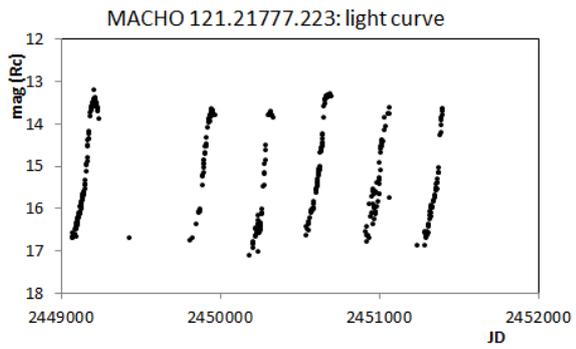
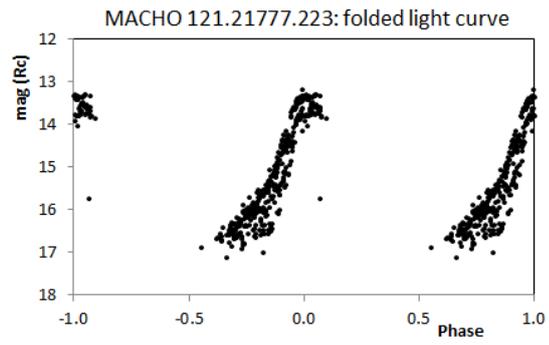
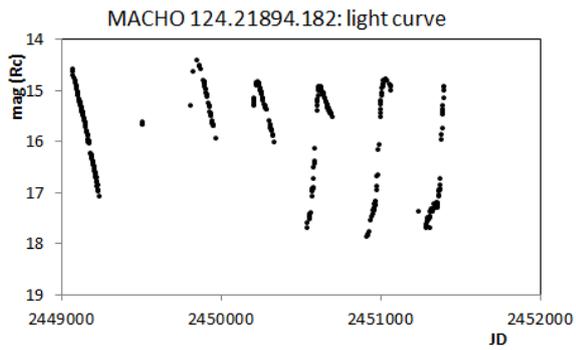
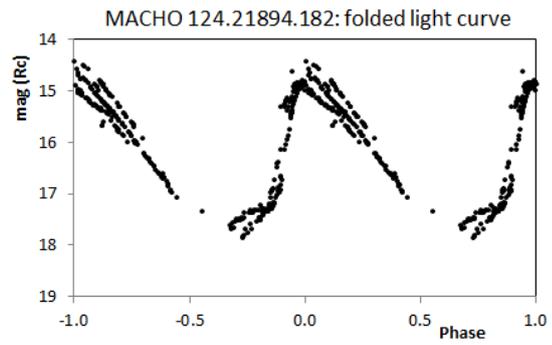
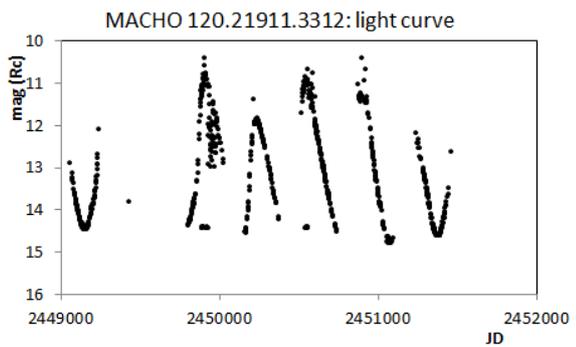
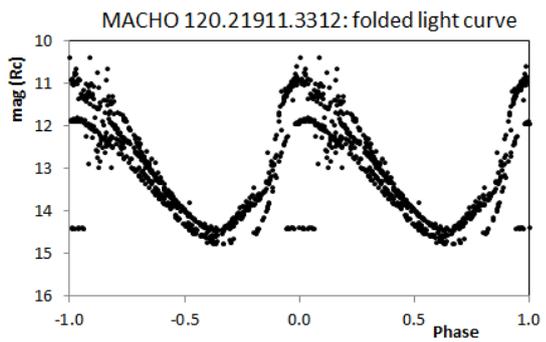

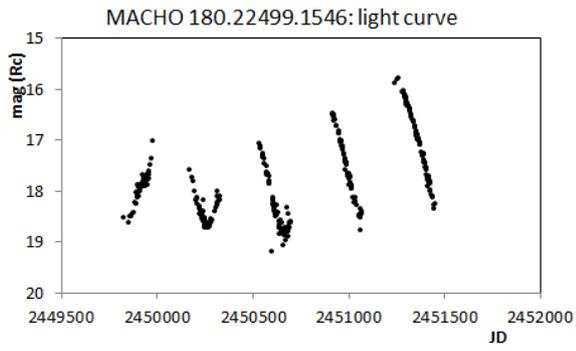
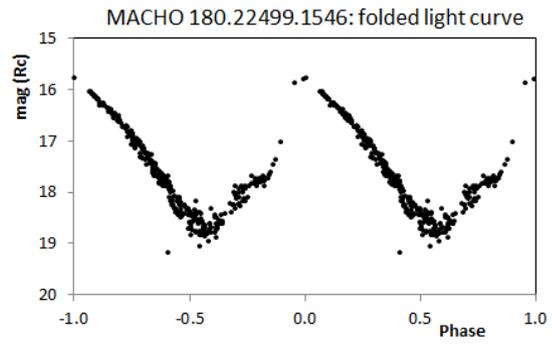
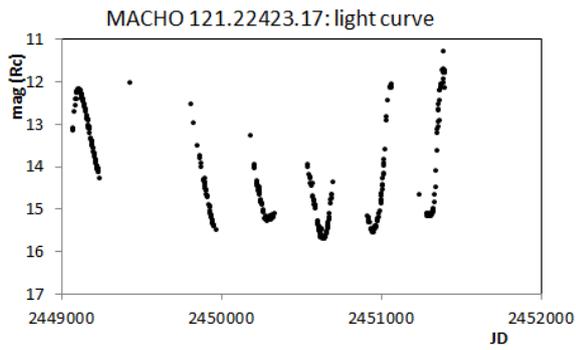
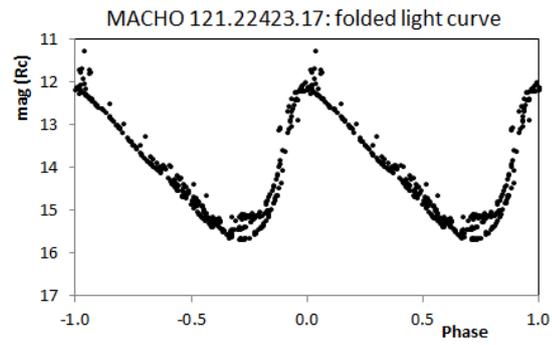
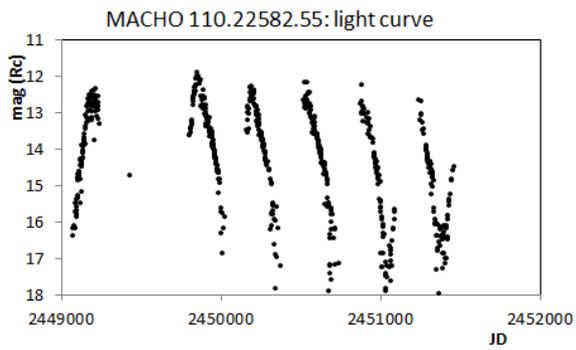
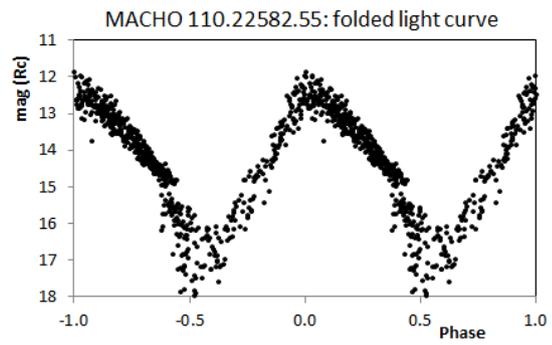
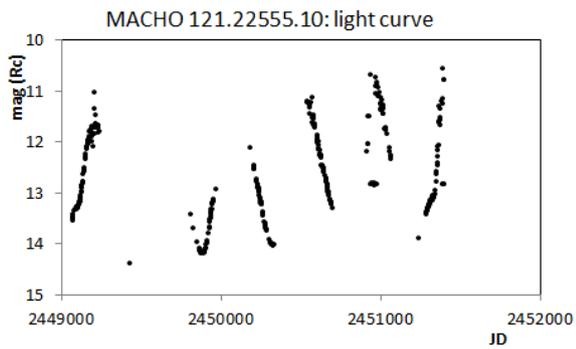
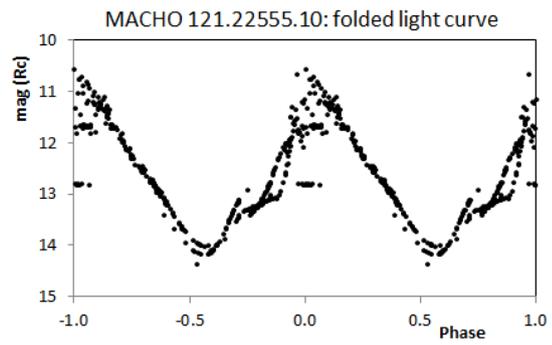
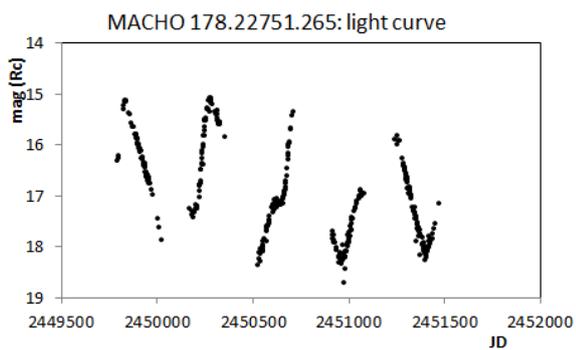
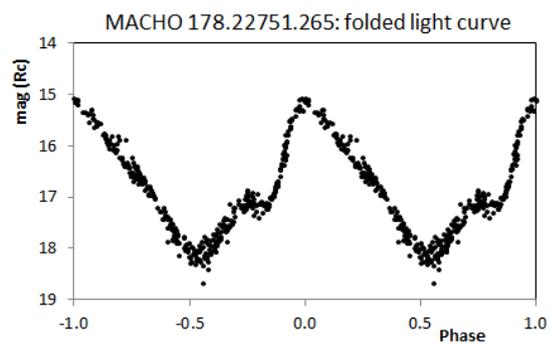

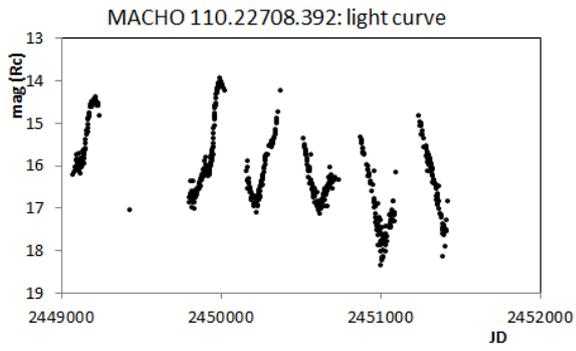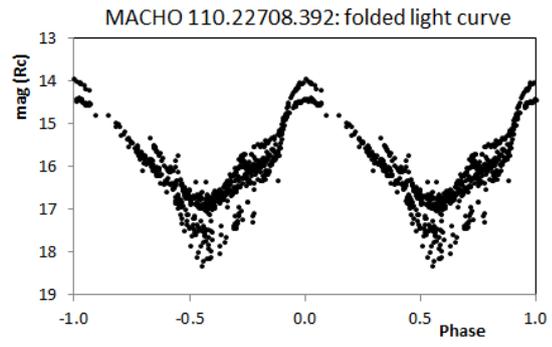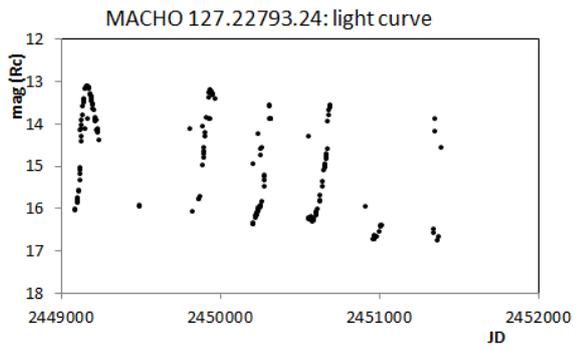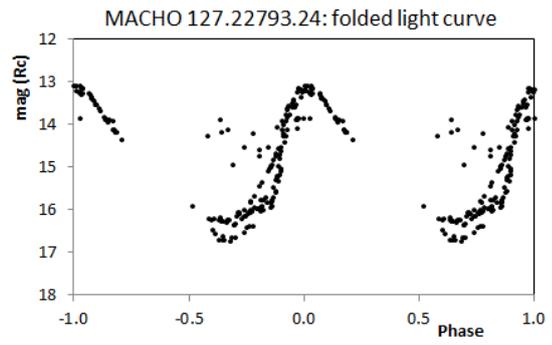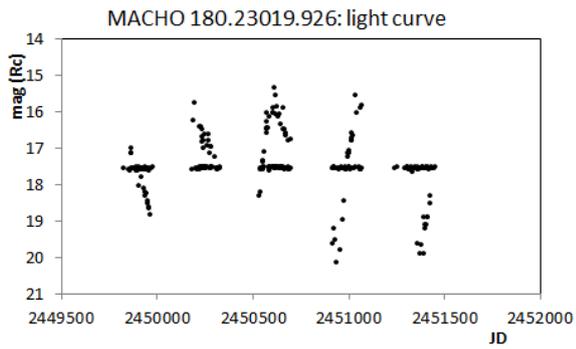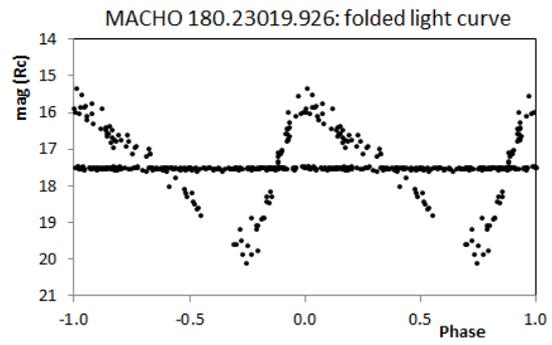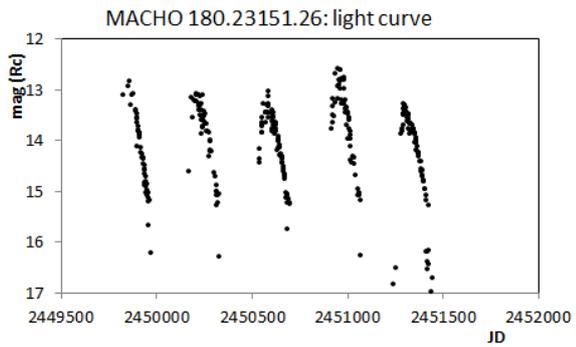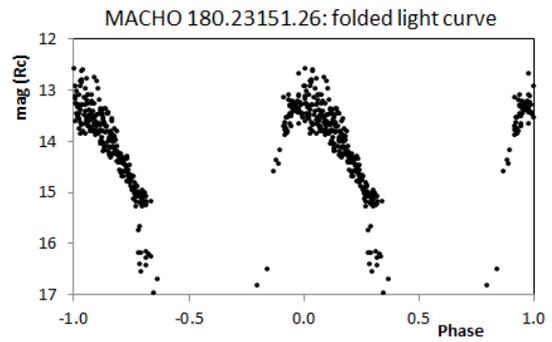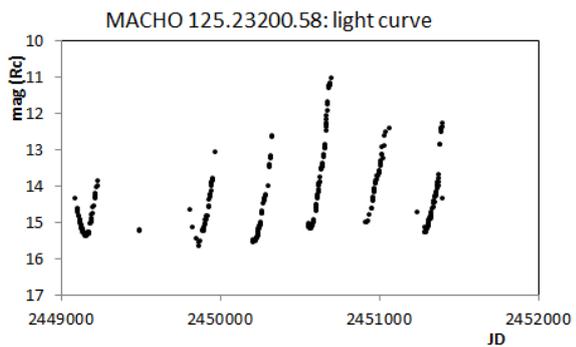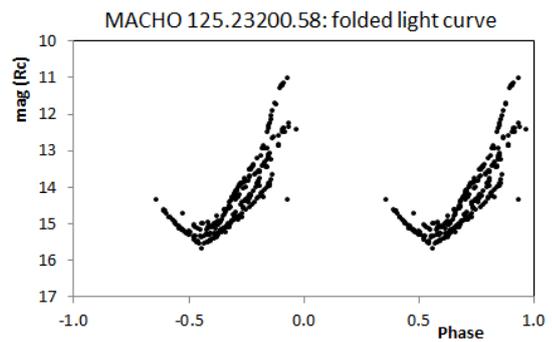

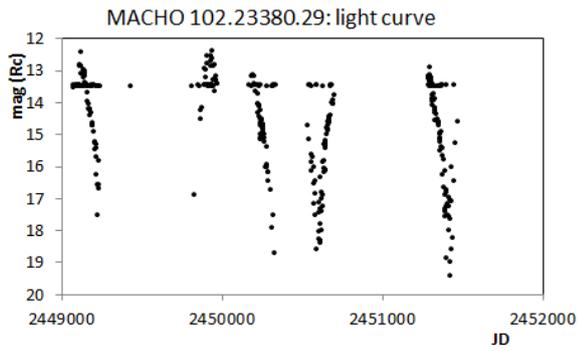
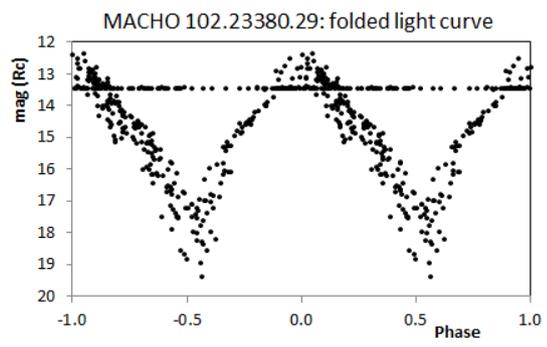
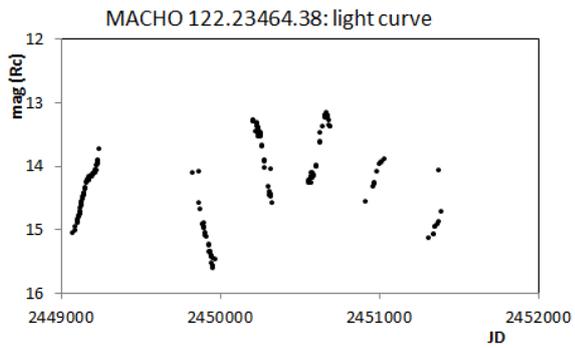
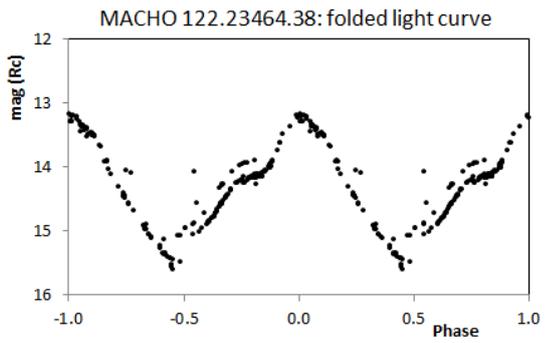
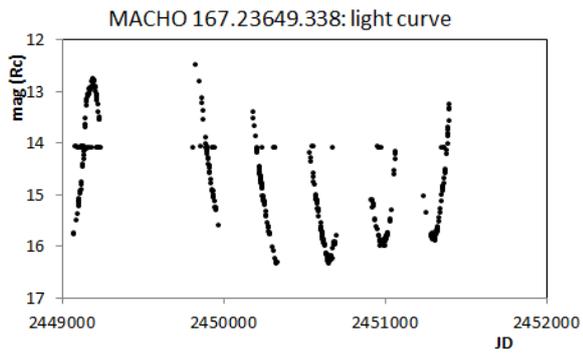
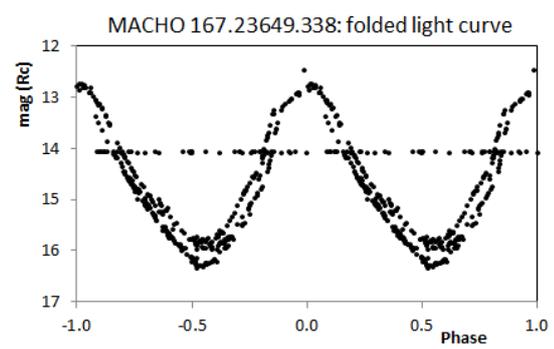
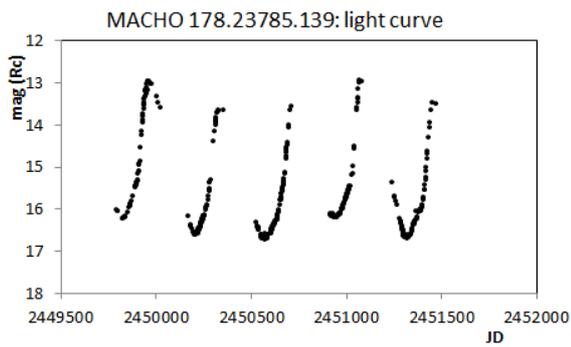
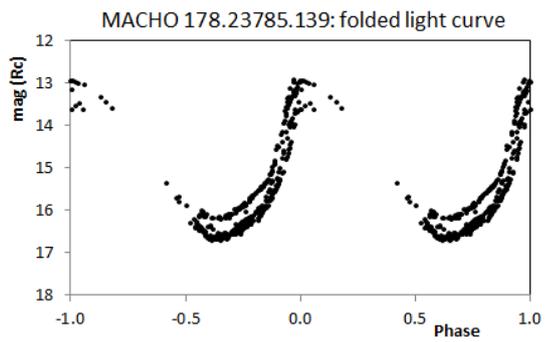
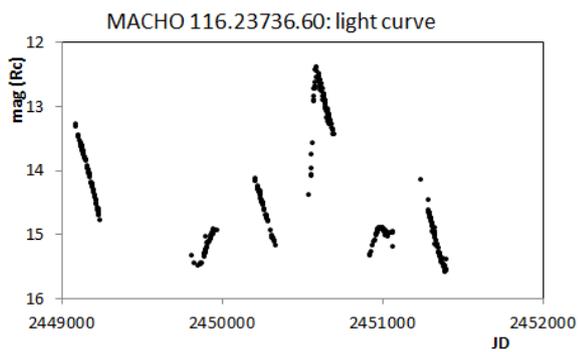
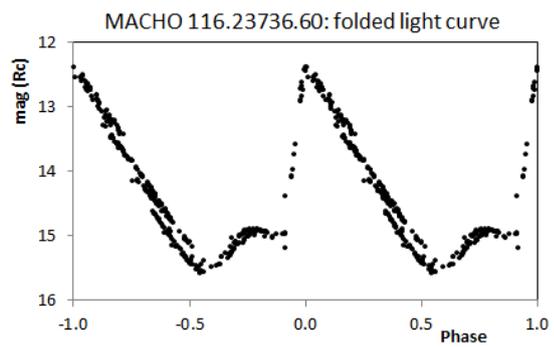

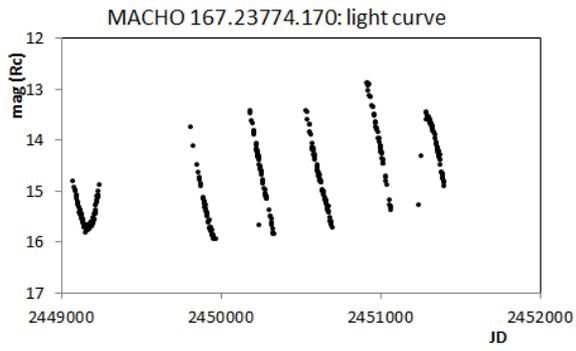
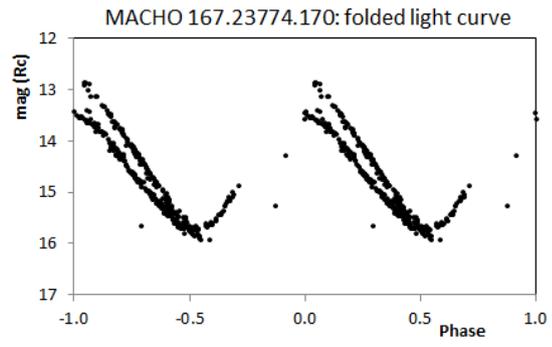
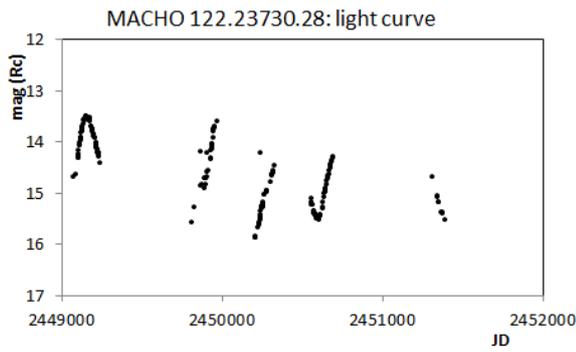
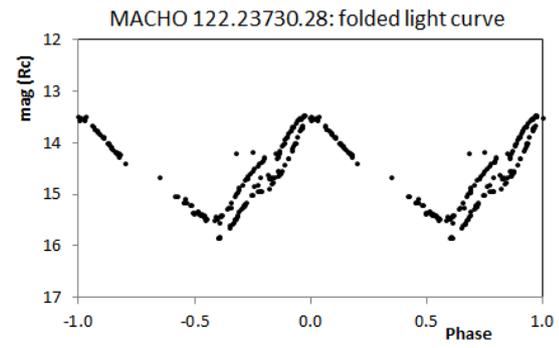
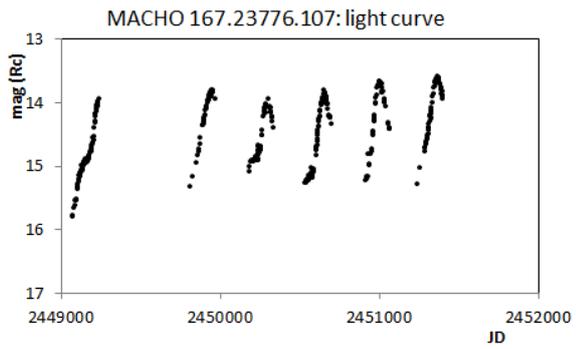
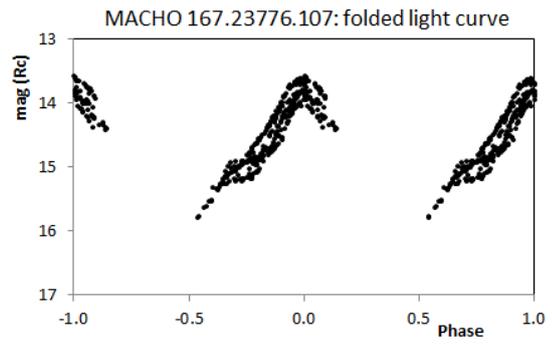
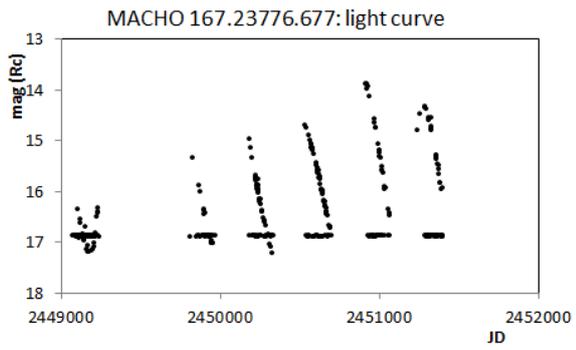
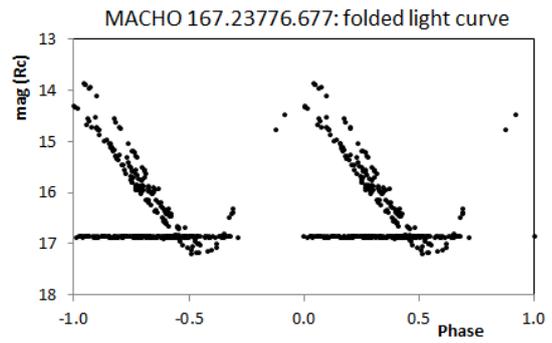
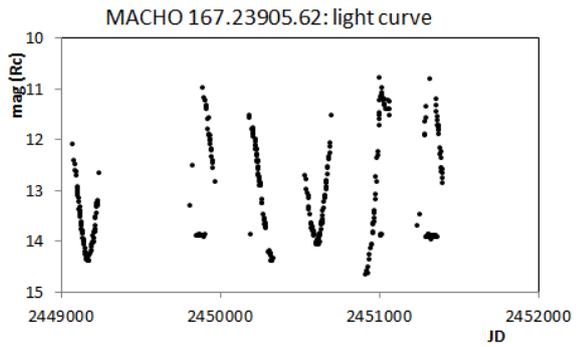
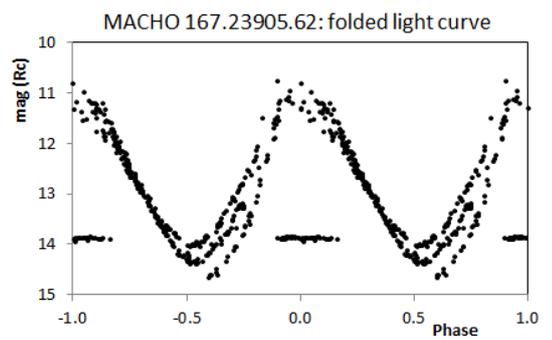

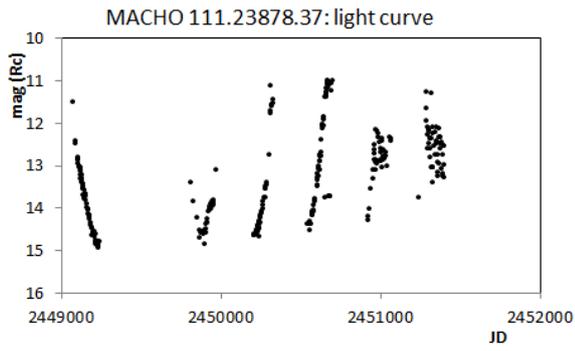
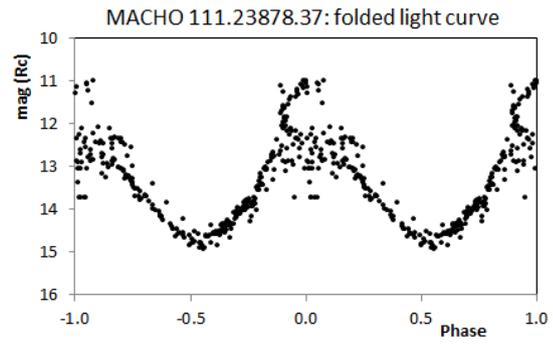
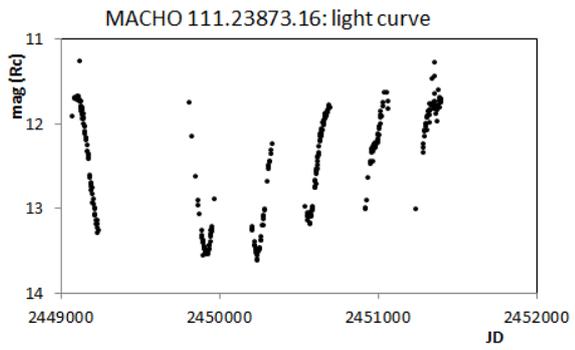
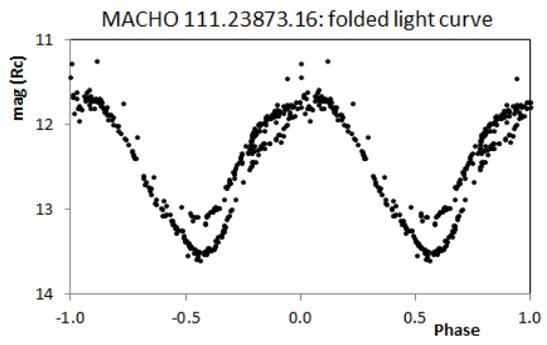
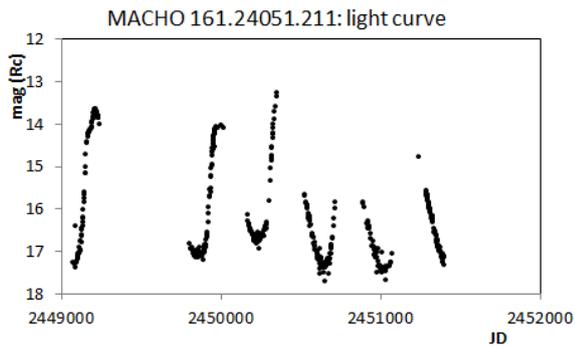
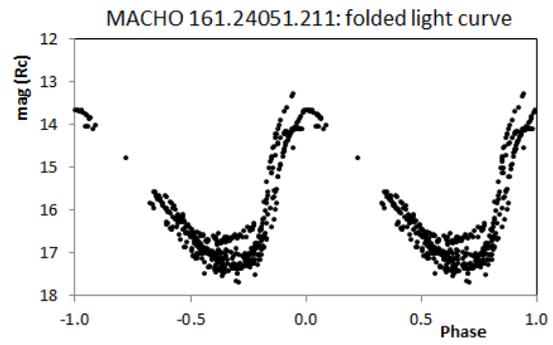
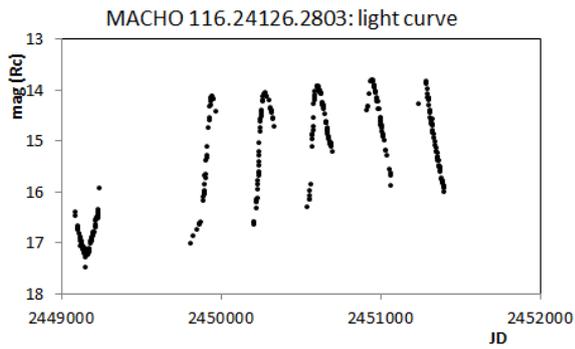
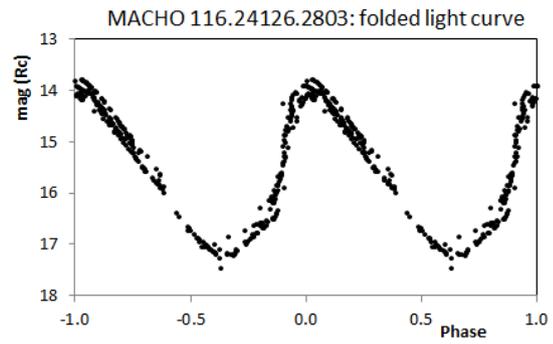
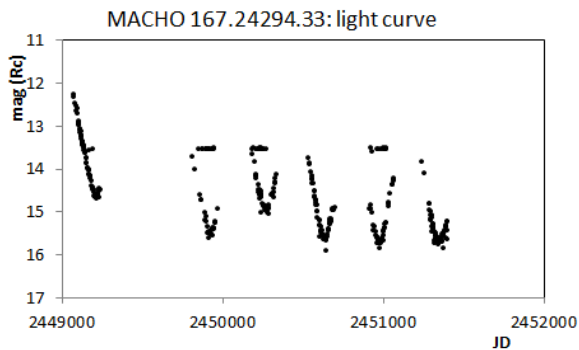
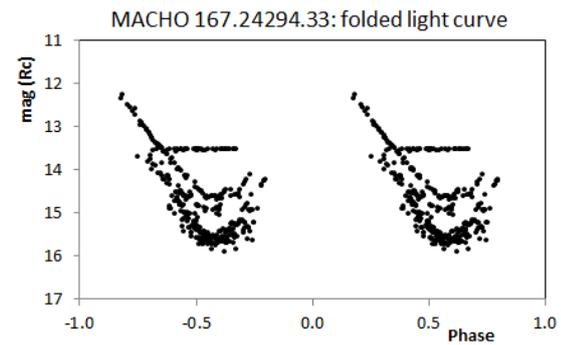

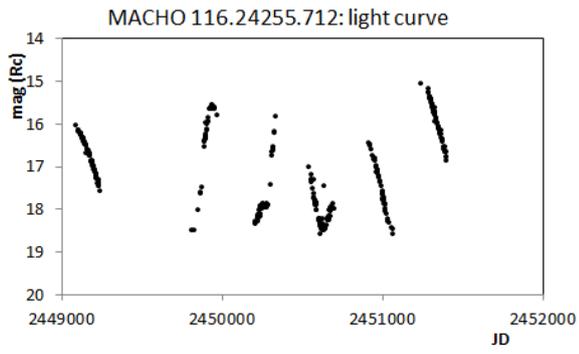
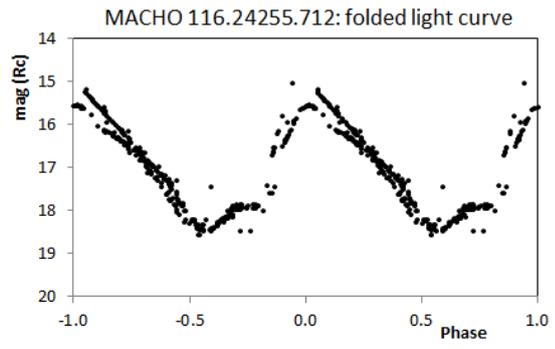
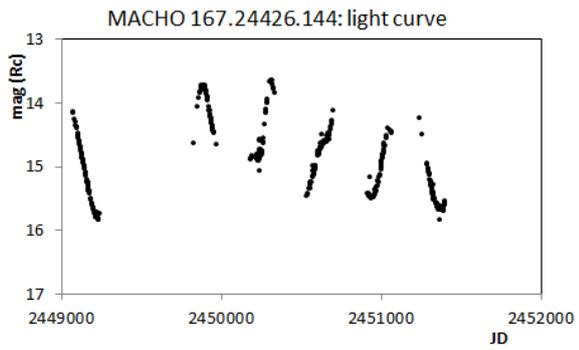
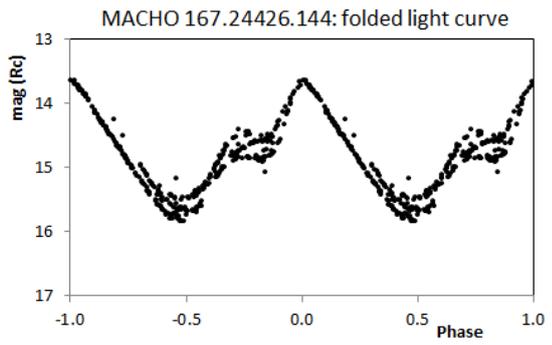
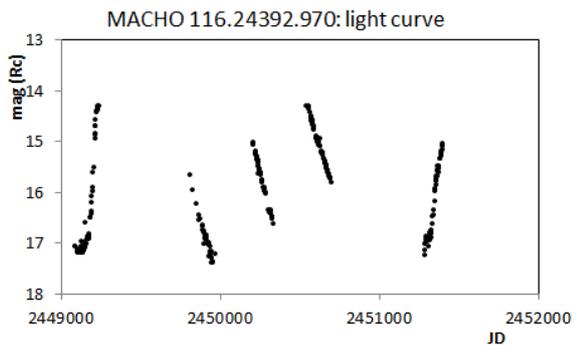
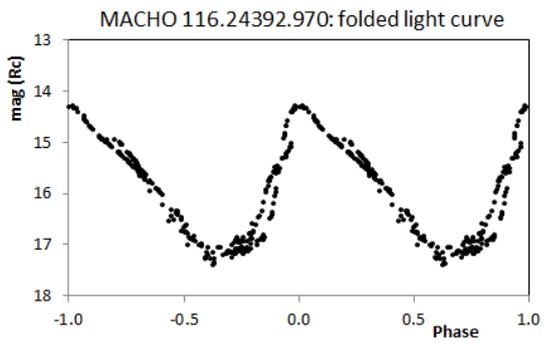
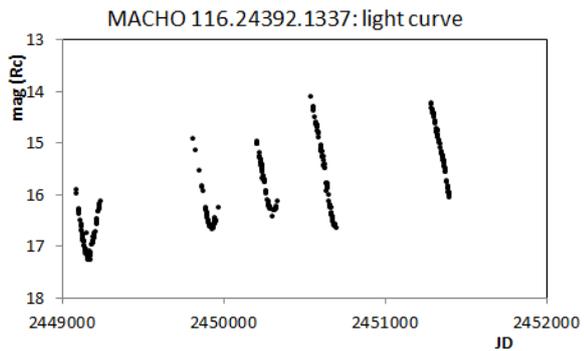
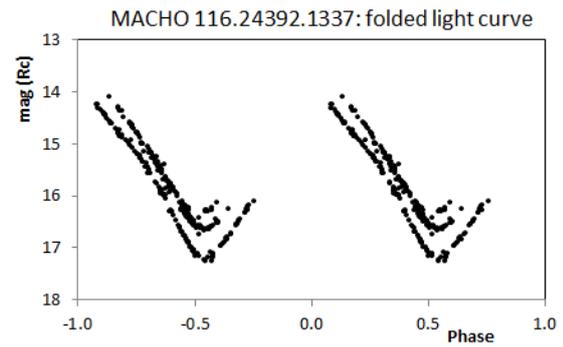
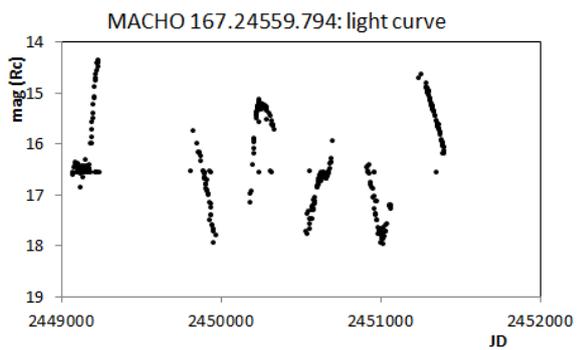
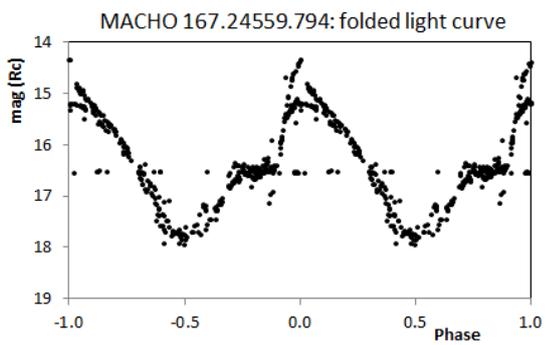

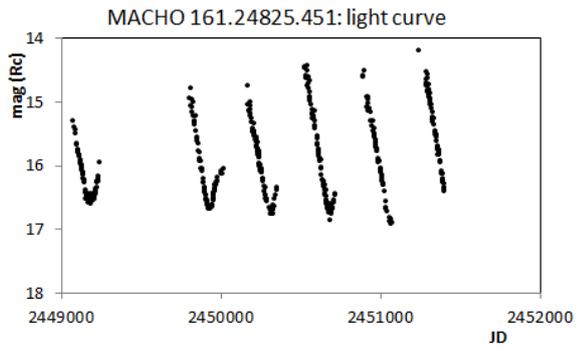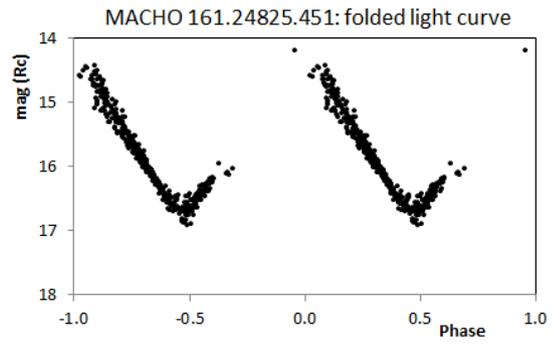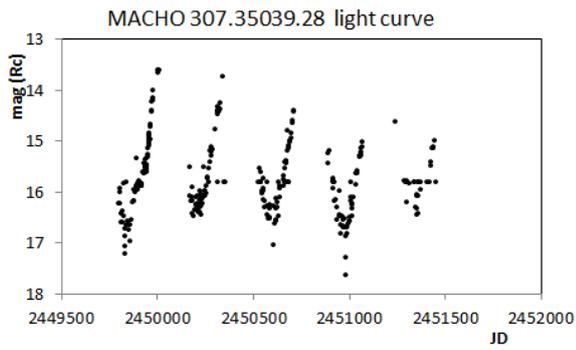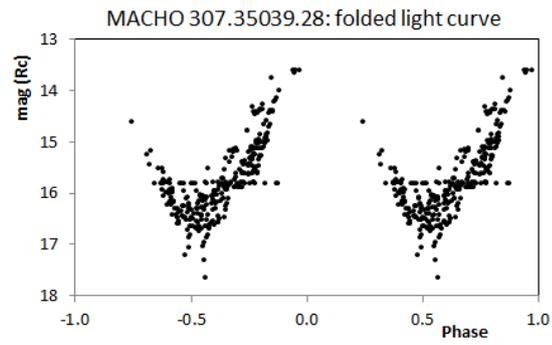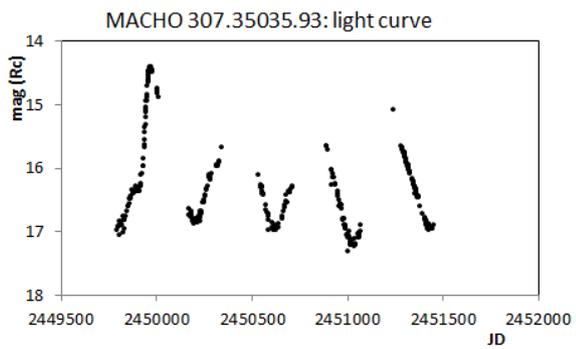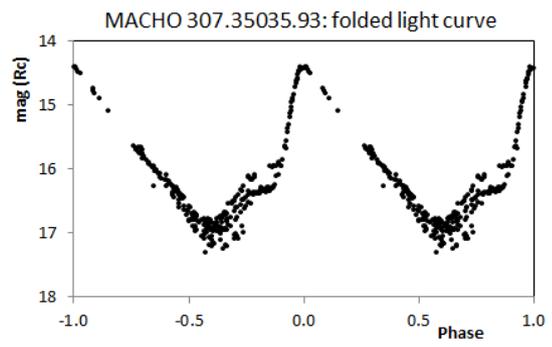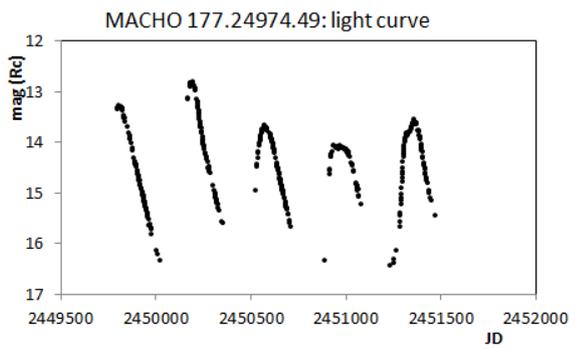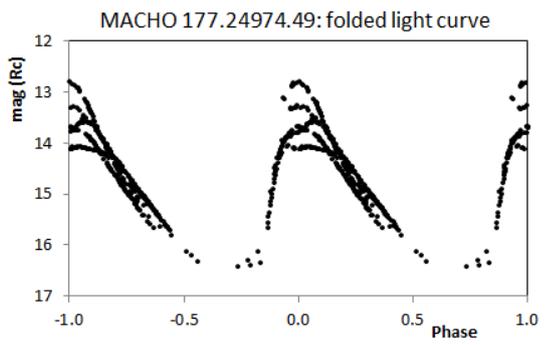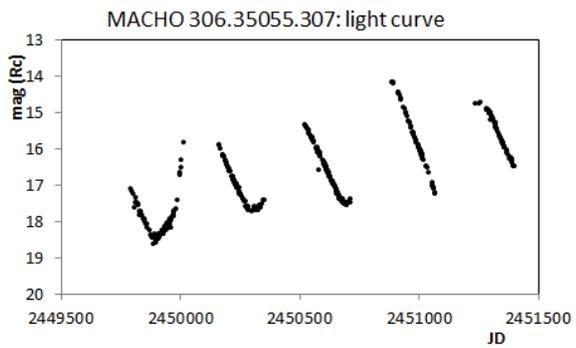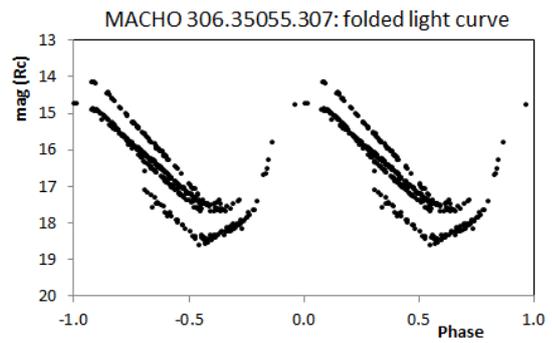

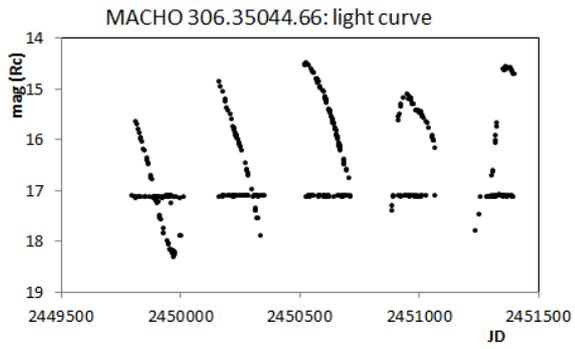
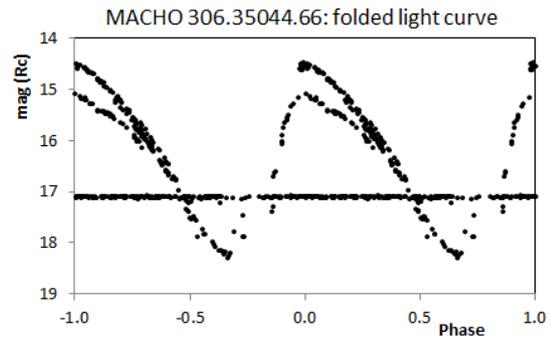
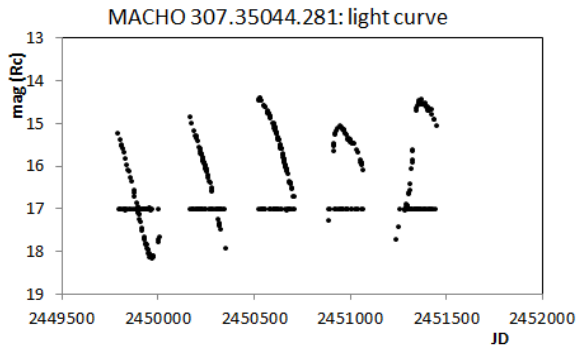
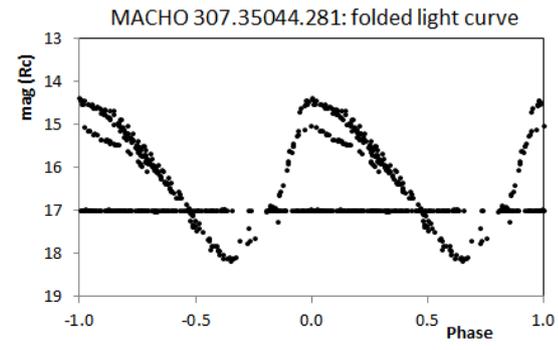
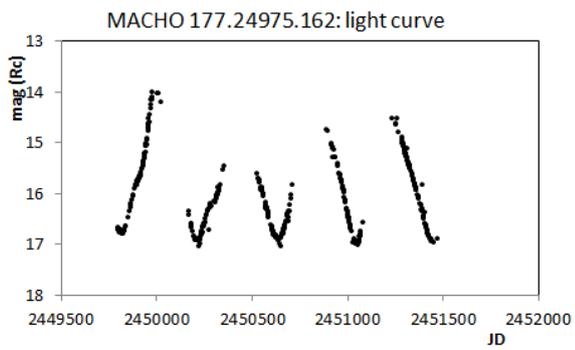
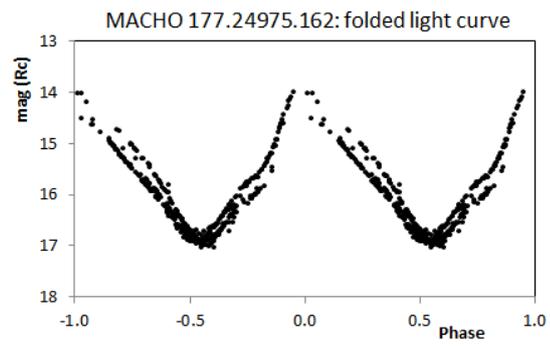
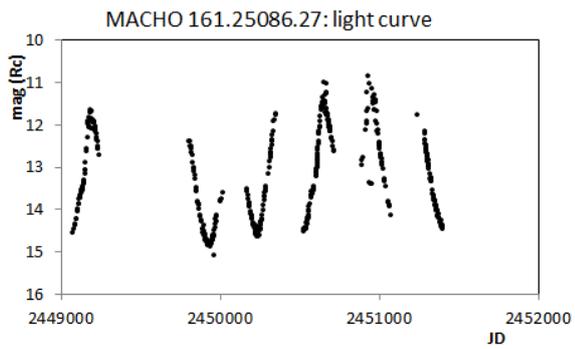
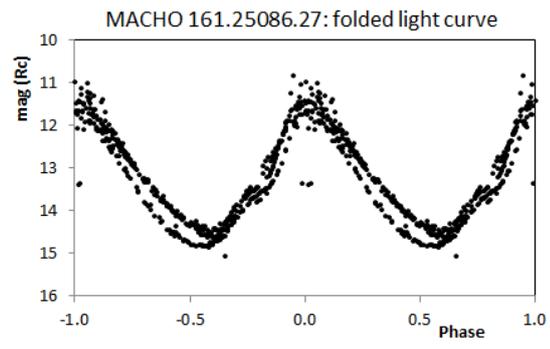
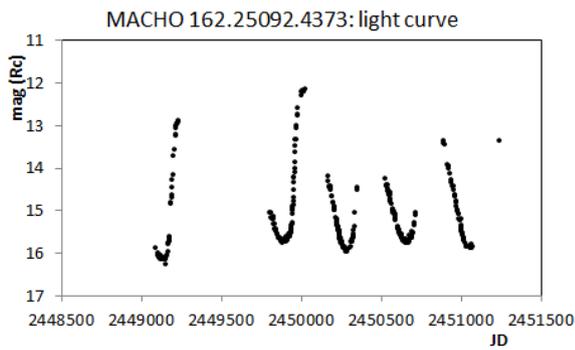
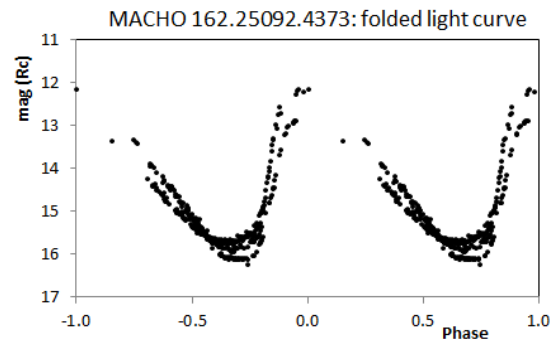

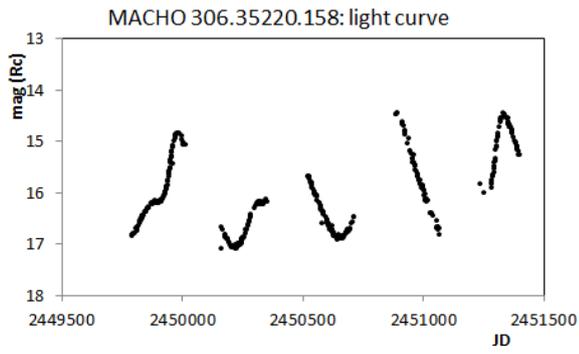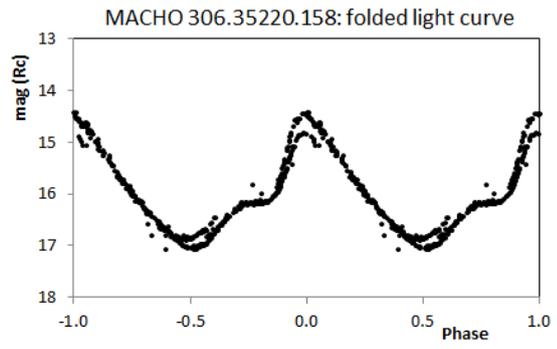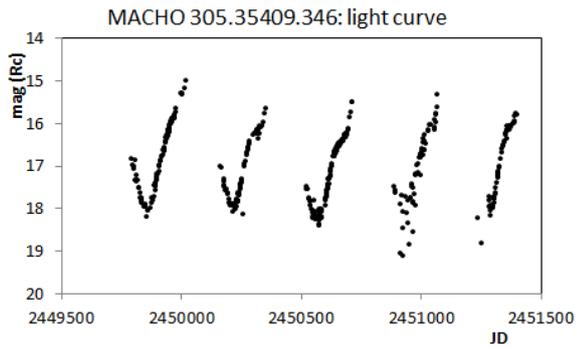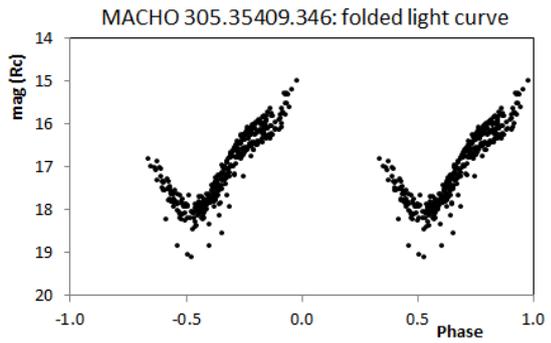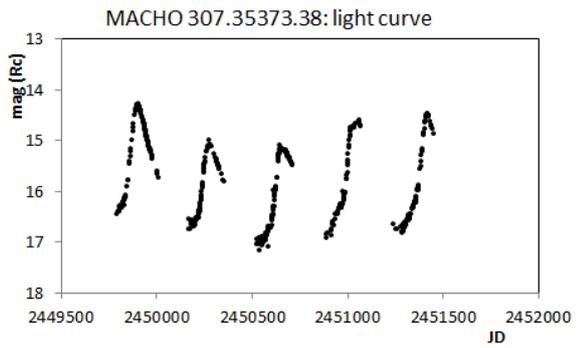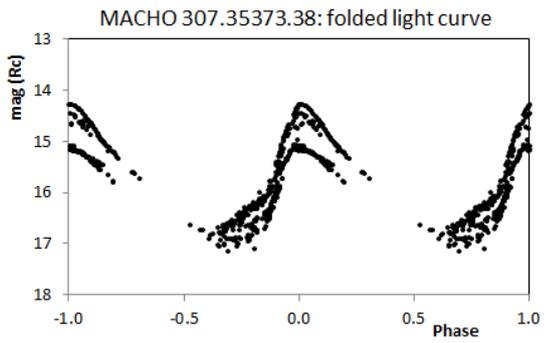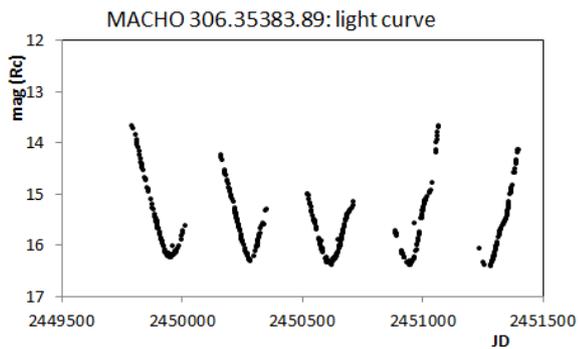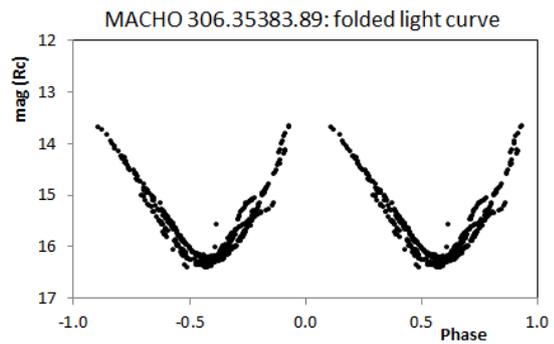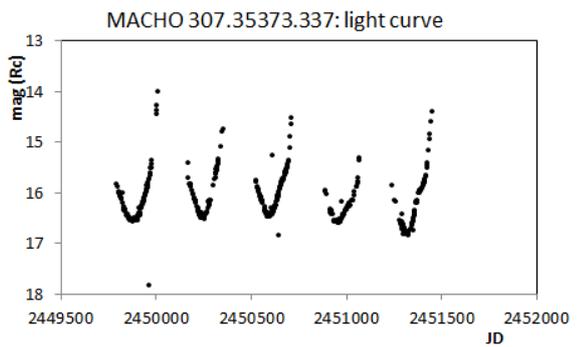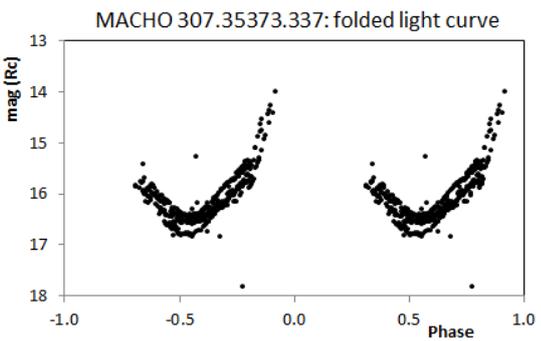

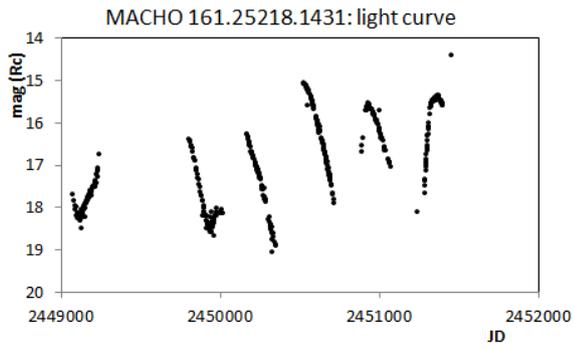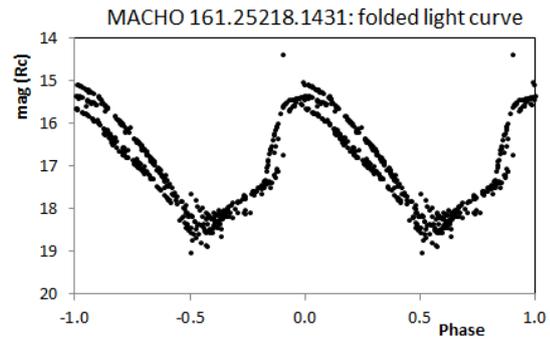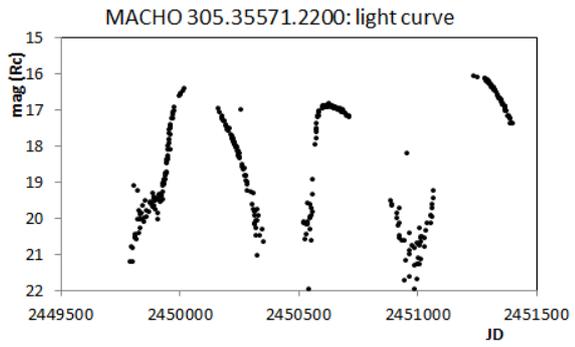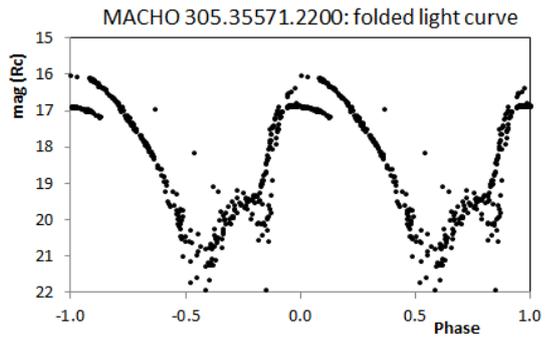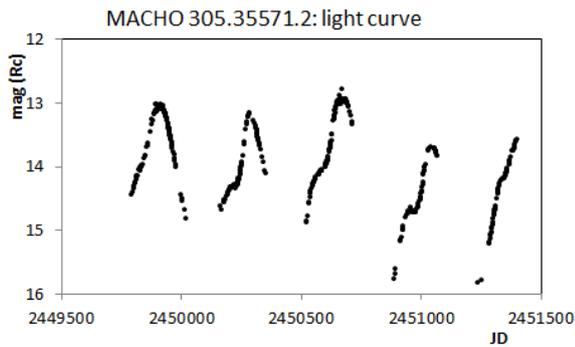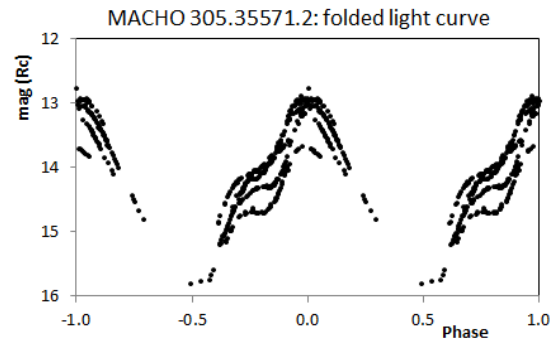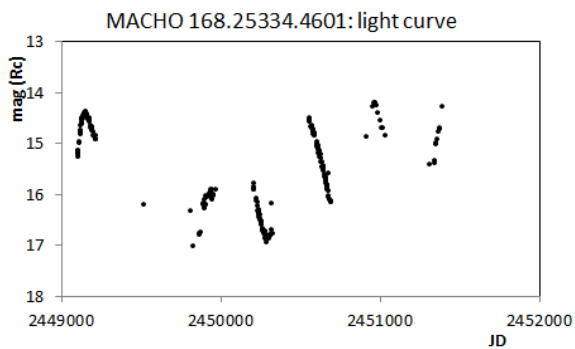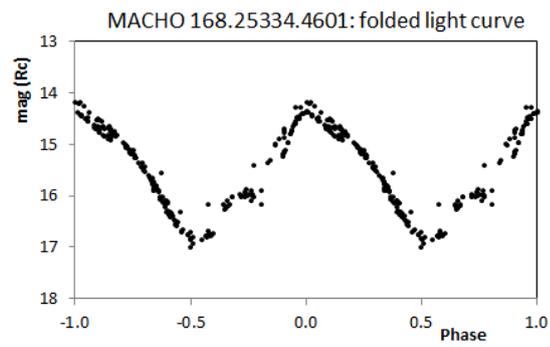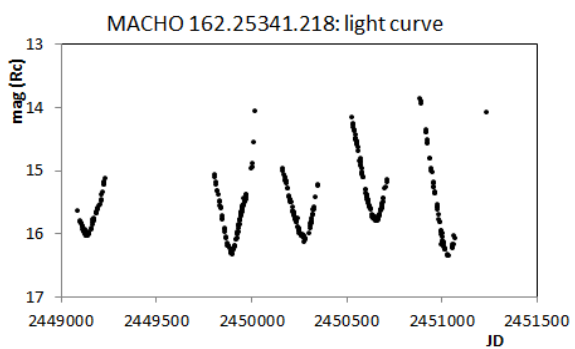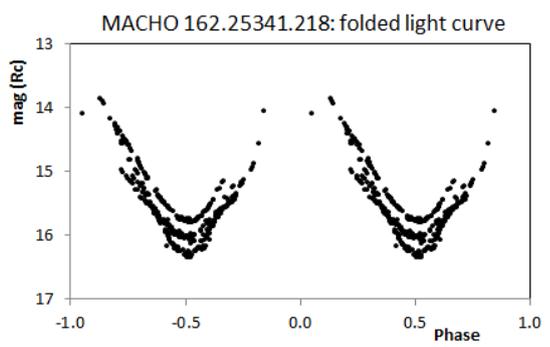

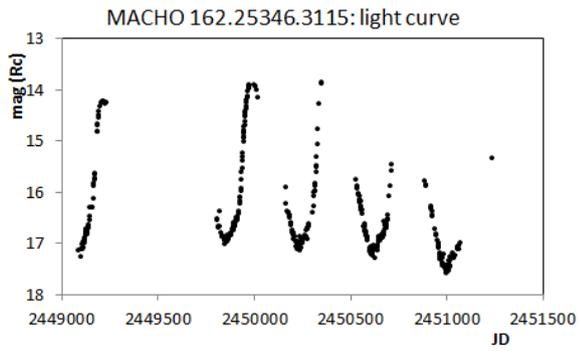
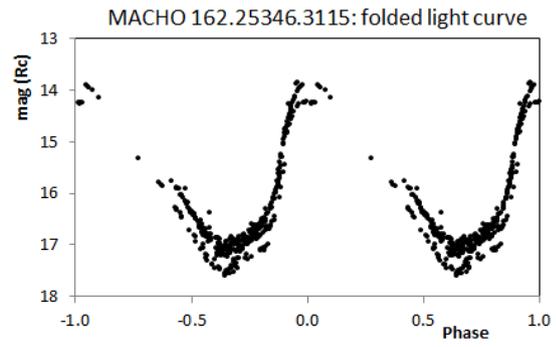
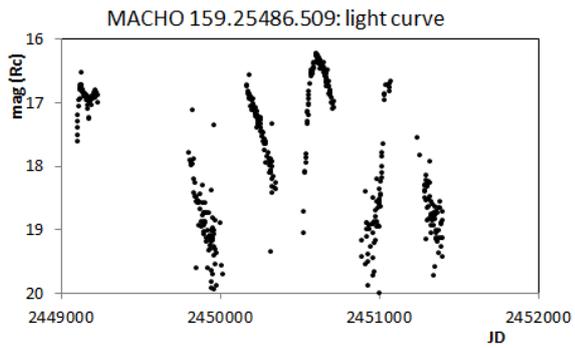
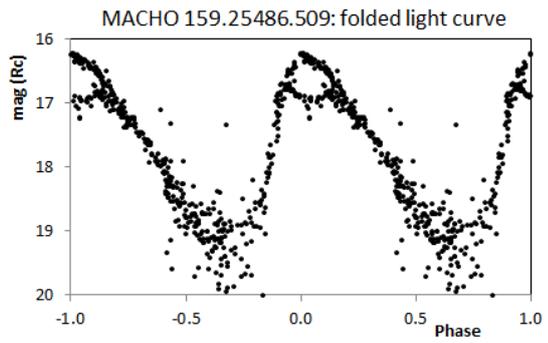
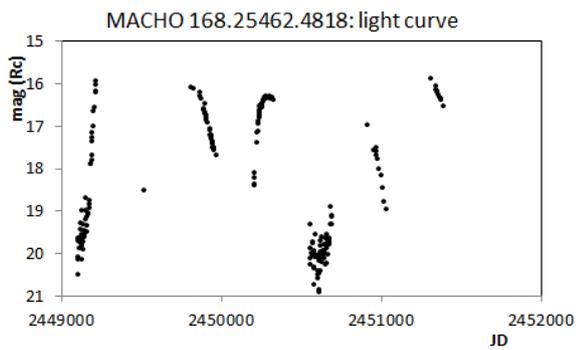
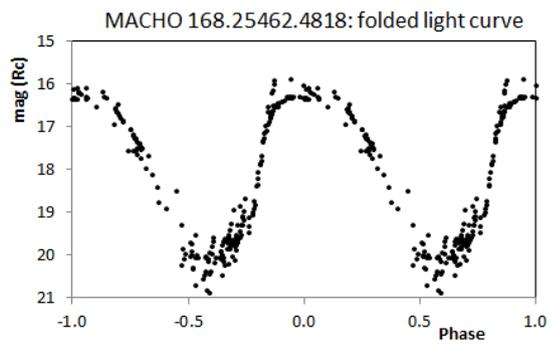
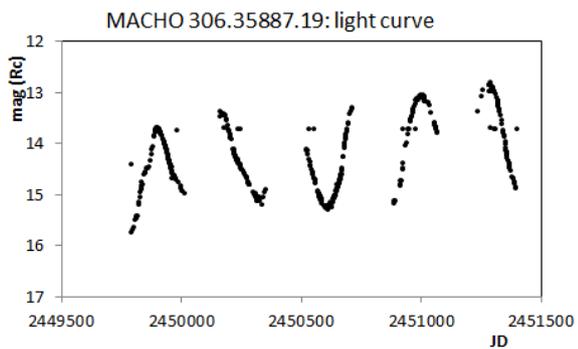
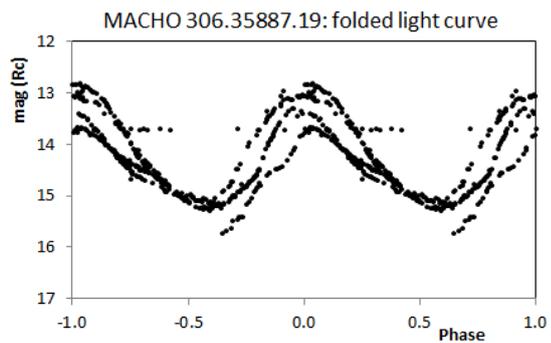
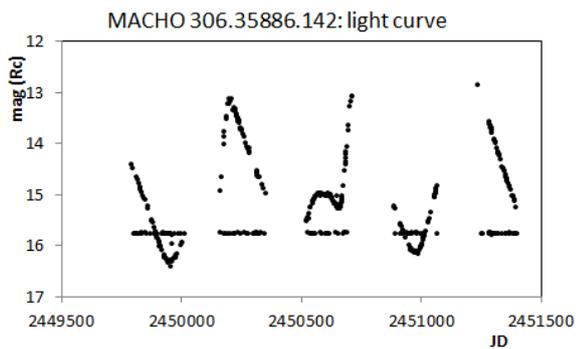
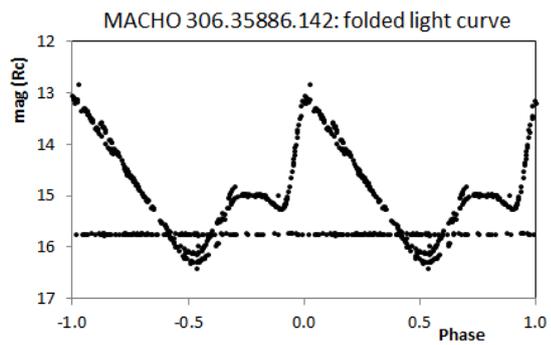

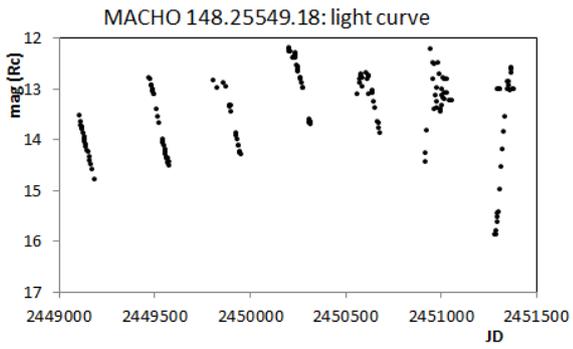
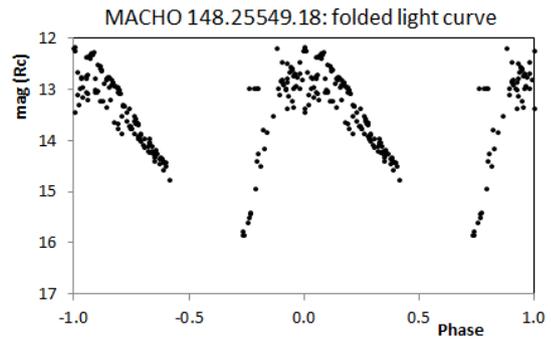
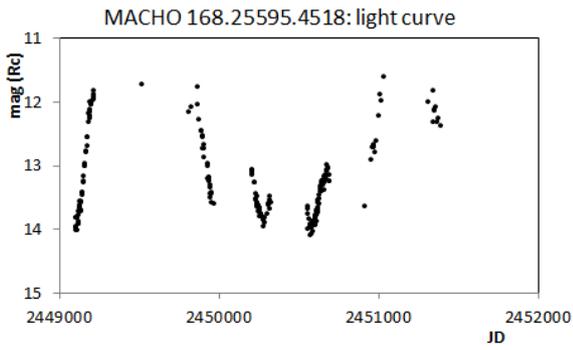
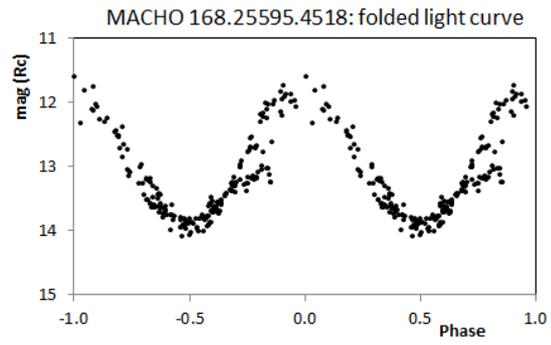
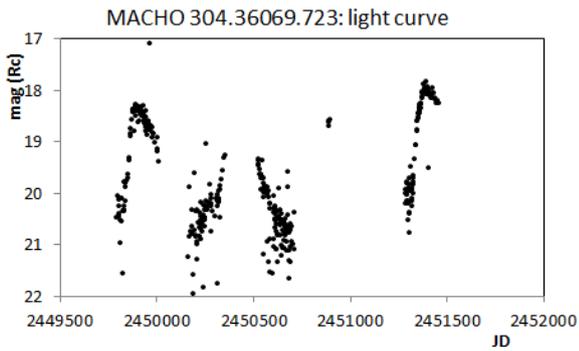
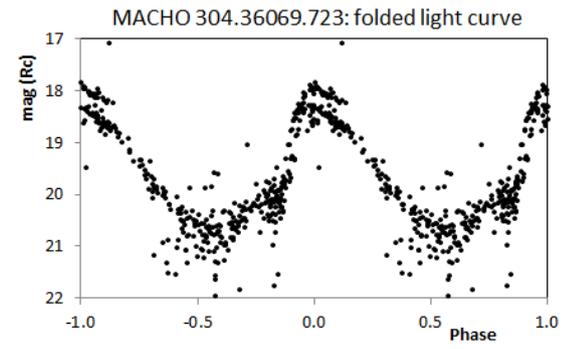
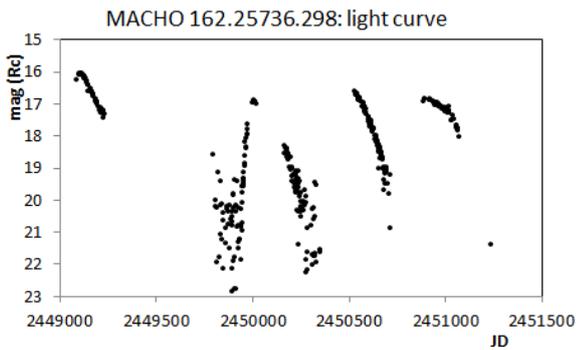
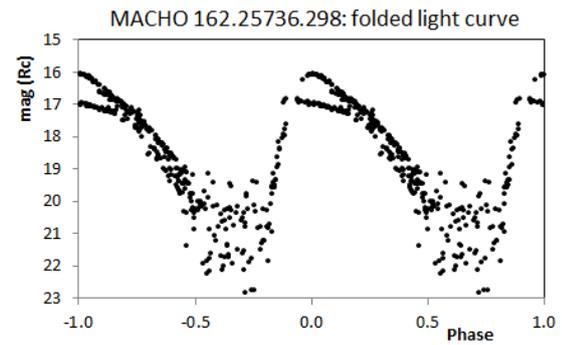
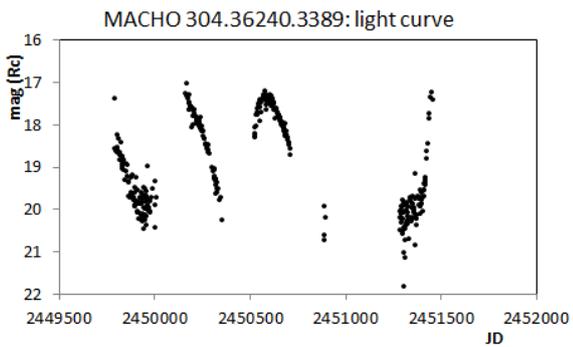
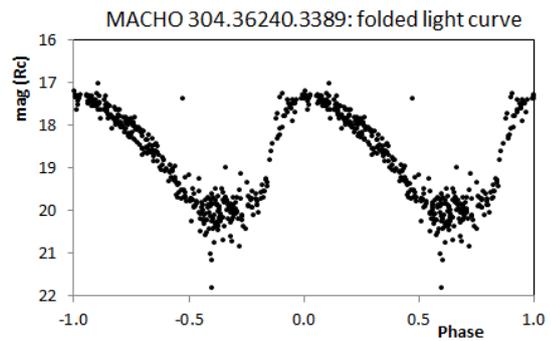

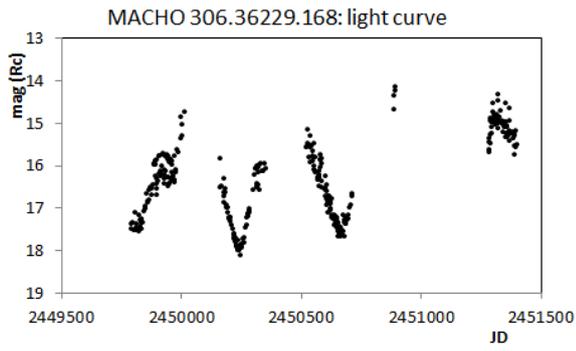 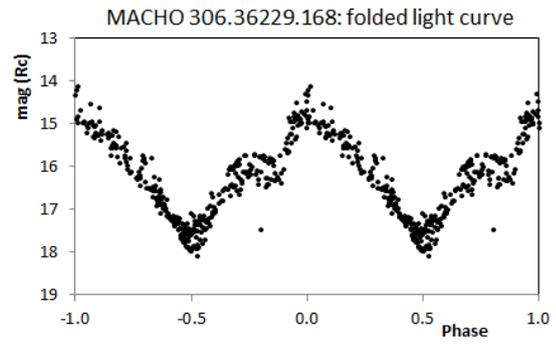
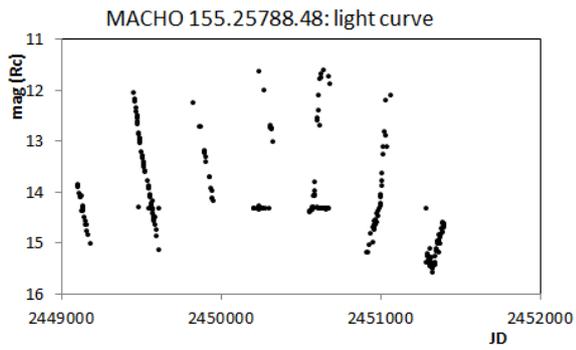 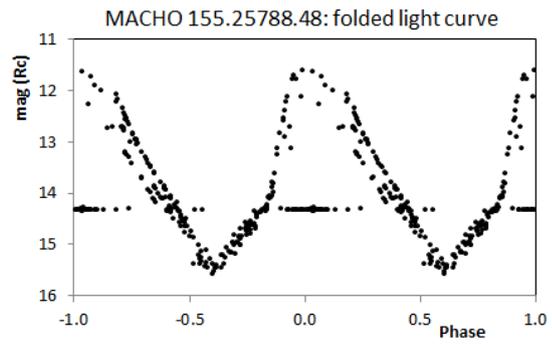
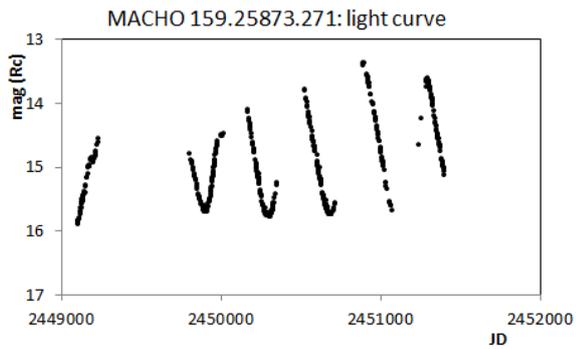 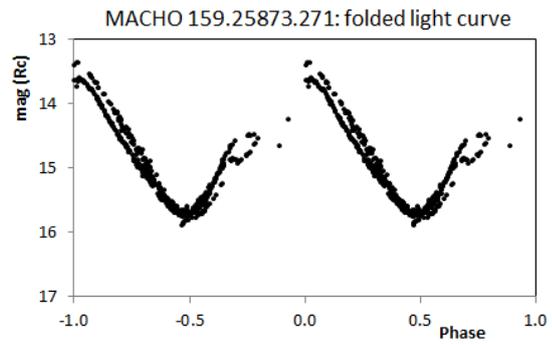
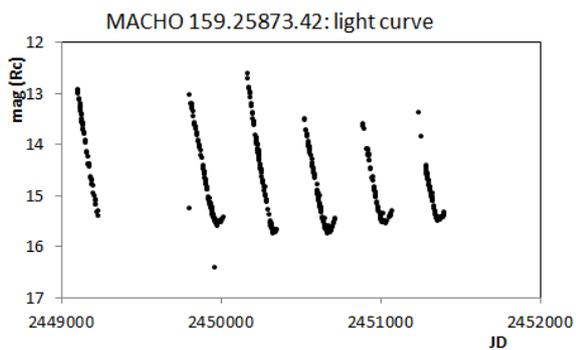 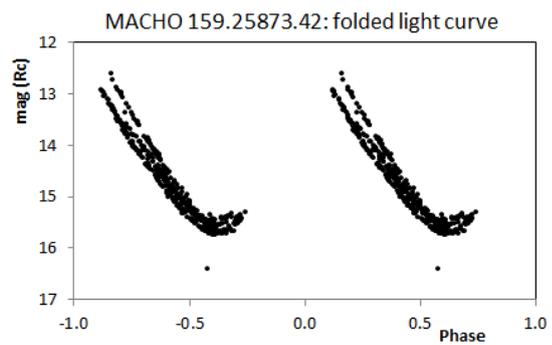
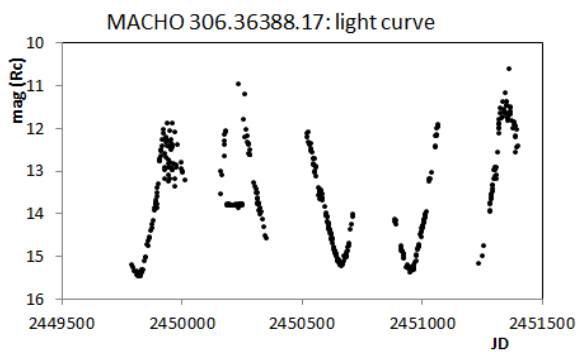 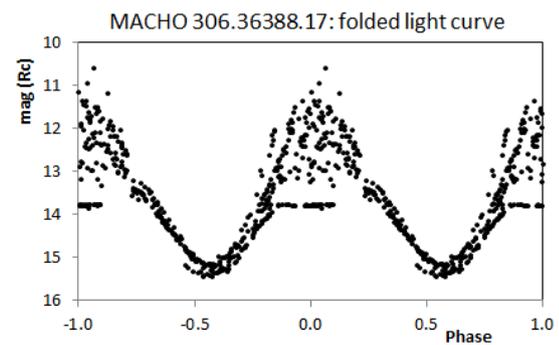

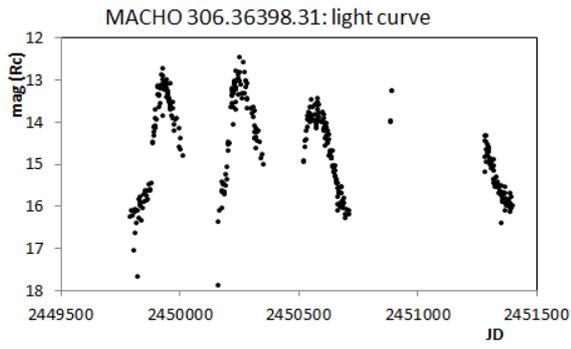
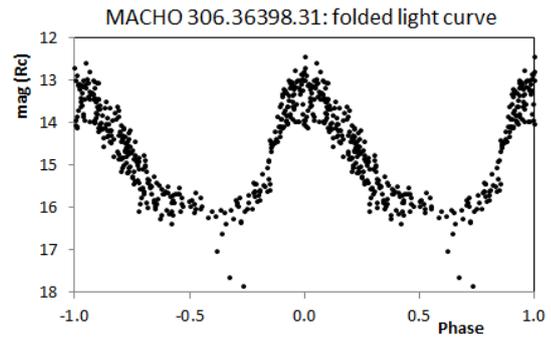
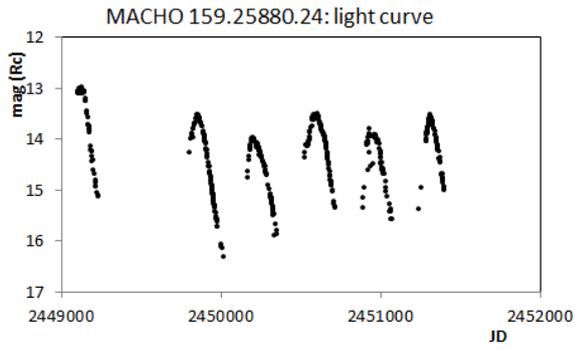
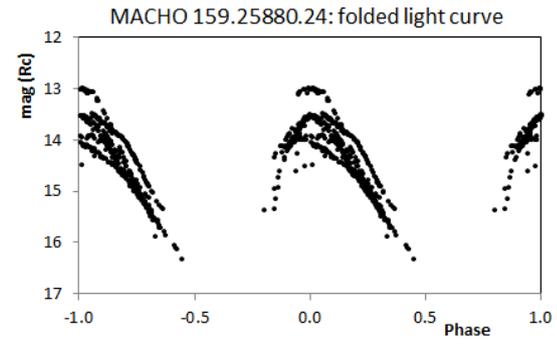
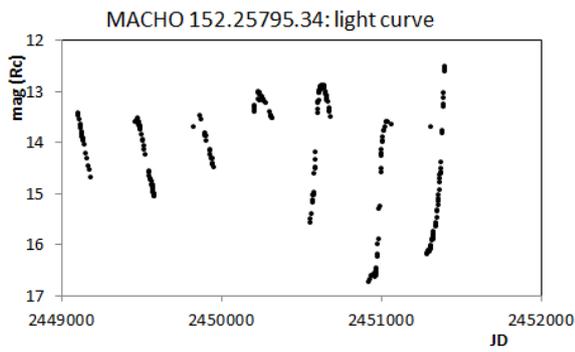
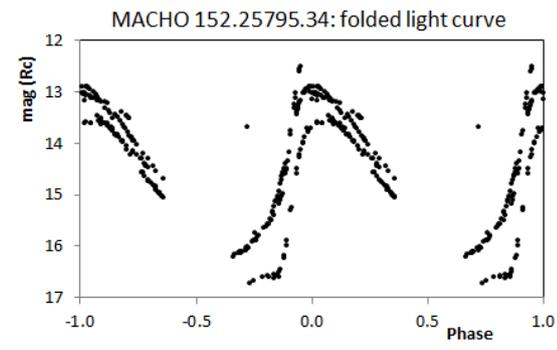
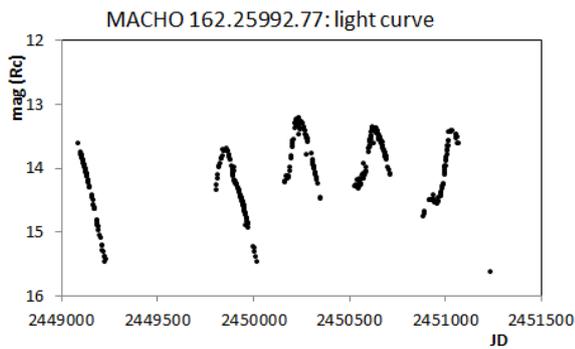
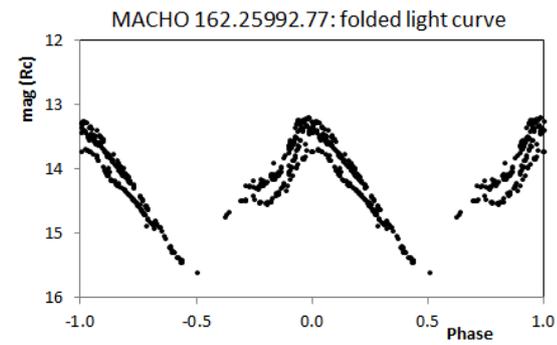
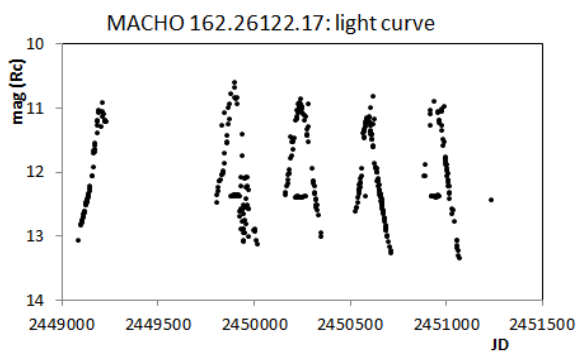
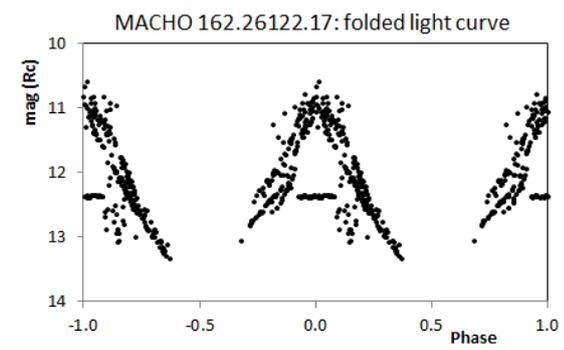

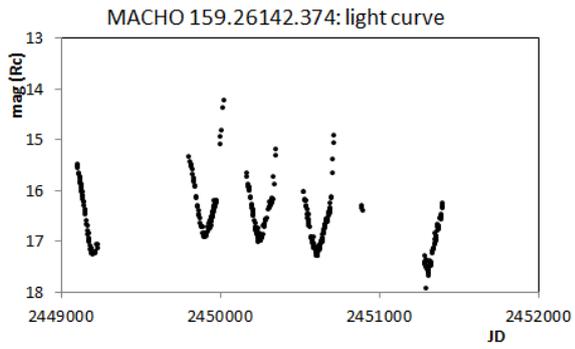
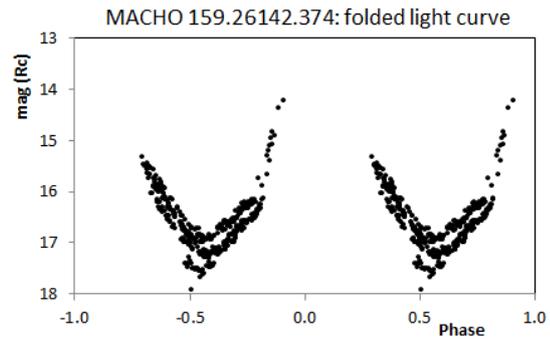
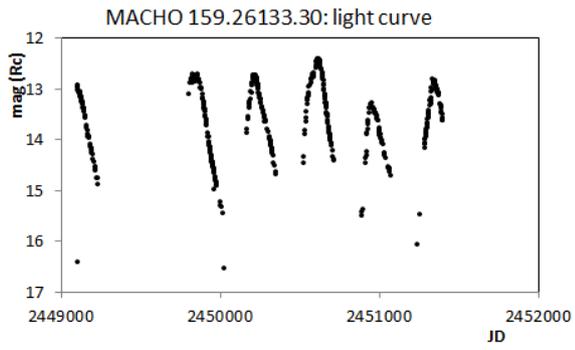
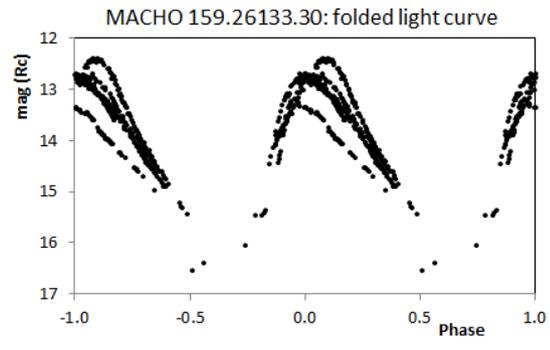
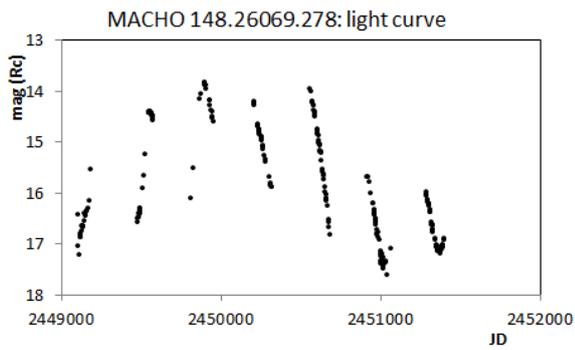
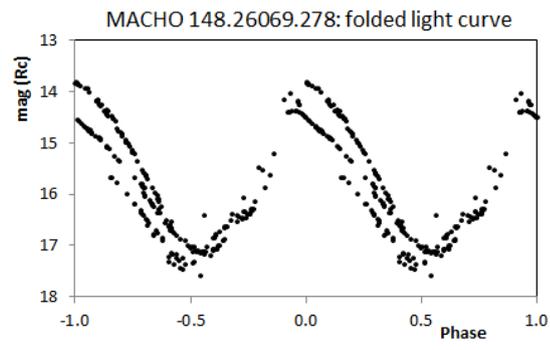
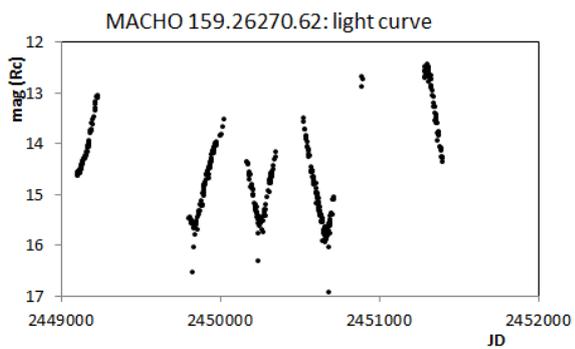
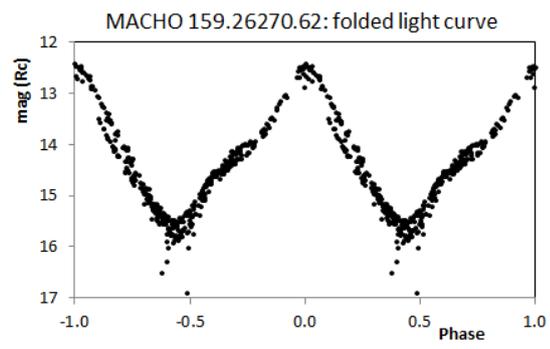
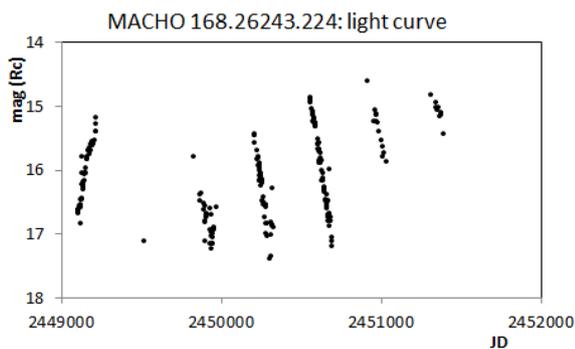
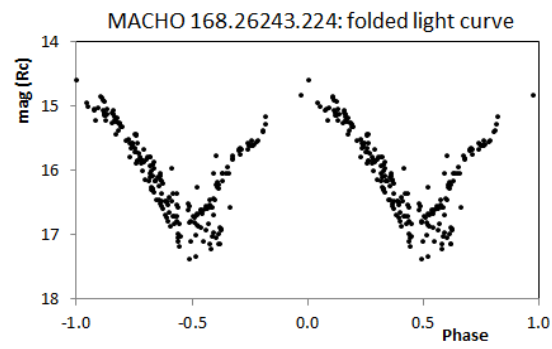

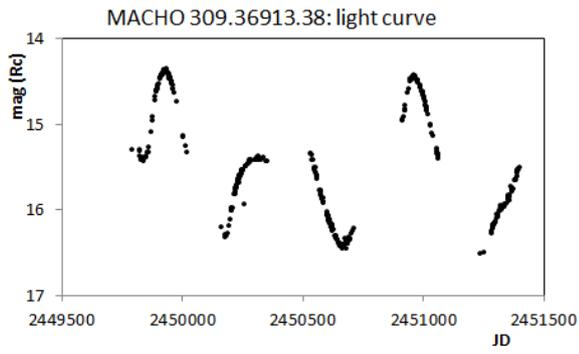
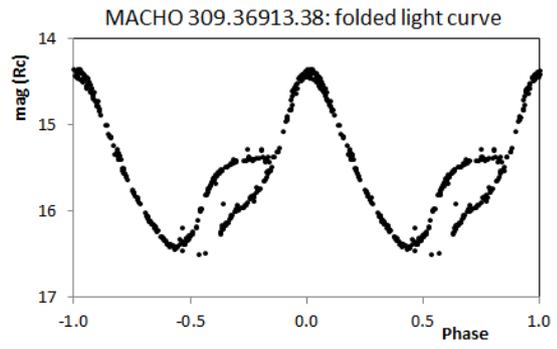
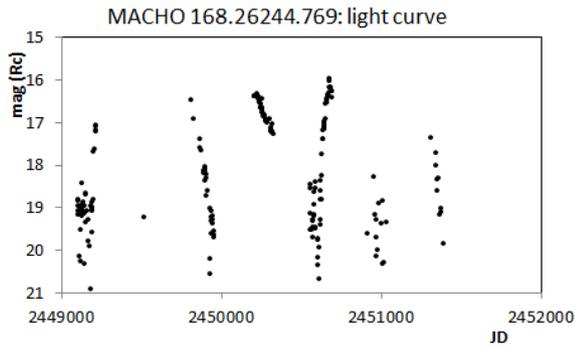
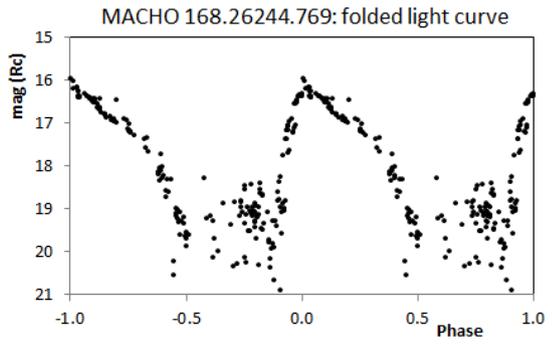
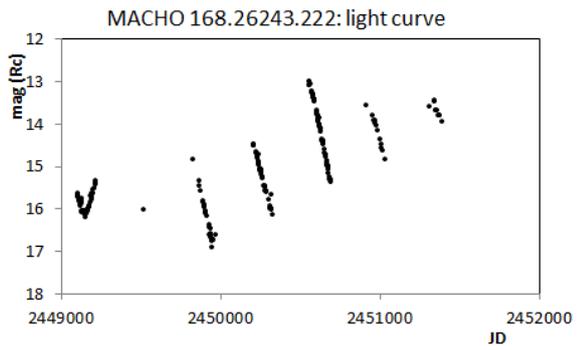
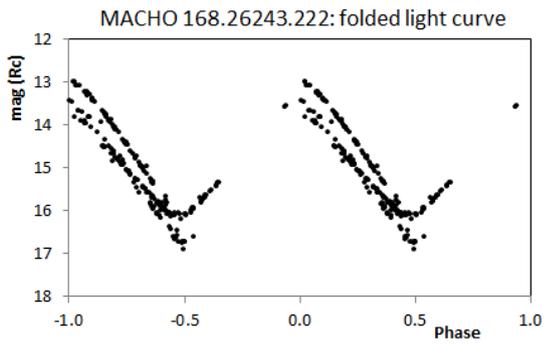
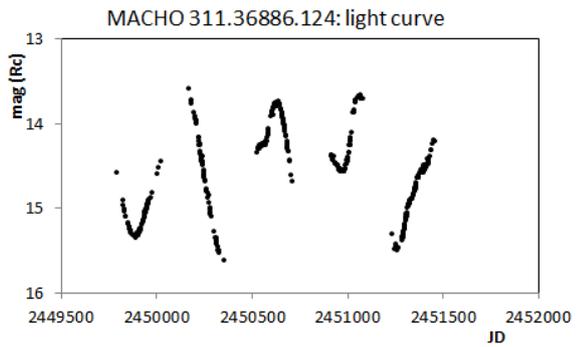
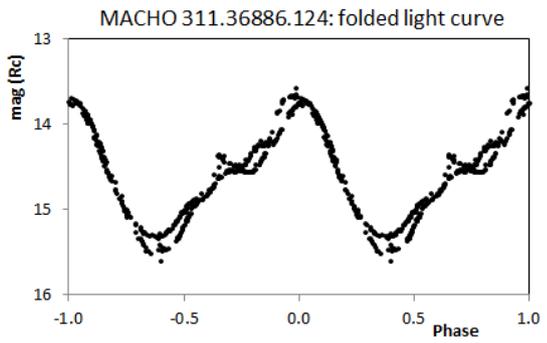
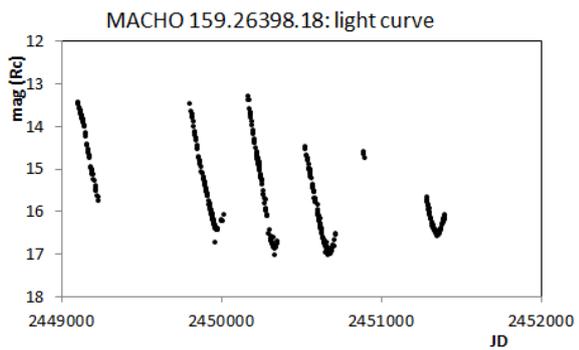
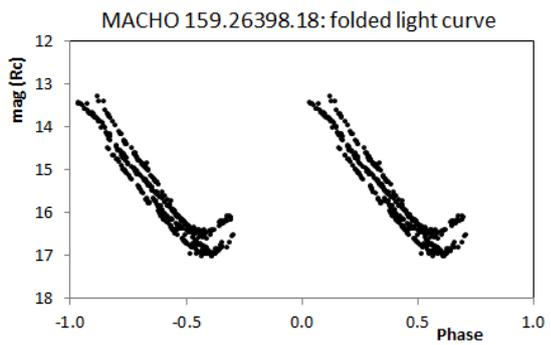

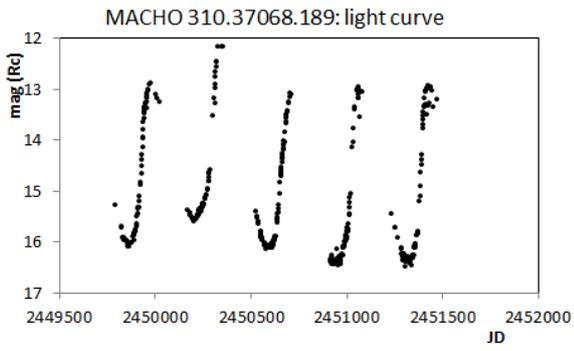
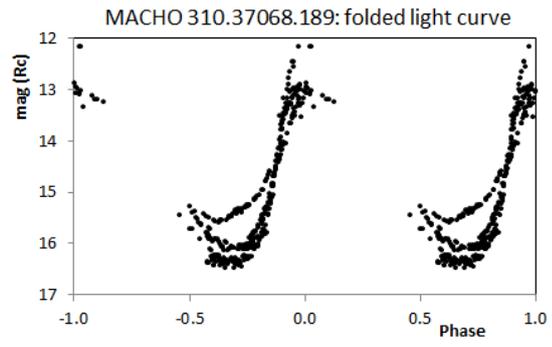
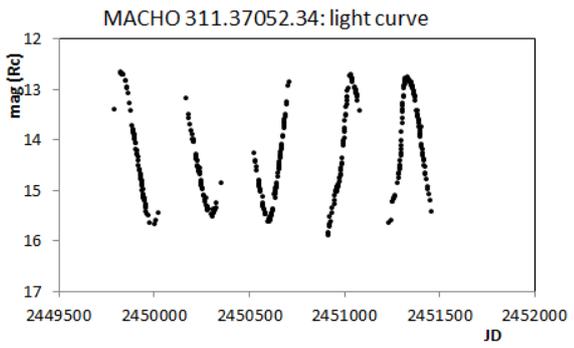
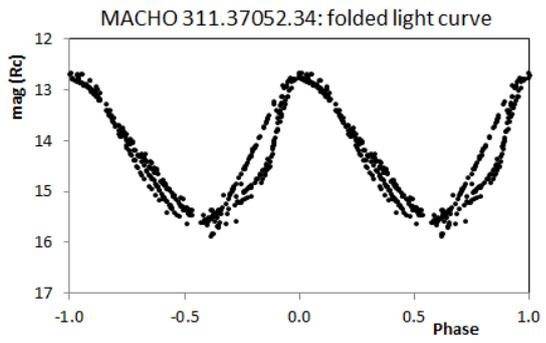
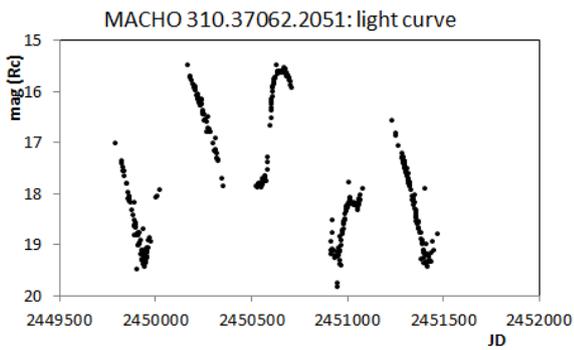
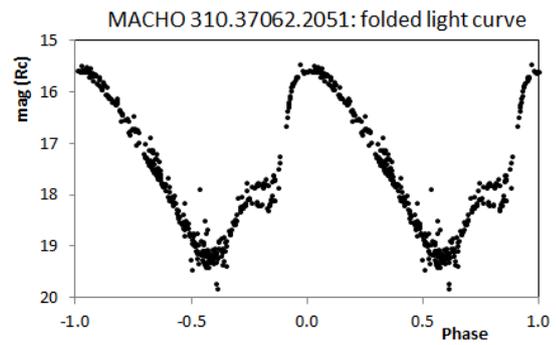
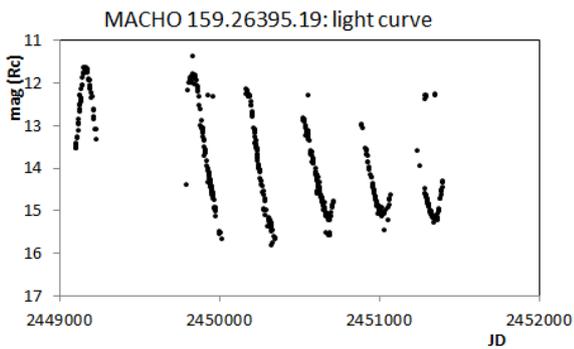
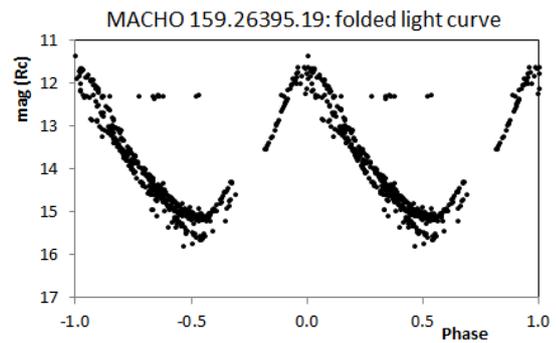
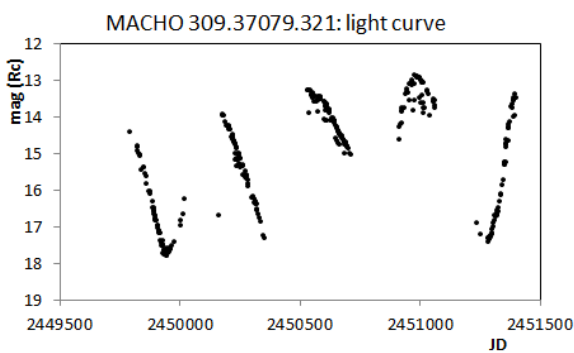
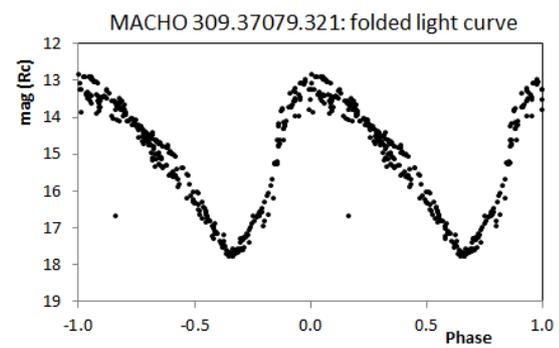

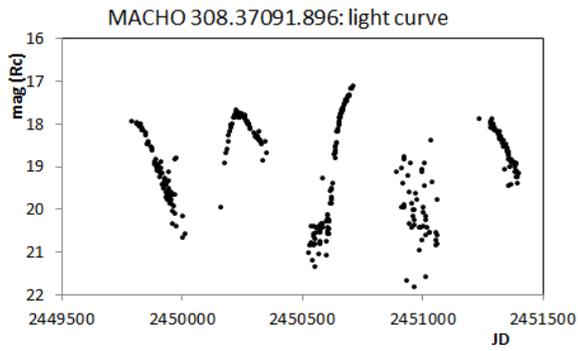
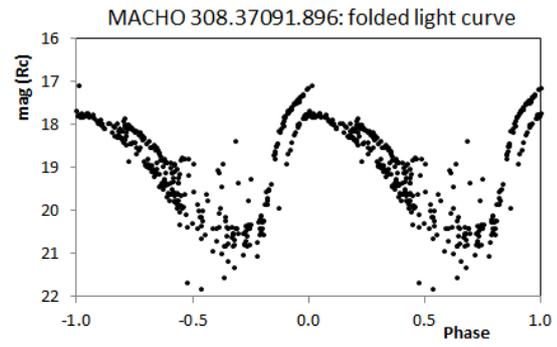
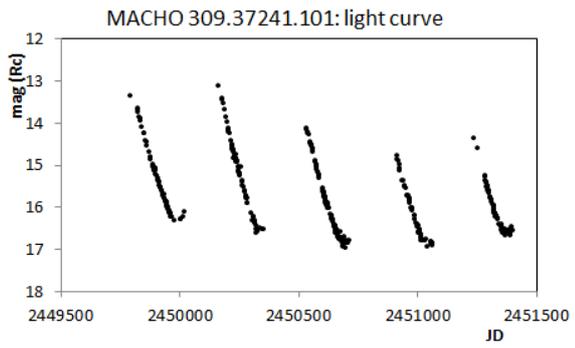
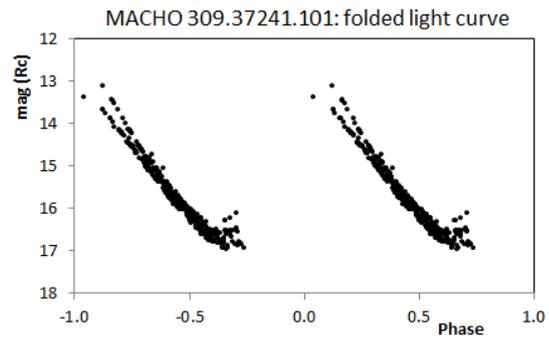
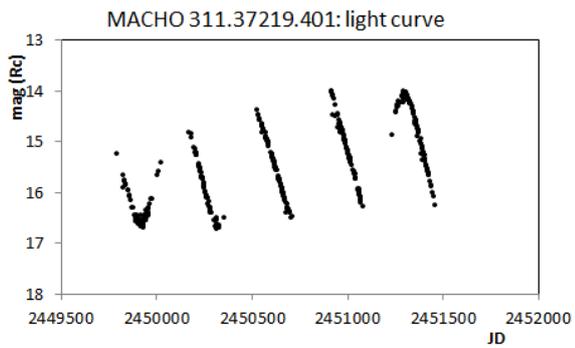
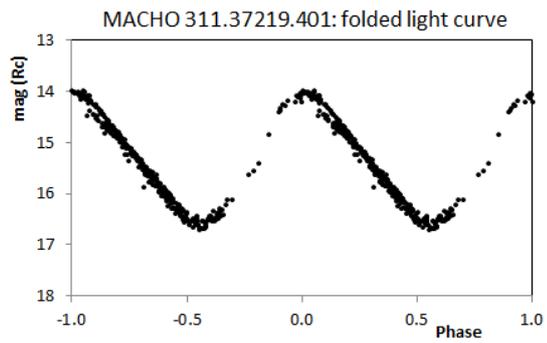
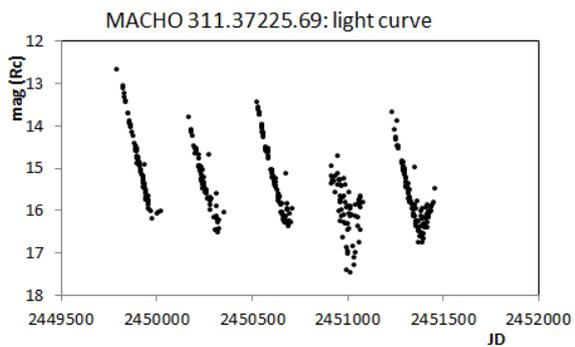
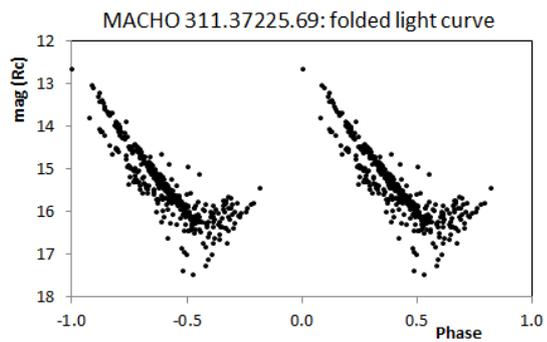
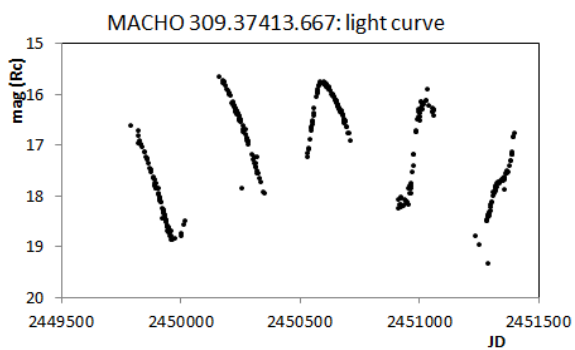
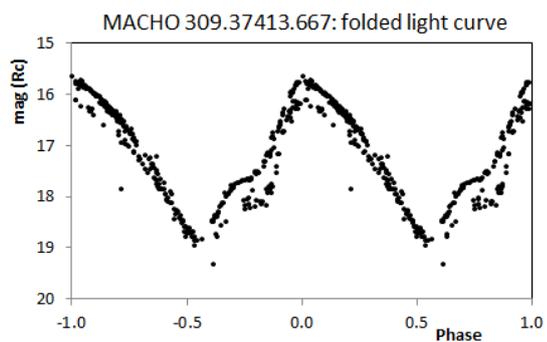

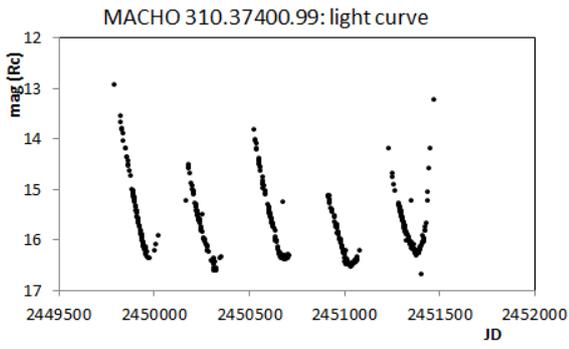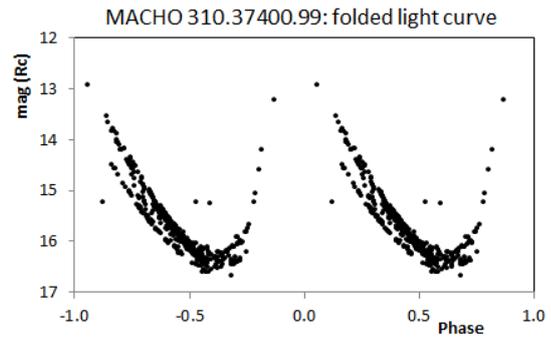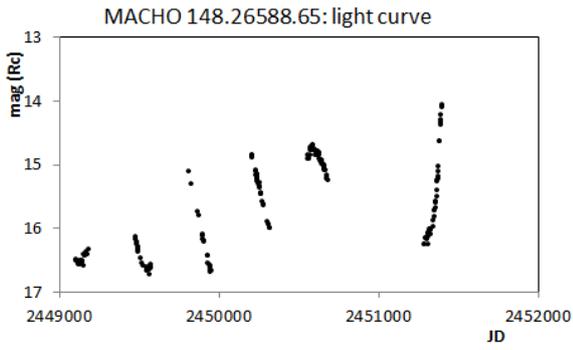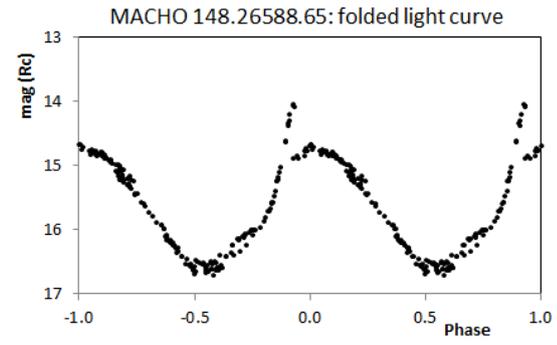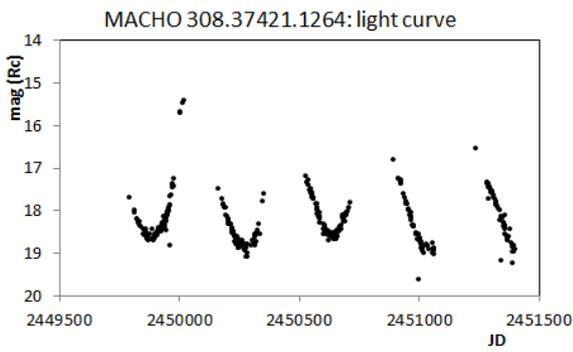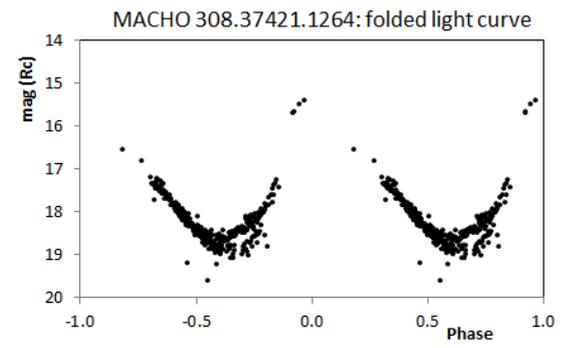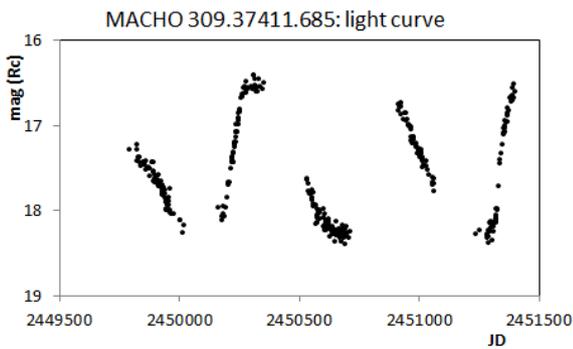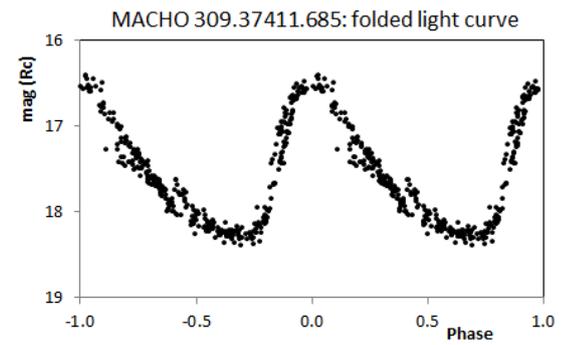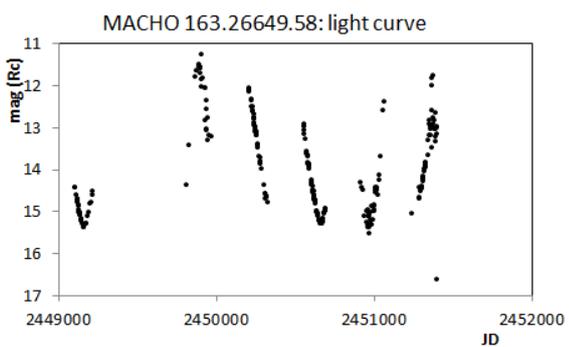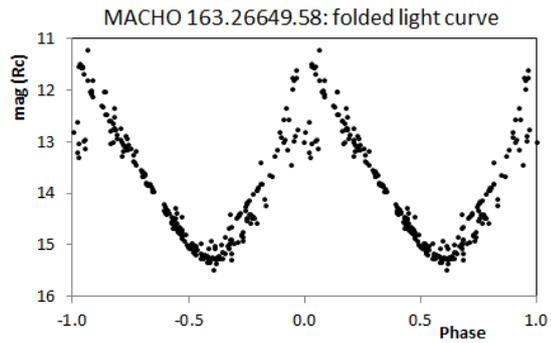

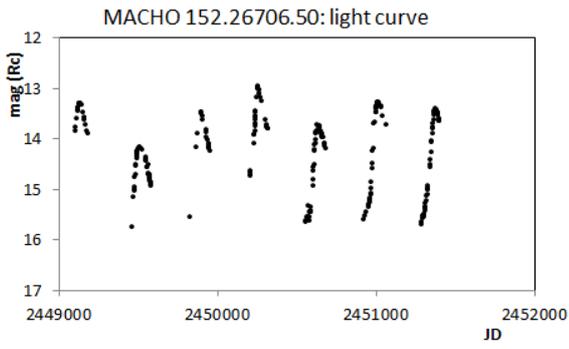
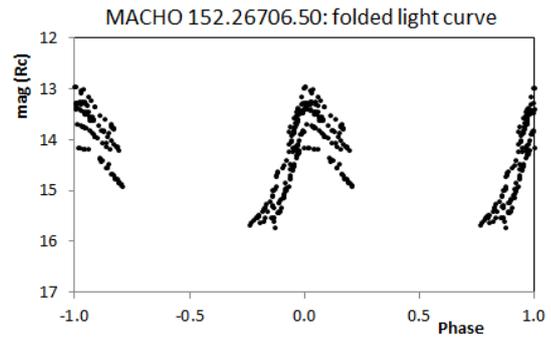
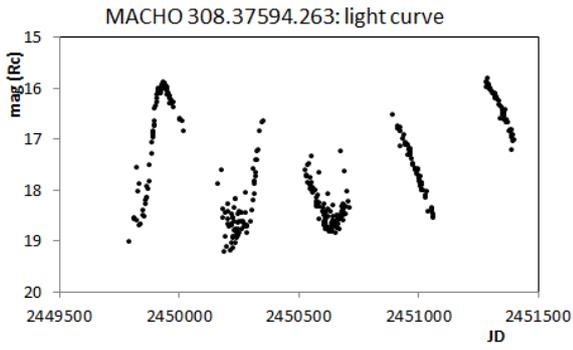
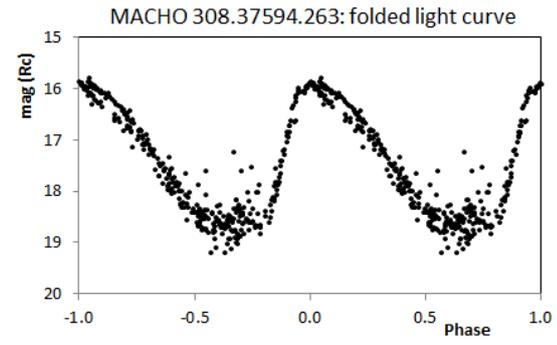
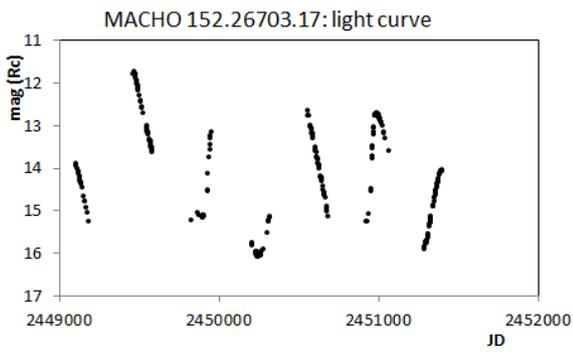
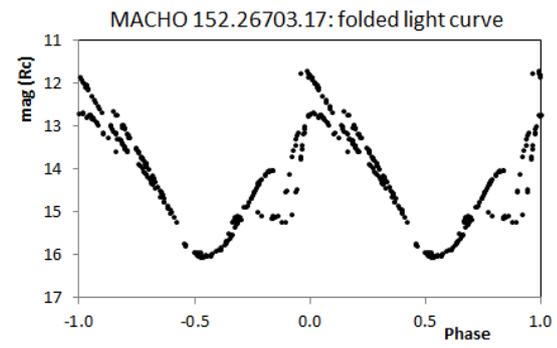
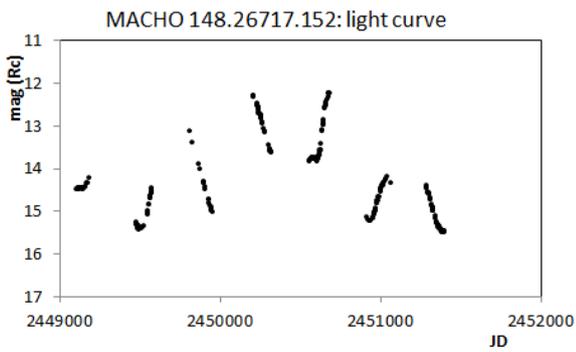
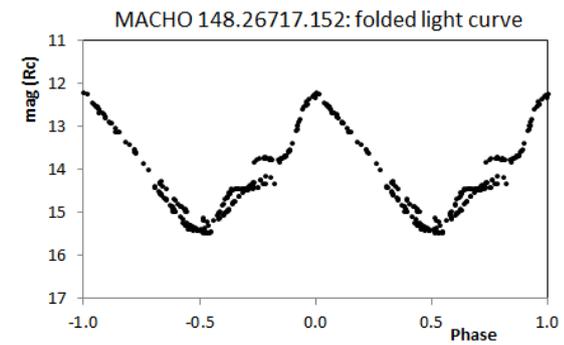
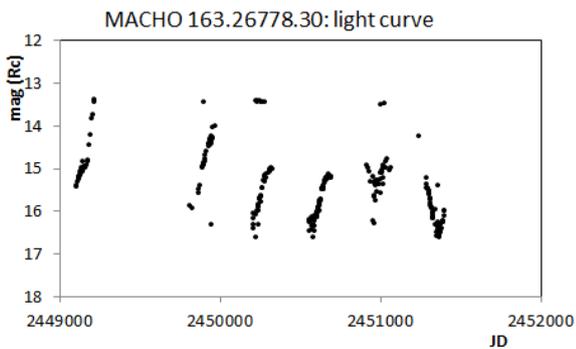
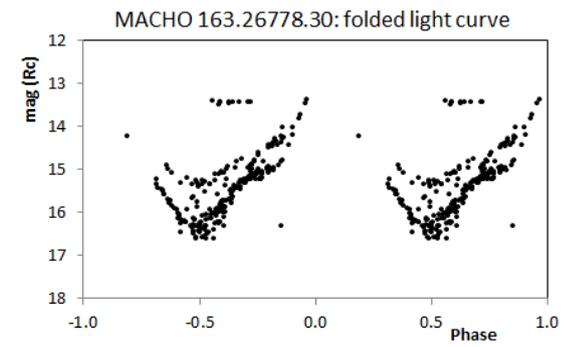

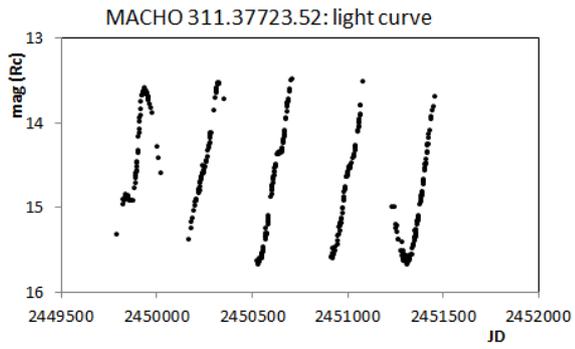
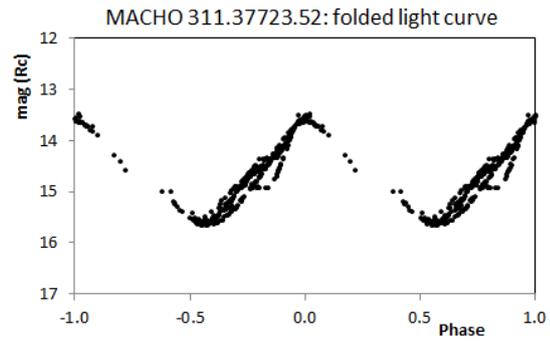
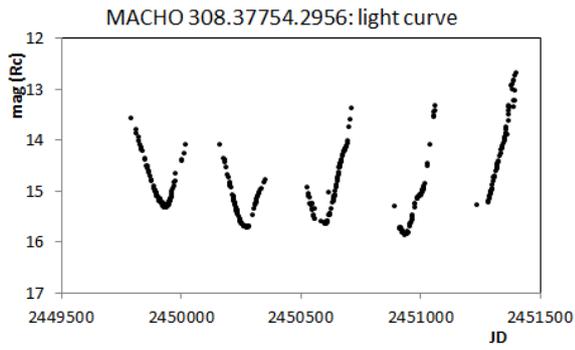
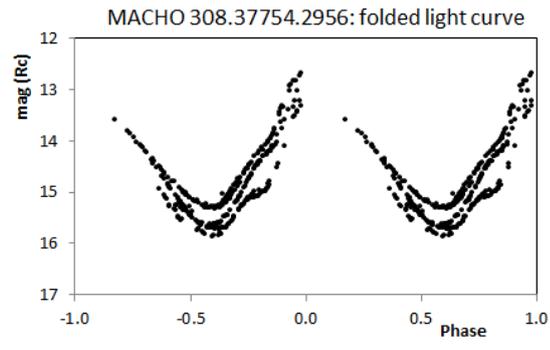
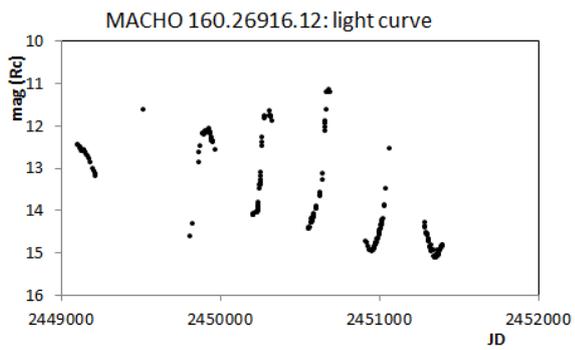
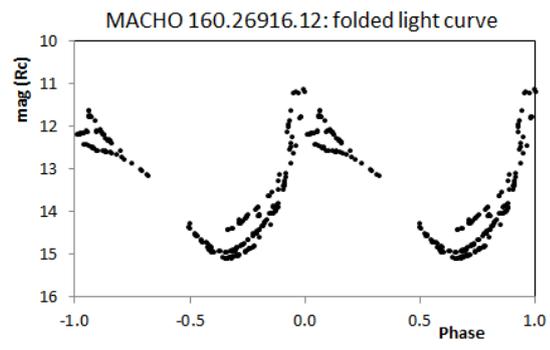
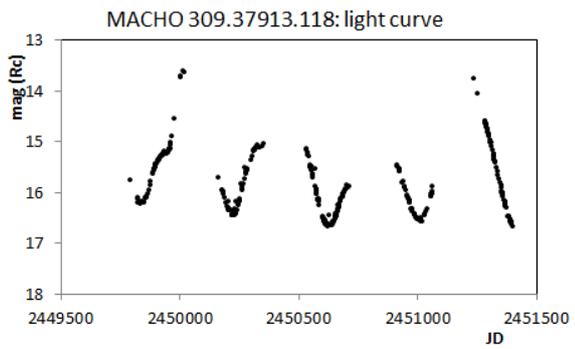
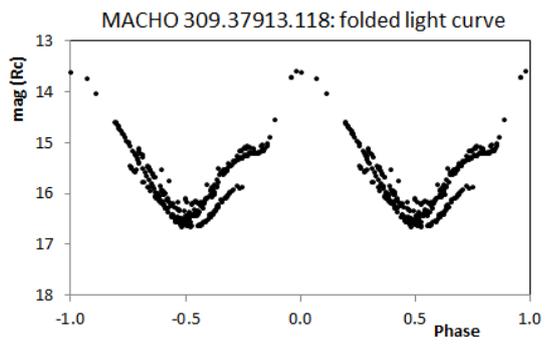
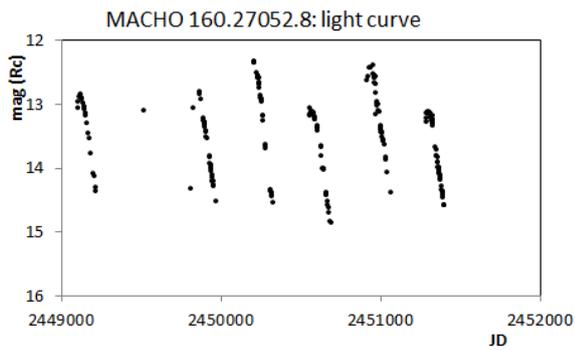
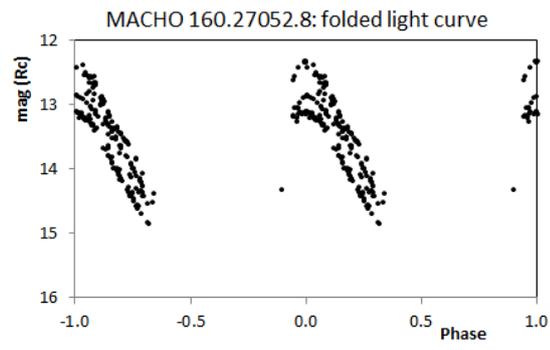

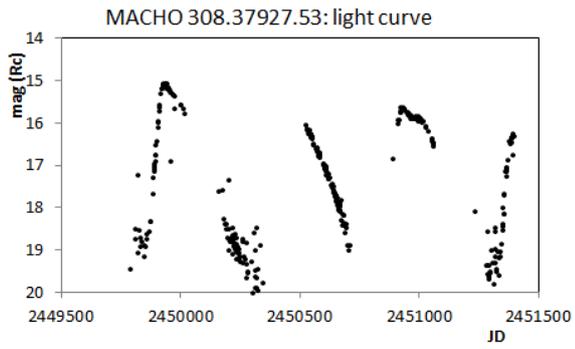
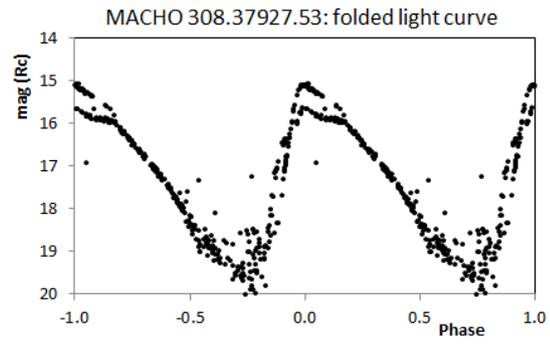
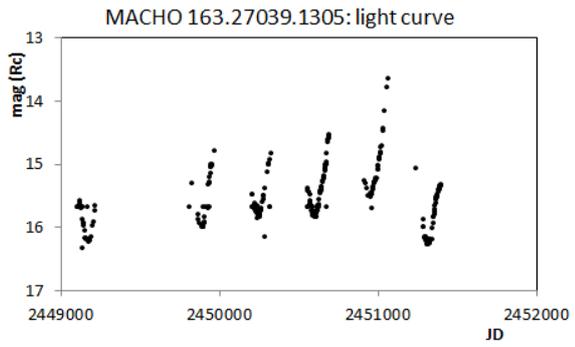
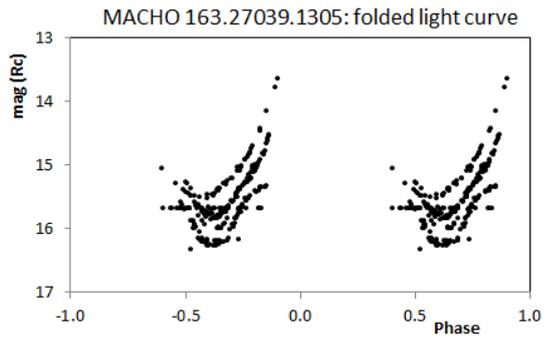
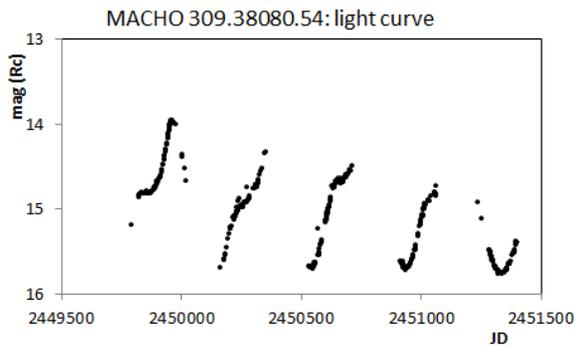
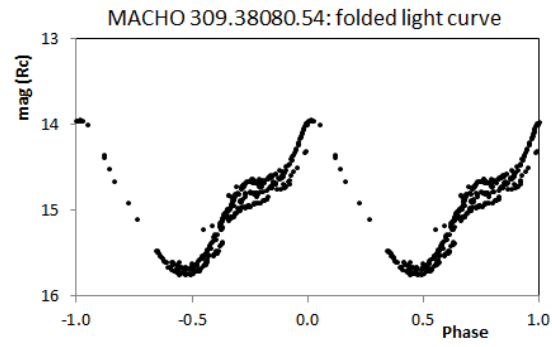
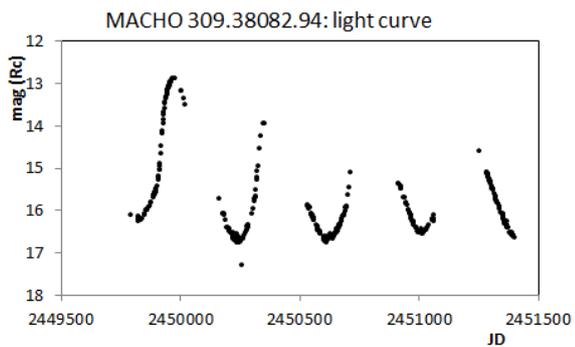
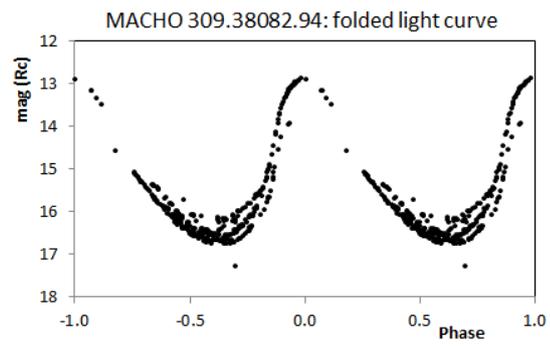
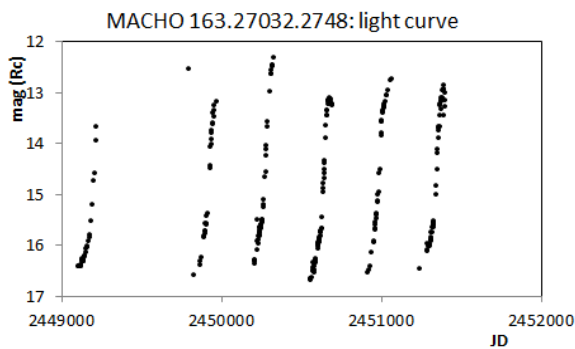
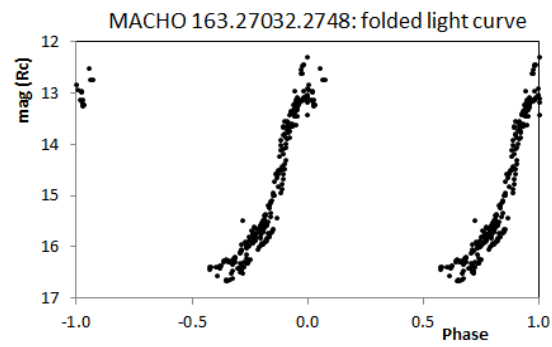

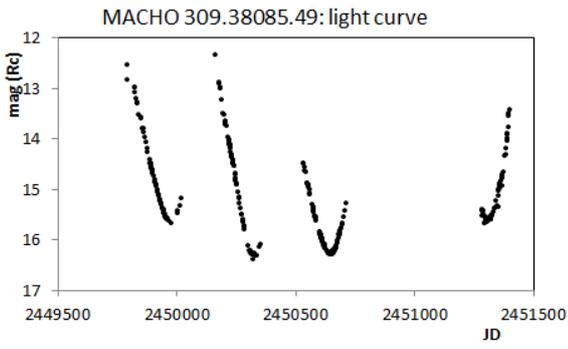
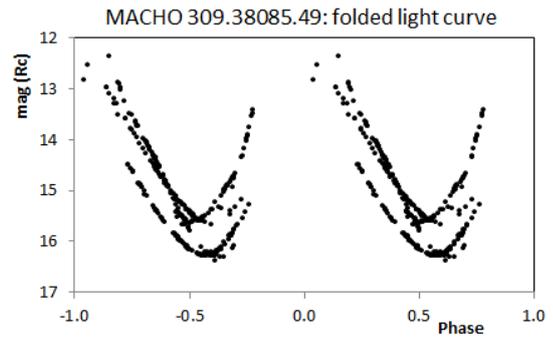
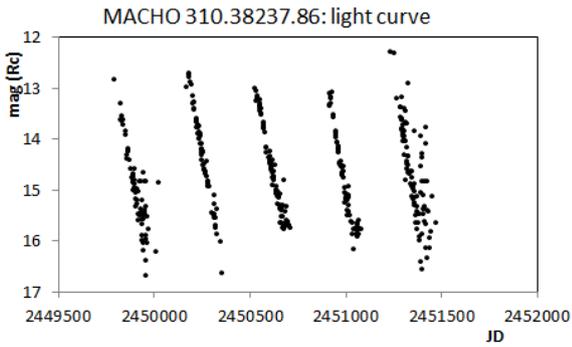
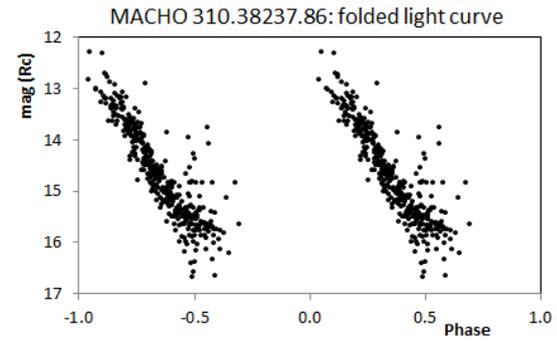
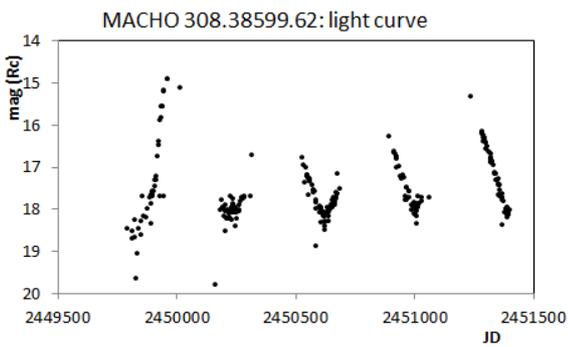
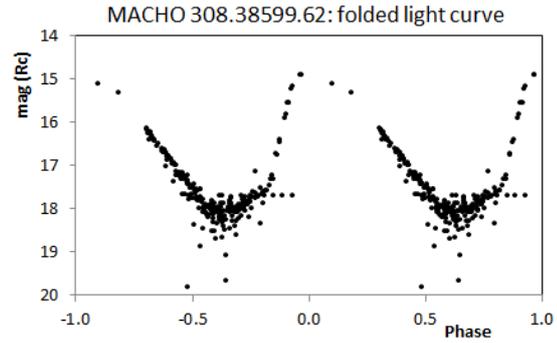
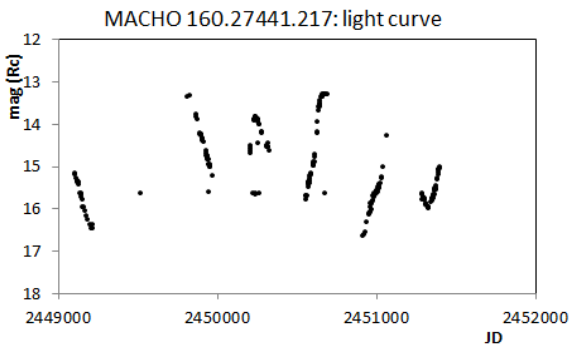
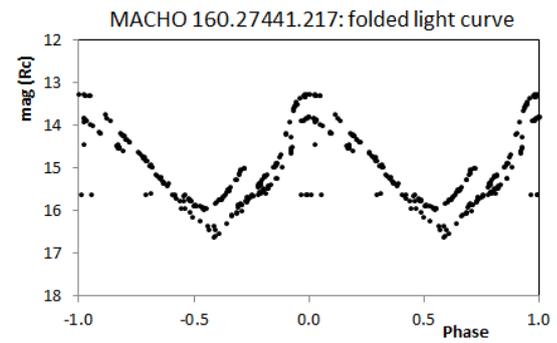
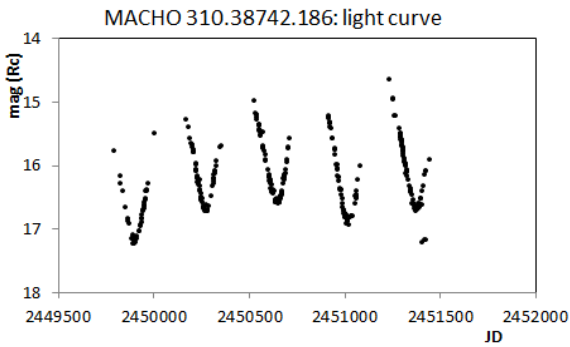
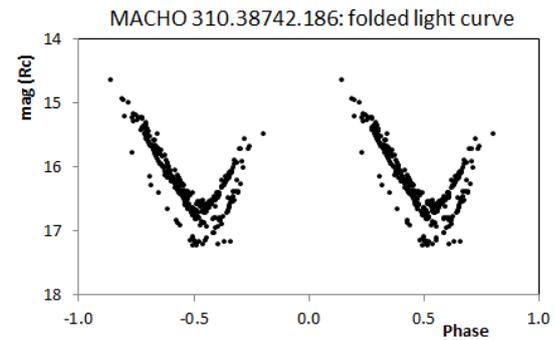

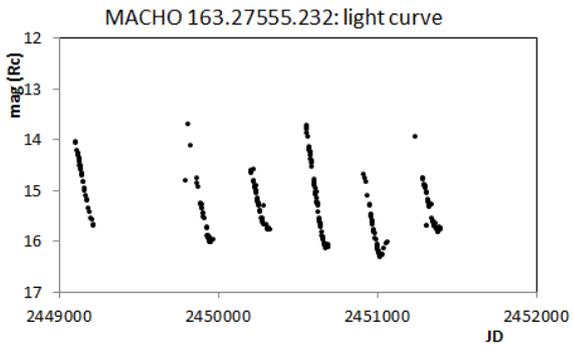
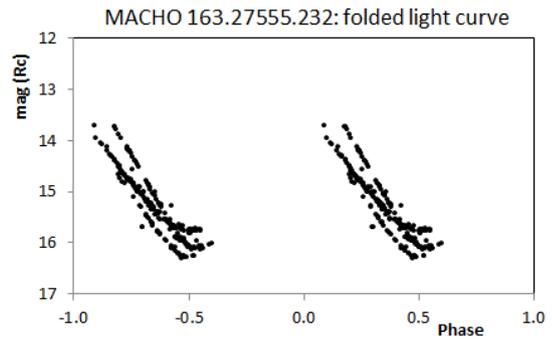
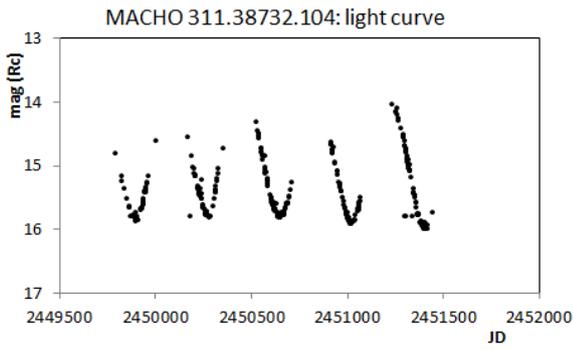
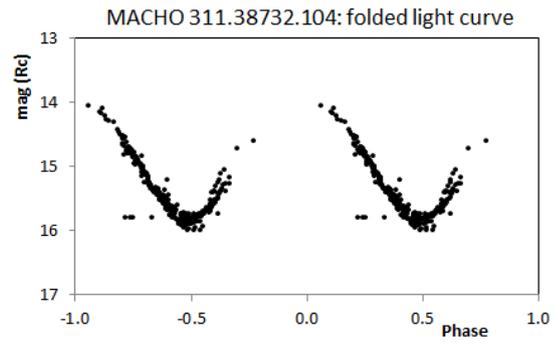
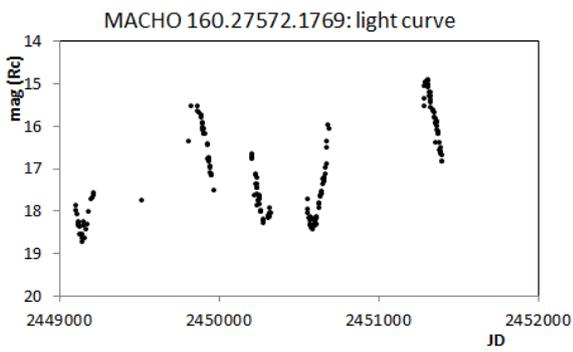
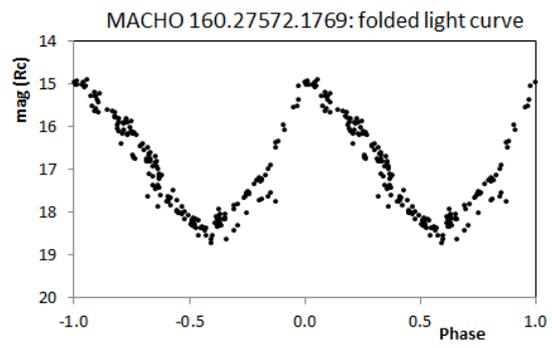
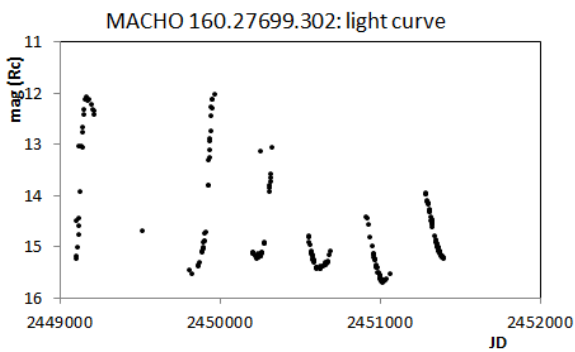
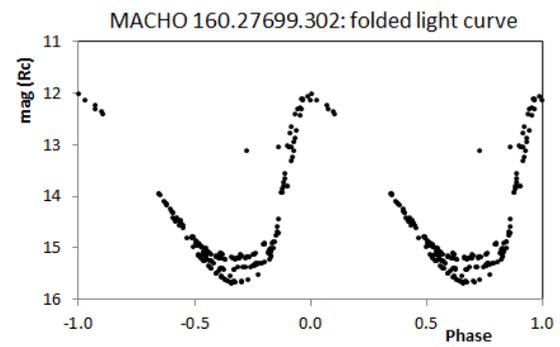
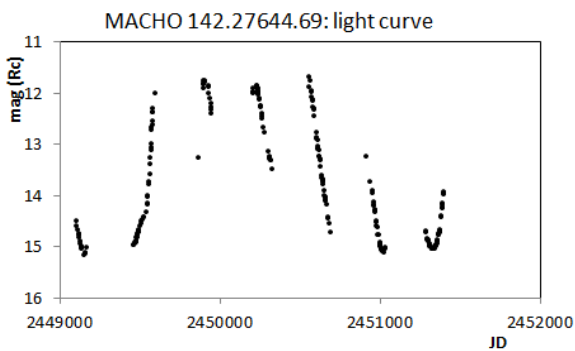
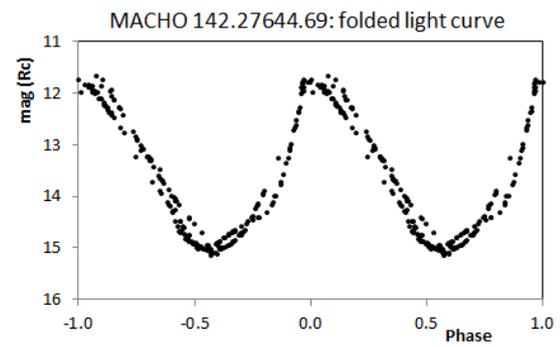

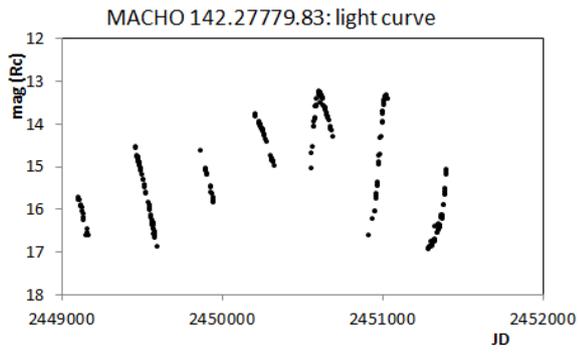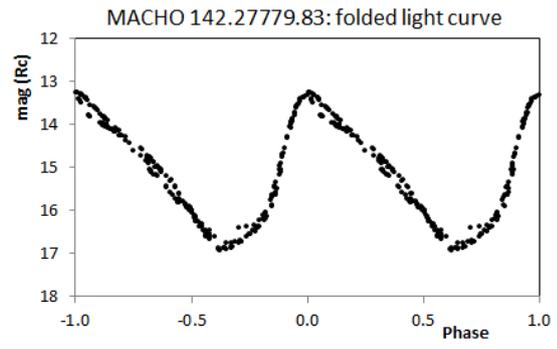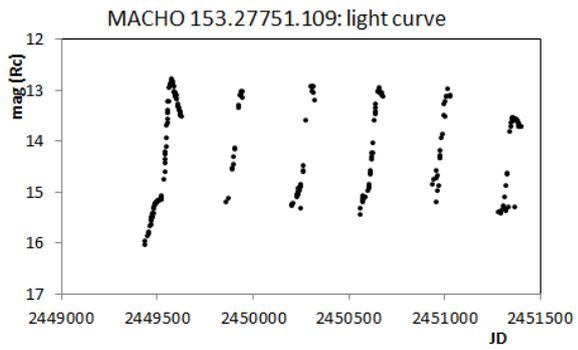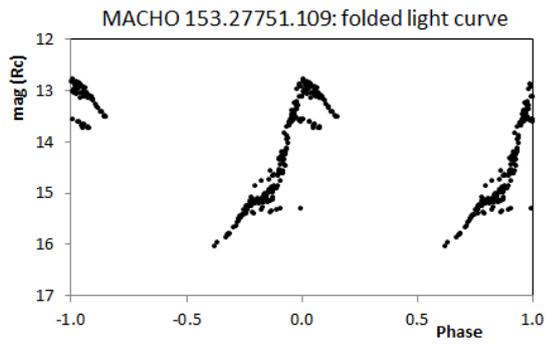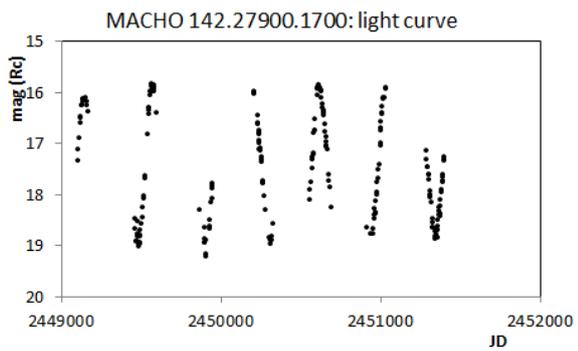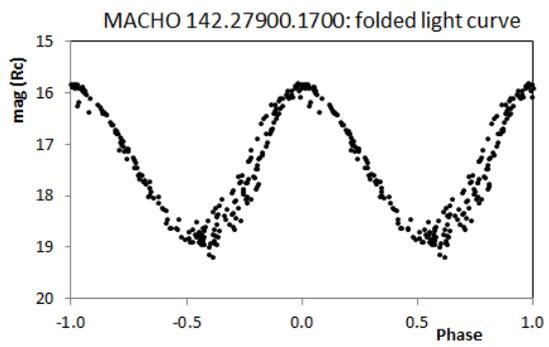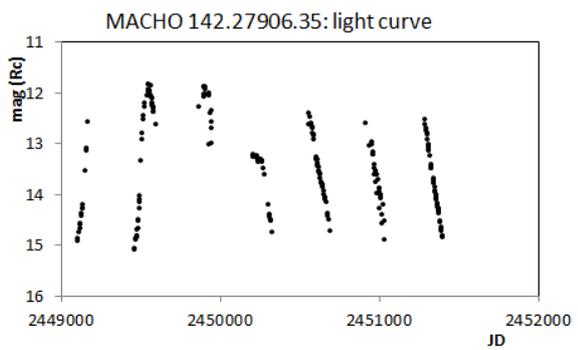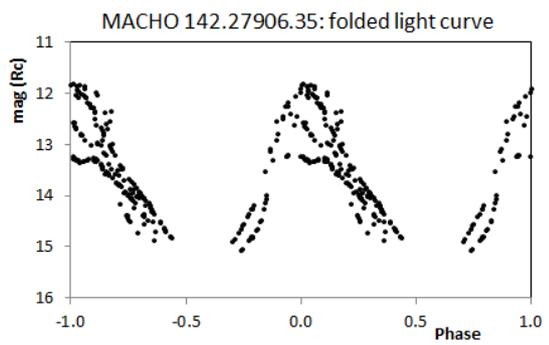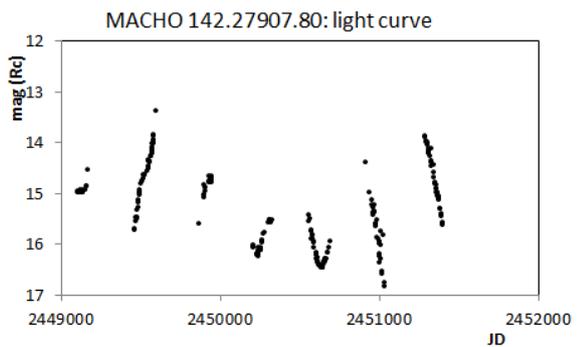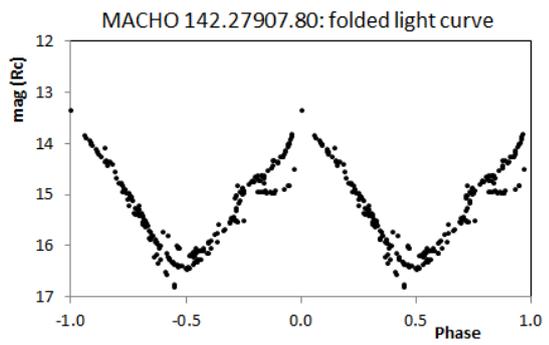

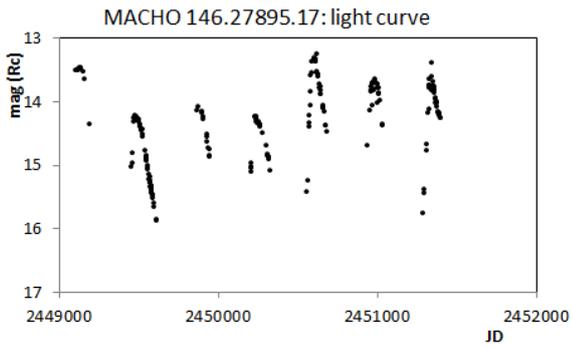
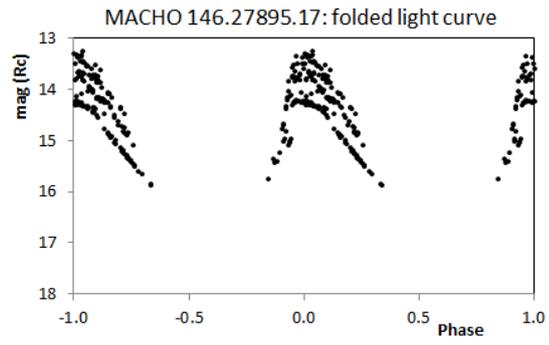
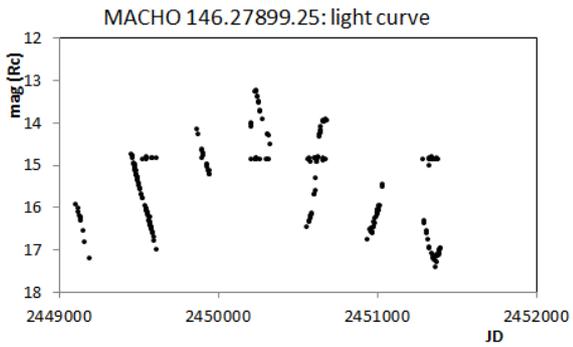
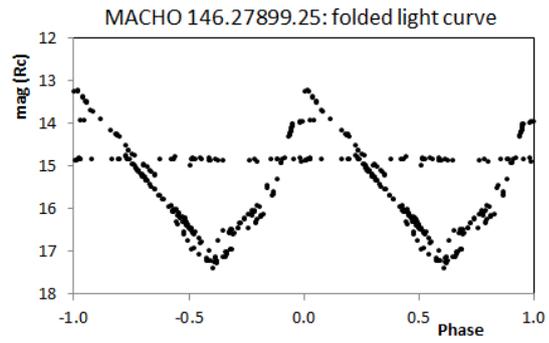
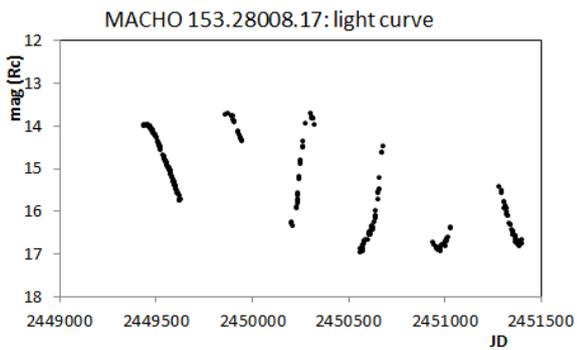
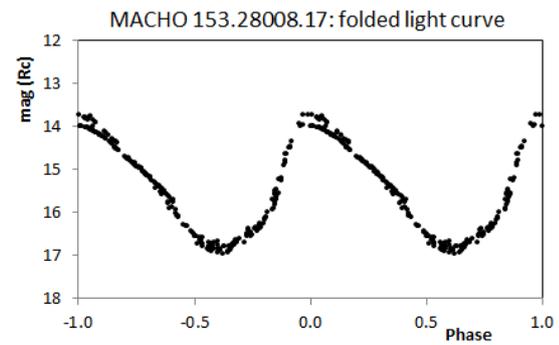
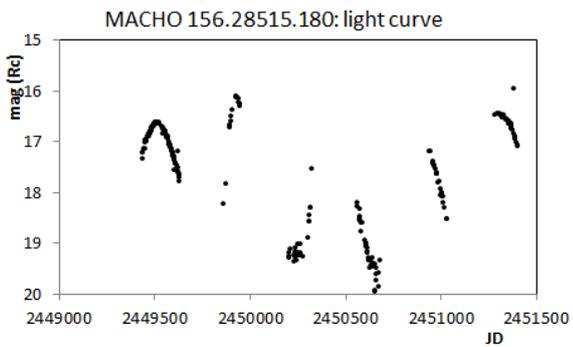
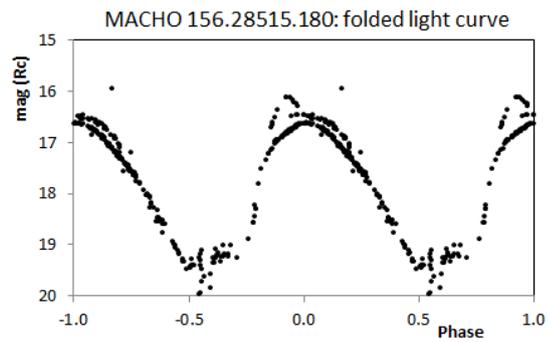
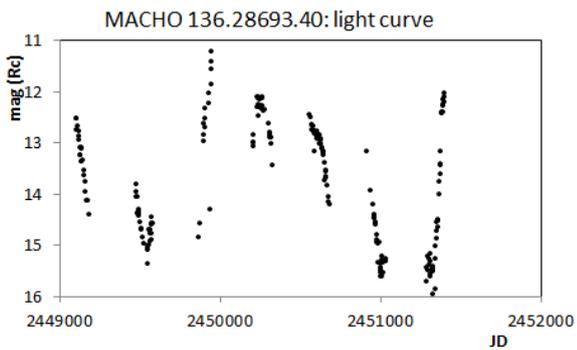
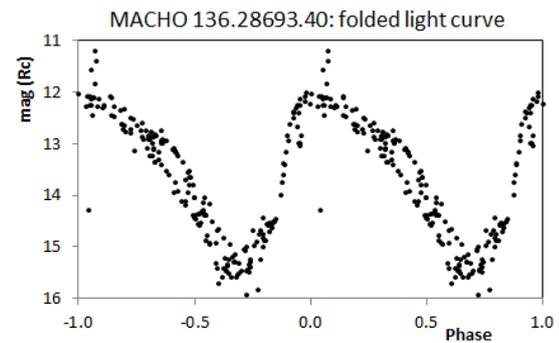

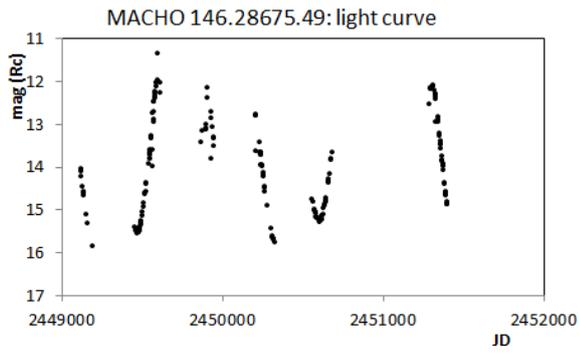
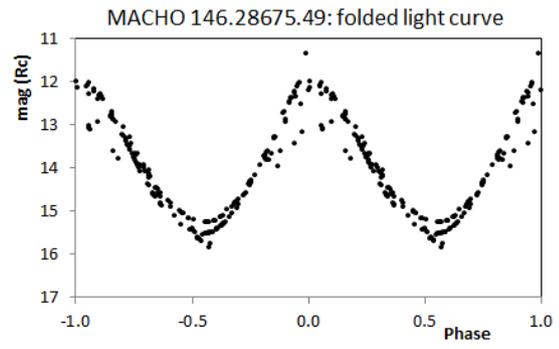
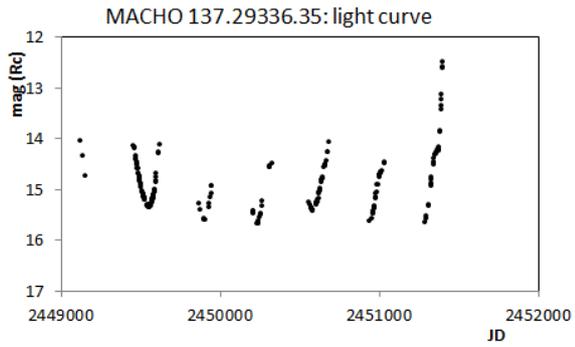
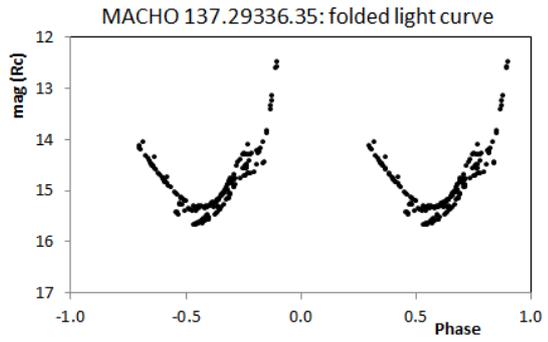
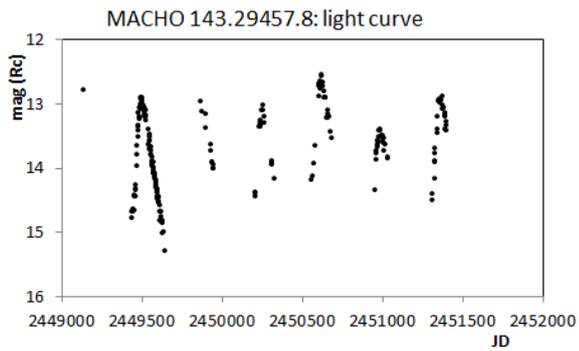
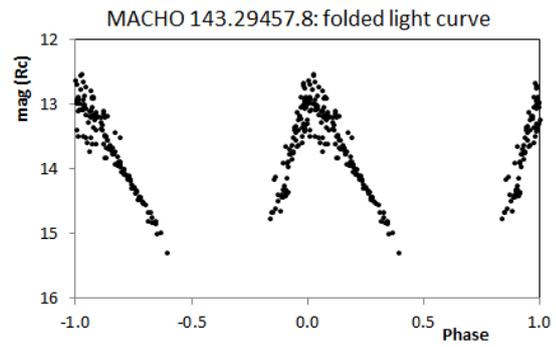
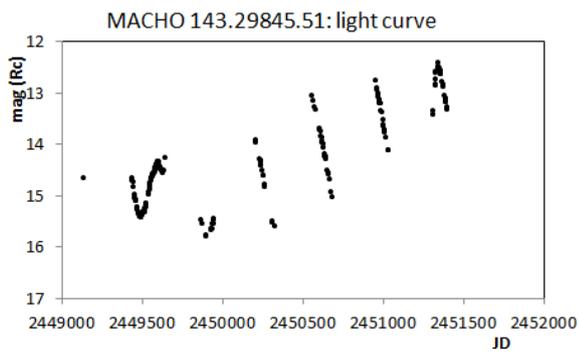
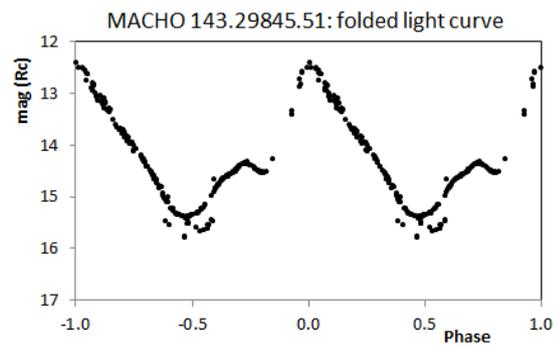
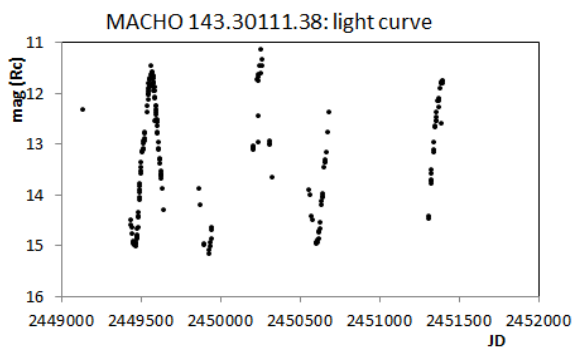
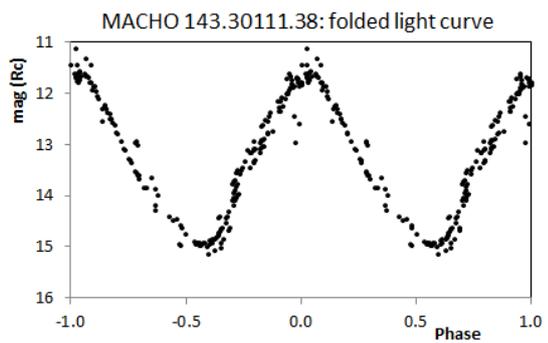

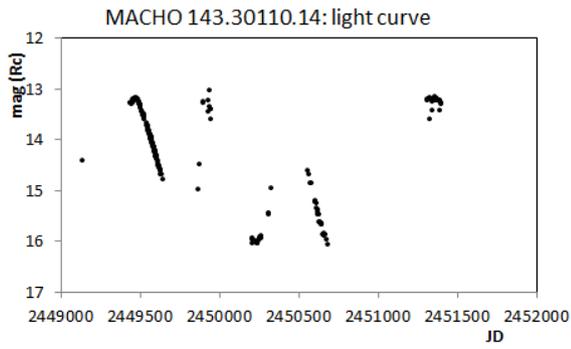
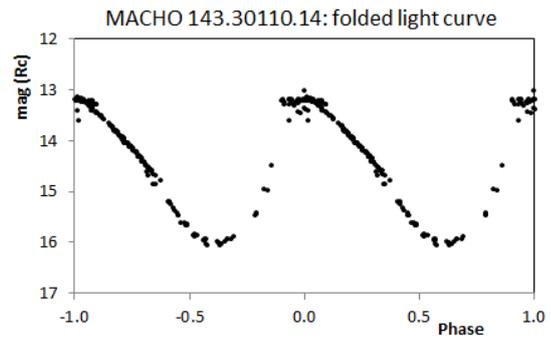
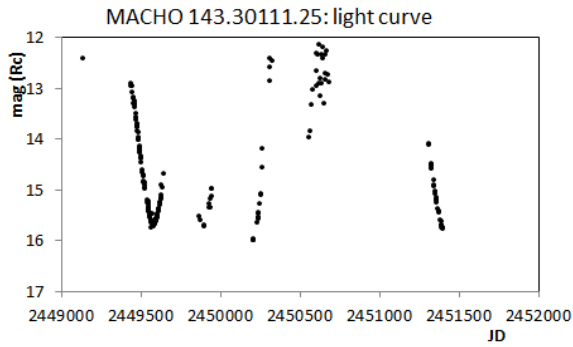
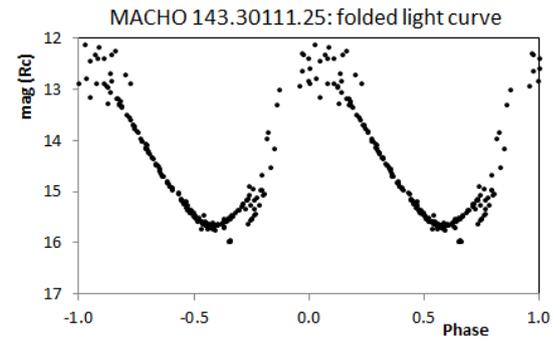
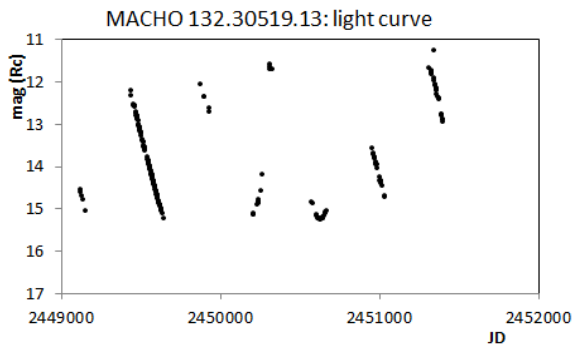
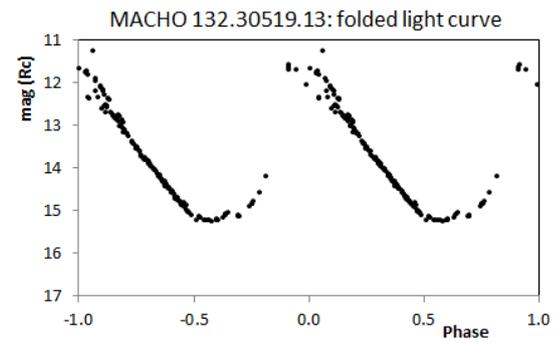
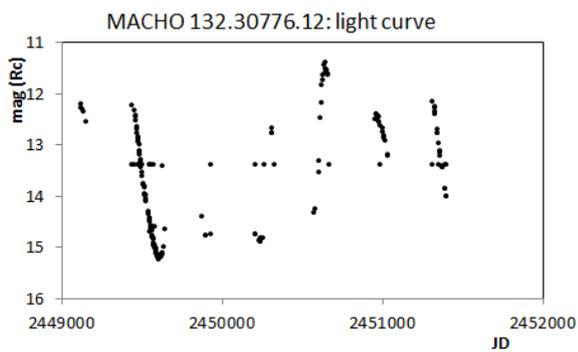
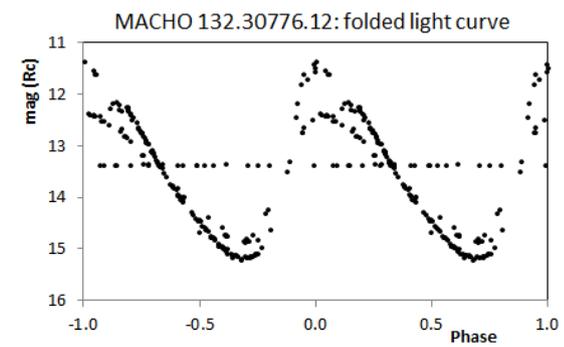
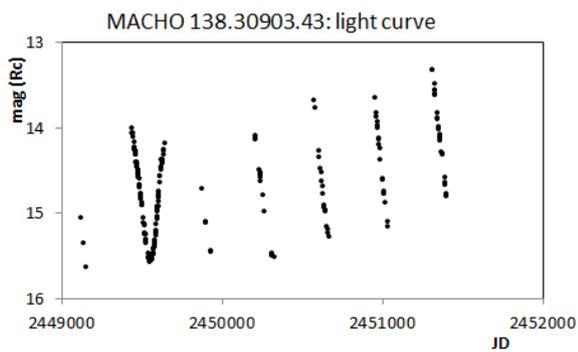
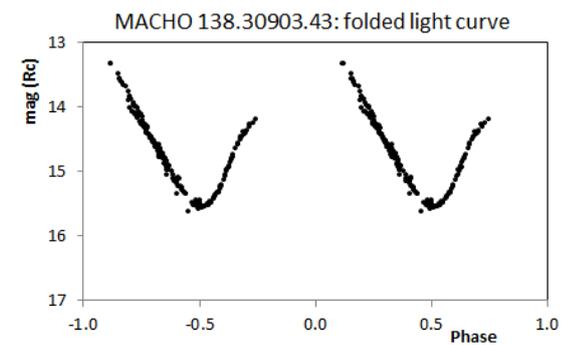

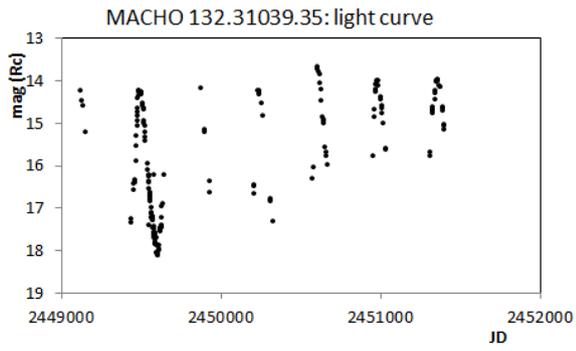
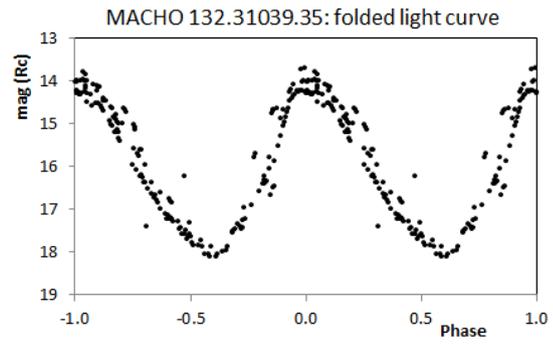
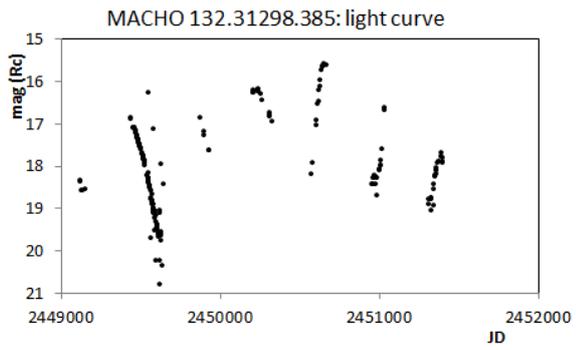
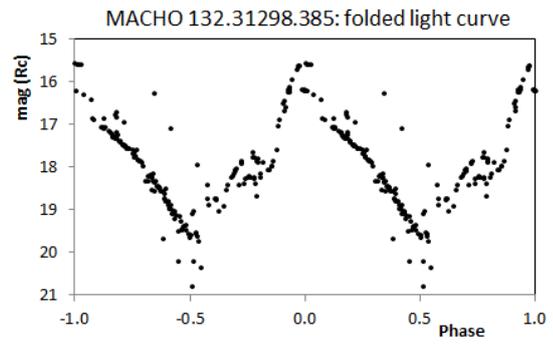
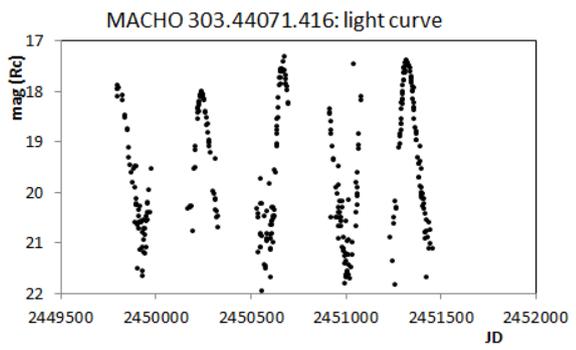
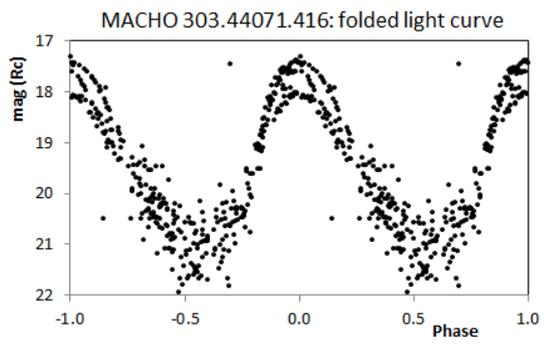
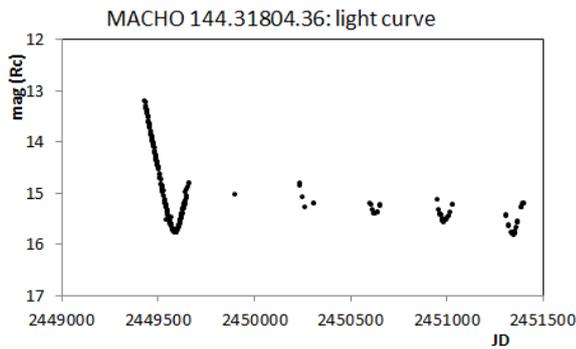
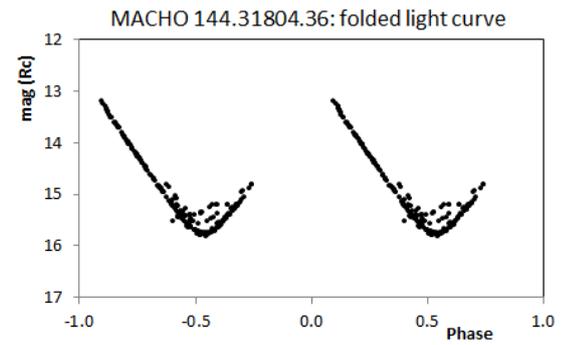
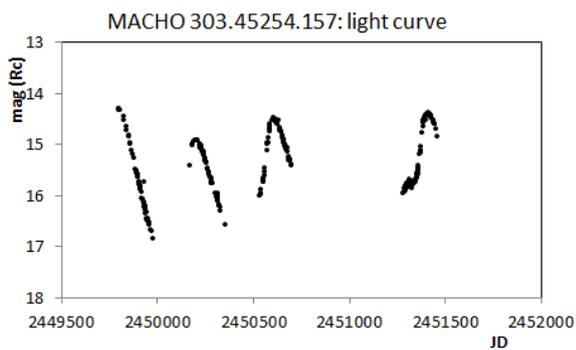
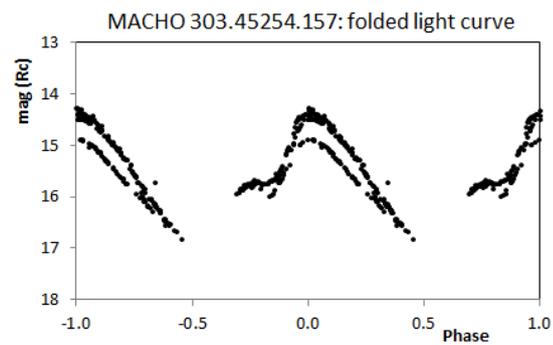

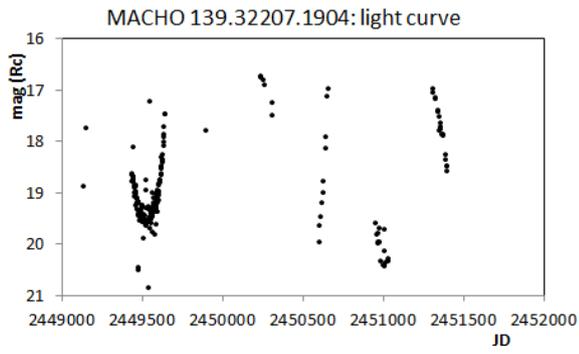
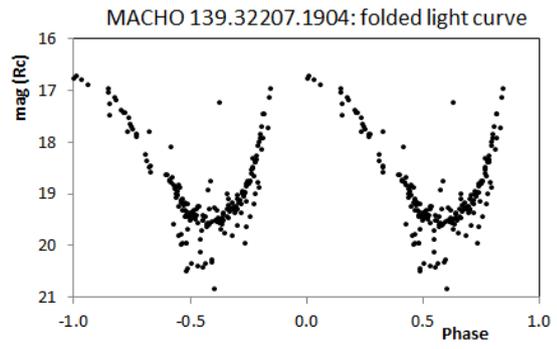
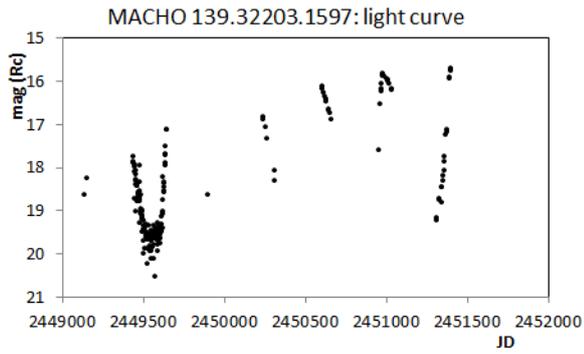
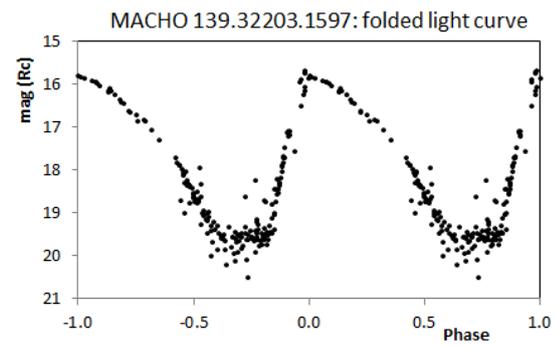
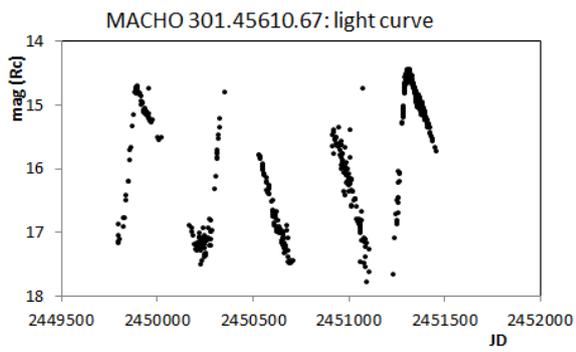
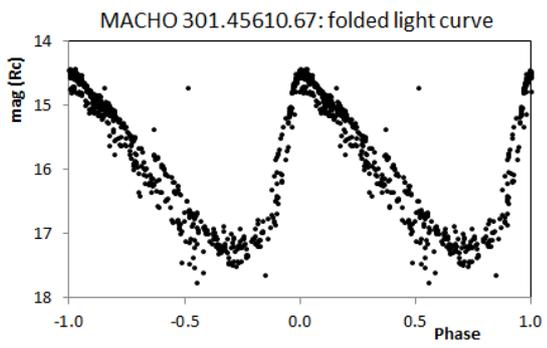
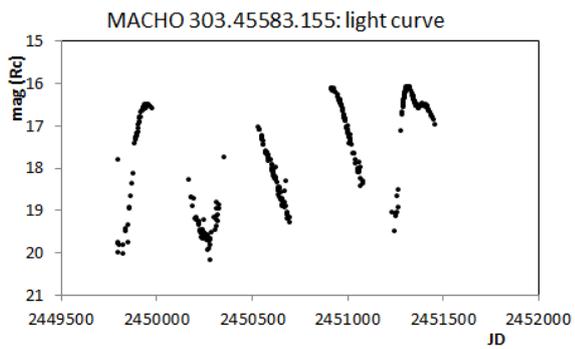
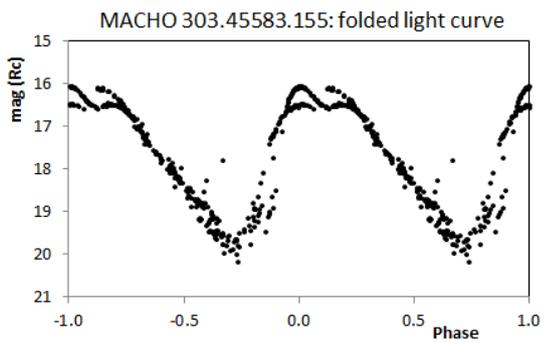
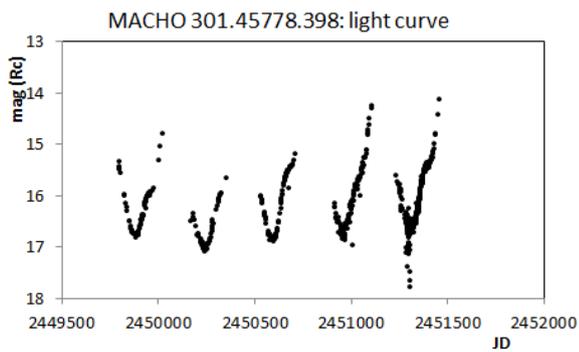
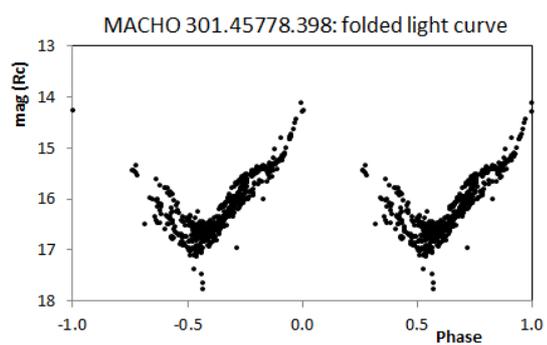

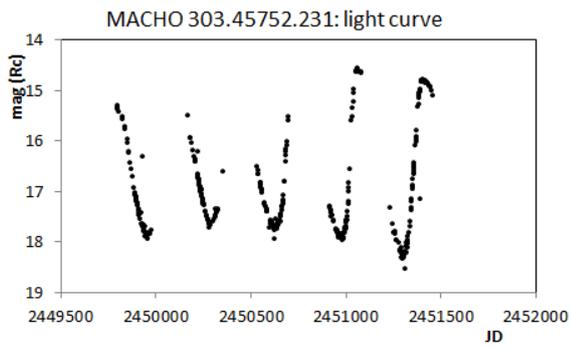
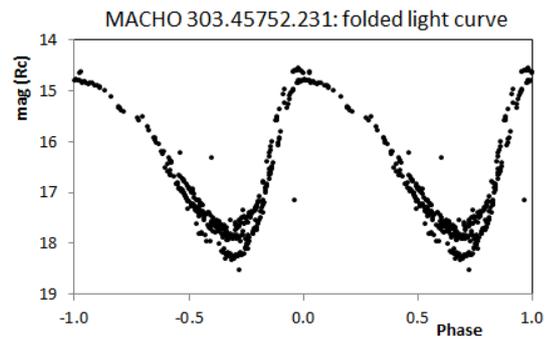
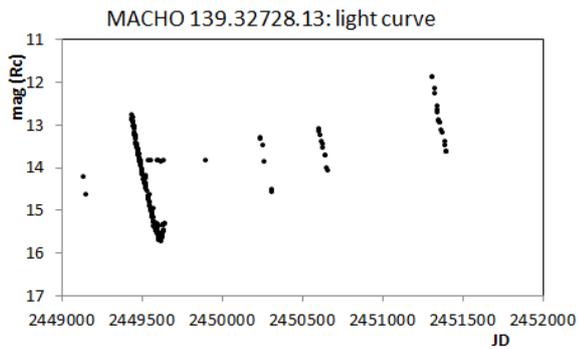
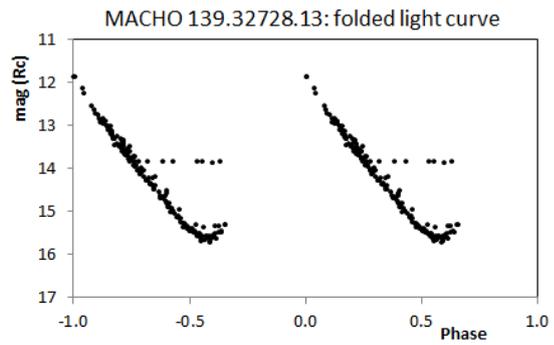
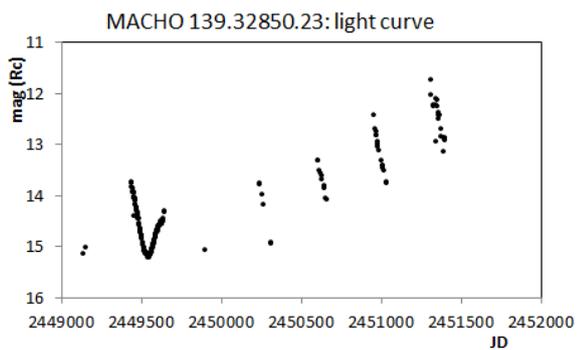
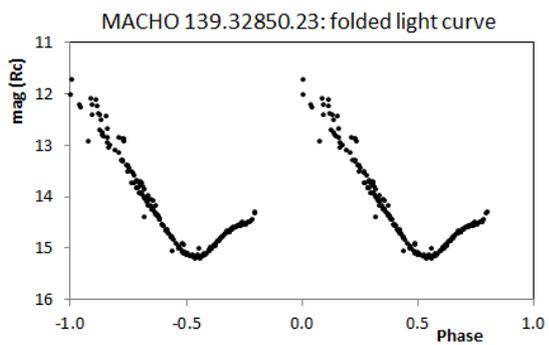
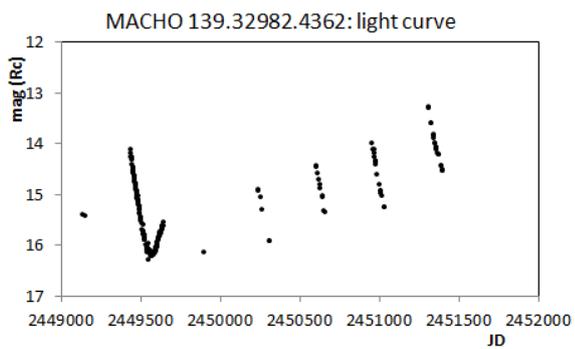
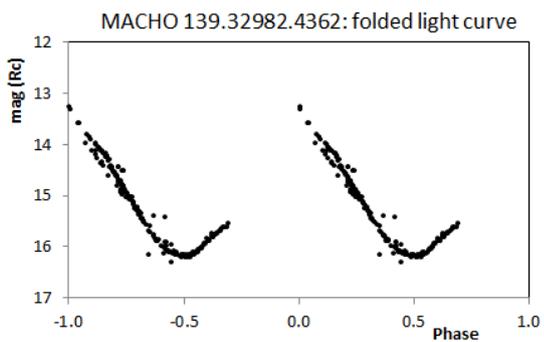
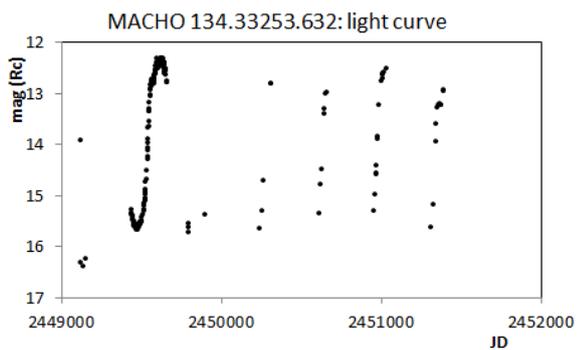
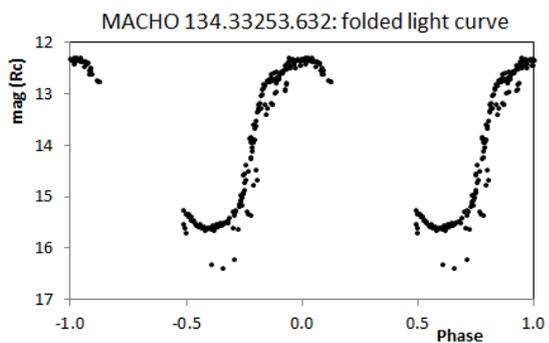

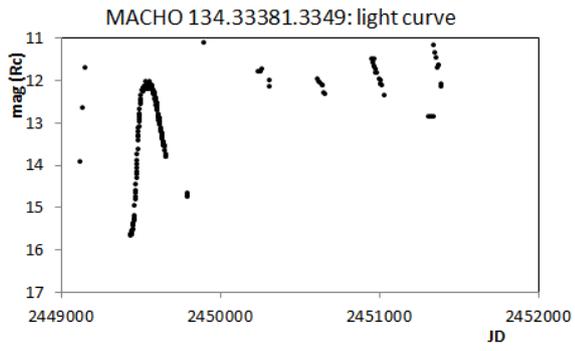
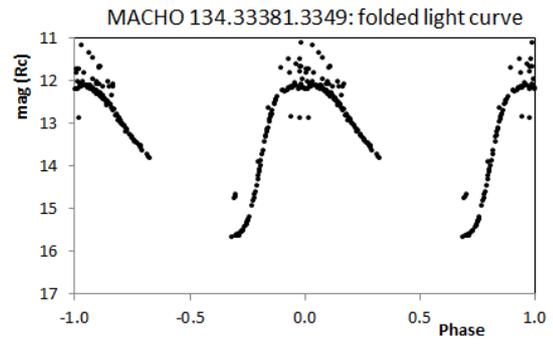
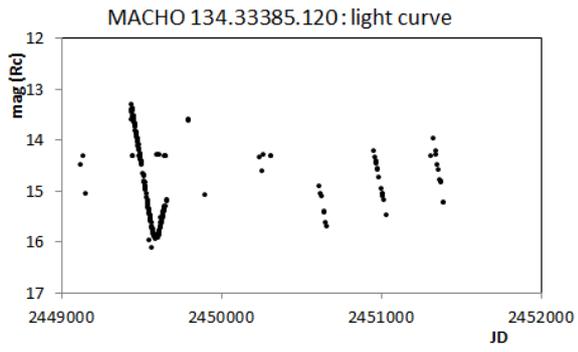
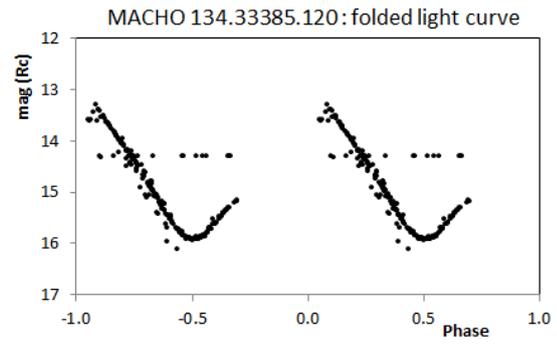